\documentclass[aps, pra, superscriptaddress, a4paper, showpacs, twocolumn, english, 10pt]{revtex4-2}

\usepackage{bbm, amsthm, bm,textcomp, nicefrac,geometry,ragged2e}
\usepackage[dvipsnames]{xcolor}
\usepackage{float}
\usepackage[bbgreekl]{mathbbol}
\usepackage{graphicx,epstopdf,color,verbatim,enumitem}
\usepackage{pbox,array}
\usepackage{mathrsfs}
\usepackage{amssymb}
\usepackage{textgreek}
\usepackage{mathtools}
\usepackage{stackrel}
\usepackage[thinlines]{easytable}
\usepackage{amsmath}
\usepackage[utf8]{inputenc}
\usepackage{tocloft}

\makeatletter
\usepackage{soul}
\usepackage[normalem]{ulem}

\usepackage{hyperref}
\usepackage[T1]{fontenc}
\usepackage[caption=false]{subfig}
\usepackage{babel}
\usepackage[linesnumbered,ruled]{algorithm2e}

\hypersetup{
    colorlinks = true,
    linkcolor = darkgray,
    citecolor = purple,
    filecolor = purple,      
    urlcolor = purple,
}

\newcommand{\ket}[1]{|#1\rangle}

\newcommand{\bra}[1]{\langle #1|}

\newcommand{\proj}[1]{\ket{#1}\!\bra{#1}}

\newcommand{\id}{\mathbb{1}}

\newcommand{\tr}{\text{Tr}}

\begin{document}

\title{
Enhancing Quantum Computation via Superposition of Quantum Gates
}
\author{Jorge Miguel-Ramiro}
    \email{Jorge.Miguel-Ramiro@uibk.ac.at}
    \thanks{this author contributed equally}
    \affiliation{Universit\"at Innsbruck, Institut f\"ur Theoretische Physik, Technikerstra{\ss}e 21a, 6020 Innsbruck, Austria}
\author{Zheng Shi}
    \thanks{This author contributed equally}
    \affiliation{Institute for Quantum Computing, University of Waterloo, Waterloo, ON N2L 3G1, Canada}
    \affiliation{Department of Physics \& Astronomy, University of Waterloo, Waterloo, ON N2L 3G1, Canada}
\author{Luca Dellantonio}
    \affiliation{Institute for Quantum Computing, University of Waterloo, Waterloo, ON N2L 3G1, Canada}
    \affiliation{Department of Physics \& Astronomy, University of Waterloo, Waterloo, ON N2L 3G1, Canada}
    \affiliation{Department of Physics and Astronomy, University of Exeter, Stocker Road, Exeter EX4 4QL, United Kingdom}
\author{Albie Chan}
    \affiliation{Institute for Quantum Computing, University of Waterloo, Waterloo, ON N2L 3G1, Canada}
    \affiliation{Department of Physics \& Astronomy, University of Waterloo, Waterloo, ON N2L 3G1, Canada}
\author{Christine A. Muschik}
    \affiliation{Institute for Quantum Computing, University of Waterloo, Waterloo, ON N2L 3G1, Canada}
    \affiliation{Department of Physics \& Astronomy, University of Waterloo, Waterloo, ON N2L 3G1, Canada}
    \affiliation{Perimeter Institute for Theoretical Physics, Waterloo, Ontario N2L 2Y5, Canada}
\author{Wolfgang D\"ur}
    \affiliation{Universit\"at Innsbruck, Institut f\"ur Theoretische Physik, Technikerstra{\ss}e 21a, 6020 Innsbruck, Austria}

\begin{abstract}
Overcoming the influence of noise and imperfections in quantum devices is one of the main challenges for viable quantum applications. In this article, we present different protocols, which we denote as ``superposed quantum error mitigation'', that enhance the fidelity of single gates or entire computations by performing them in coherent superposition. Our results demonstrate that via our methods, significant noise suppression can be achieved for most kinds of decoherence and standard experimental parameter regimes. Our protocols can be either deterministic, such that the outcome is never post-selected, or probabilistic, in which case the resulting state must be discarded unless a well-specified condition is met. By using sufficiently many resources and working under broad assumptions, our methods can yield the desired output state with unit fidelity. Finally, we analyze our approach for gate-based, measurement-based and interferometric-based models, demonstrating the applicability in all cases and investigating the fundamental mechanisms they rely upon.

\vspace{30pt}
\end{abstract}

\maketitle


%
\section{Introduction}
\label{sec:Introduction}

Quantum computing \cite{Steane1998,nielsen_chuang_2010} is one of the most remarkable applications of the emergent quantum technologies \cite{Preskill2018,national2019quantum,Baker2021}, capable of solving problems whose solutions are inaccessible with classical devices \cite{Gyongyosi2019}. Despite the variety of approaches towards quantum computation, both from the conceptual \cite{national2019quantum,Gyongy2019} and the experimental \cite{national2019quantum,OBrien2007} sides, decoherence and noise coming from imperfect apparatuses \cite{Preskill2018,Bharti2022} jeopardize the processes. Significant effort has been invested in minimizing or correcting several sources of noise, developing quantum error correction codes \cite{Steane96,Dur2007,Roffe19}, and fault-tolerant quantum computation \cite{Shor96,Raussendorf2012}. However, despite all these efforts, many practical applications \cite{Hassija2020,Bharti2022} are still out of reach for the current devices, and quantum supremacy has only been demonstrated \cite{Arute2019} for tailored problems with limited practical use.

A bottleneck encountered with quantum error correction codes is that they generally require excessive resources to be successfully employed in state-of-the-art apparatuses \cite{Bharti2022}. It is therefore of paramount importance to develop new techniques that can lower the detrimental effects of decoherence while maintaining low computational costs. 

In Ref.~\cite{prl_us} we design a family of protocols, denoted as ``superposed quantum error mitigation'' (SQEM), based on performing computations in coherent superposition. Ideas in the same spirit have been proved to be advantageous when performing superposition of paths or causal orders  \cite{Gisin_2005, Araujo2014, Procopio2015, Abbott2020, Chiribella2019, Caleffi2020, Chiribella2021, Rubino2021}. We achieve error mitigation by applying the desired computation in superposition, such that it either affects the  input or some auxiliary state. The superposition is generated with the assistance of a control register and auxiliary systems that become correlated with the input. A measurement of these registers collapses the state of the system and effectively leads to error mitigation. This allows for significant noise reduction for both single gates and whole computations. Here, we provide additional information, analytical results and numerical simulations supporting the results in Ref.~\cite{prl_us}. 

Specifically, we introduce basic implementations of our protocols for gate-based (GB-) \cite{nielsen_chuang_2010} and measurement-based (MB-) \cite{Briegel2001,Oneway2005,Briegel2009} quantum computation (QC). The former relies on the application of unitary operations (chosen from a set of elementary gates), while the latter processes highly entangled states (called resource states) via single qubit measurements. Examples of GB-QC include the standard quantum circuit model \cite{Elementary1995,nielsen_chuang_2010} and adiabatic quantum computation \cite{Albash2018}, whereas the most important instance of MB-QC is the one-way quantum computer \cite{Briegel2001,Briegel2009,Walther2005}.

We analyze the underlying mechanisms of our protocols, and introduce different extensions that allow for further enhancing the protocol performance by increasing the number of auxiliary states in the superposition. The user may specify a priori this number of auxiliary qubits for the correction process. A lower number generally reduces the precision improvement of the desired computation, yet may be required to overcome hardware limitations. Importantly, our techniques are designed to work with all quantum hardware and software currently under development \cite{Preskill2018}. We also consider an alternative interferometric implementation (IB-QC) of our schemes, where we make use of different effective paths to create the coherent superposition at the basis of the fidelity enhancement.

Our SQEM protocols reduce the effects of decoherence by carrying out the computation in a coherent superposed fashion.
In their simplest version, our protocols are probabilistic, meaning that the enhancement is achieved contingent on the outcome of a (set of) measurement(s). However, we also demonstrate how it is possible to combine our approach with optimized correcting operations to obtain a deterministic advantage.
The main results of this work, which complement Ref.~\cite{prl_us}, are:
\begin{itemize}
    \item We provide analytical and numerical demonstrations that noise mitigation can be achieved via our protocols in probabilistic and deterministic ways, for any GB- or MB-QC implementation, and when the additional resources are also noisy.
    \item We introduce an alternative approach, called ``nested SQEM'',  that maximizes error mitigation when many auxiliary subsystems are available. We derive asymptotic relations which ensure that, under broad assumptions, a desired unitary can be perfectly implemented on any input state with arbitrary noise.
    \item We introduce an interferometric-based (IB-QC) SQEM implementation that relies on other available degrees of freedom to substitute the auxiliary subsystems that are employed in the other SQEM protocols.
    \item We provide a detailed theoretical and numerical analysis for all SQEM implementations that we propose.
\end{itemize}

The paper is structured as follows. In Sec.~\ref{sec:Background} we review the concepts and tools that are employed throughout the work. We introduce the problem setting and the general idea of our SQEM schemes in Sec.~\ref{sec:Settings}. The GB-QC, MB-QC, and interferometric implementations are introduced in Sec.~\ref{sec:standardgatebased}, Sec.~\ref{sec:enhancedMBQC} and Sec.~\ref{sec:vacuum} respectively, supported with analytical and numerical performance results. In particular, we provide a detailed analytical protocol analysis in Sec.~\ref{sec:standardgatebased} for GB-QC, where we also analyze possible extensions, such as the nested protocol or the use for quantum memories. Most of the results and conclusions can be extended to the MB-QC setting. We summarize and conclude in Sec.~\ref{sec:conclusions}.

\section{Background}
\label{sec:Background}
In this section, we summarize the relevant elements that are required for the development of our protocols. These include the formalisms used to describe noise processes, the noise models considered, as well as a brief review of the related literature. 

\subsection{Noise channels and computational fidelity}
\label{sec:Noise_Fidelity}
We review and summarize the mathematical description of noise affecting a quantum state. This description will then serve to characterize the fidelity of a given computation, and eventually the improvement resulting from the application of our SQEM protocols. Importantly, while the noise description is crucial to understanding our methods and quantifying their effectiveness, on a real device one can adopt our approaches without any knowledge of the noise affecting the experimental apparatus.

\subsubsection{Operator sum representation and process matrix} \label{sec:operatorsum}
A mathematical description of decoherence can be formulated on the basis of the Stinespring theorem \cite{nielsen_chuang_2010}. The idea is that to accurately characterize the evolution of  an open quantum system, one must take into account its interaction with the so-called environment (i.e., another inaccessible system) that steers the computation away from the desired result. Therefore, the composite state is an element of $\mathcal{H}_{\rm s} \otimes \mathcal{H}_{\rm e}$, where $\mathcal{H}_{\rm s}$ and $\mathcal{H}_{\rm e}$ are the Hilbert spaces of the system of interest and the environment, respectively. One then assumes that at an initial time  the quantum state is pure and separable, $\rho_{\rm in} = \rho_{\rm in}^{\rm s} \otimes \rho_{\rm in}^{\rm e} = \rho_{\rm in}^{\rm s} \otimes \proj{e_{0}}$, where $\rho_{\rm in}^{\rm e} = \proj{e_{0}}$ is a generally unknown environmental state. 

Since the system and the environment together form a closed system, at a later time the composite density matrix $\rho_{\rm out}$ is obtained by applying a unitary operator $U_{\rm se}$ to $\rho_{\rm in}$, i.e. $\rho_{\rm out} = U_{\rm se} (\rho_{\rm in}^{\rm s} \otimes \rho_{\rm in}^{\rm e}) U_{\rm se}^{\dagger}$. Tracing out the environment, it is then possible to find
\begin{equation}
\label{eq:Krausrepresentation}
\begin{split}
    \rho^{\rm s}_{\rm out}  & = \tr_{\rm e} \left\lbrace U_{\rm se} (\rho_{\rm in}^{\rm s} \otimes \proj{e_{0}}) U_{\rm se}^{\dagger}\right\rbrace 
    \\ 
    & = 
    \sum_{i} \bra{e_{i}} U_{\rm se} (\rho_{\rm in}^{\rm s} \otimes \proj{e_{0}}) U_{\rm se}^{\dagger} \ket{e_{i}} 
    \\
    & = 
     \sum_{i} K_{i} U_{\rm s} \rho_{\rm in}^{\rm s} U_{\rm s}^{\dagger} K_{i}^{\dagger},\\
    \end{split}
\end{equation}
where ``$\tr_{\rm e}$'' indicates partial trace over subsystem ``${\rm e}$'', $\lbrace \ket{e_{i}} \rbrace$ is an orthonormal basis of the environment and $K_{i} U_{\rm s} =\bra{e_{i}} U_{\rm se} \ket{e_{0}}$. These operators $K_{i}$ acting on subsystem ``${\rm s}$'' are usually known as Kraus operators \cite{nielsen_chuang_2010} and fulfill the completeness relation $\sum_{i} K_{i}^{\dagger} K_{i}= \id$, with $\id$ being the identity operator. In this work, we indicate the map described in Eq.~\eqref{eq:Krausrepresentation} with ${\cal E}_{U_{\rm s}}$, such that $\rho^{\rm s}_{\rm out} = {\cal E}_{U_{\rm s}} \left( \rho^{\rm s}_{\rm in} \right)$.

This description  allows us to analyze any quantum channel as a unitary evolution in a larger Hilbert space, such that according to the Stinespring theorem \cite{nielsen_chuang_2010}
\begin{equation}\label{eq:Stinespring0}
\left|\psi\right\rangle \left|\epsilon_{0}\right\rangle _{\epsilon}  \rightarrow \sum_{j} U_{\rm se} K_{j}  \left|\psi\right\rangle \otimes\left|j\right\rangle _{\epsilon},
\end{equation}
where the subscript $\epsilon$ denotes the environment into which the information of the noise is leaked out during the evolution. Observe again how by tracing out this environment one recovers the Kraus description of the noise, Eq.~\eqref{eq:Krausrepresentation}. In this work, we call the description based on the Stinespring theorem the environmental description. In fact, as the name suggests, by explicitly including the environment we can characterize the system with a larger pure state as in Eq.~\eqref{eq:Stinespring0} rather than a density matrix.
For more information, see App.~\ref{sec:appendixMB-QC111}.

Equivalently to the Kraus decomposition in Eq.~\eqref{eq:Krausrepresentation}, it is possible to describe a quantum channel in the canonical representation, also known as \textit{process matrix} representation \cite{nielsen_chuang_2010}, i.e.
\begin{equation}\label{eq:generalmapintro}
    \rho^{\rm s}_{\rm out} = \sum_{i,j} \lambda_{ij} \sigma_{i} \big( U_{\rm s} \rho^{\rm s}_{\rm in}U_{\rm s}^{\dagger} \big) \sigma_{j}^{\dagger}.
\end{equation}
Here, $\sigma_{i}$ are, for all $i$ and an $m$-input state $\rho^{\rm s}_{\rm in}$, tensor products of $m$ Pauli operators $\lbrace \id, Z,X,Y \rbrace$. For clarity, in this work we associate with $i=0$ the identity $\sigma_0 = \id^{\otimes m}$, and we call $\lambda_{00} = p_{\rm ne}$ the probability not to have an error when the input state is a Bell state.

Any channel in the Kraus decomposition can be brought into the canonical representation [See Eq.~\eqref{eq:generalmapintro}] by writing each Kraus operator in the Pauli basis
\begin{equation}\label{eq:krausinpauli}
    K_{i}= \sum_j \alpha_{i,j}  \sigma_j,
\end{equation}
where $\sum_{i,j} |\alpha_{i,j}|^2=1$. We can then directly relate these coefficients to the process matrix coefficients, such that
\begin{equation}\label{eq:def_lambda_coef}
    \lambda_{mn}=\sum_j \alpha_{jm} \alpha^{*}_{jn}.
\end{equation}
Both noise representations are therefore equivalent; we sometimes employ the latter, as there are circumstances where it is simpler to work with Pauli matrices. 

\subsubsection{Computational fidelity}

This general mathematical formalism can now be applied to describe a noisy quantum computation acting on an input state $\rho_{\rm in}^{\rm s}$. For simplicity, let us consider the noiseless case first. Indicating with $U_{\rm s}$ the unitary characterizing the whole computation, owing to the absence of decoherence we can express $U_{\rm se}$ in Eq.~\eqref{eq:Krausrepresentation} as $U_{\rm se} = U_{\rm s} \otimes U_{\rm e}$, where $U_{\rm e}$ describes the free evolution of the environment. The Kraus operators $K_{i}$ then become $K_{i} = U_{\rm s} \alpha_{i}$, with $\alpha_{i}$ being complex numbers such that $\sum_i \lvert \alpha_{i}\rvert^2 = 1$. Therefore, we obtain $\rho_{0}^{\rm s}  = U_{\rm s} \rho_{\rm in}^{\rm s} U_{\rm s}^{\dagger}$, where the subscript ``$0$'' has been included to indicate the absence of decoherence. As expected, in the noiseless case the output state $\rho_{0}^{\rm s}$ of the computation corresponds to the state obtained by applying $U_{\rm s}$ to the initial state $\rho_{\rm in}^{\rm s}$. 

When the computation is noisy, on the other hand, one can generally not further simplify the output state $\rho_{\rm out}^{\rm s}$ in Eq.~\eqref{eq:Krausrepresentation}. However, it is still possible to quantify the decoherence through the state fidelity \cite{miszczak2008sub, jozsa1994fidelity}
\begin{equation}
\label{eq:fid}
    F = \tr\left\lbrace \left(\sqrt{\rho_{0}^{\rm s}} \rho_{\rm out}^{\rm s} \sqrt{\rho_{0}^{\rm s}}\right)^{\frac{1}{2}} \right\rbrace ^2,
\end{equation}
which yields one if and only if the noise does not affect the computation. In the case that the input state is pure, $\rho_{\rm in}^{\rm s} = \proj{\psi_{\rm in}}_{\rm s}$, the ideal output can be expressed as $\ket{\psi_{\rm out}}_{\rm s} = U_{\rm s} \ket{\psi_{\rm in}}_{\rm s}$ and Eq.~\eqref{eq:fid} becomes 
\begin{equation}\label{eq:fid_simple}
    F = \left\langle \psi_{\rm out} \right| \rho_{\rm out}^{\rm s} \left|\psi_{\rm out}\right\rangle_{\rm s},
    \end{equation}
where we recall that $\rho_{\rm out}^{\rm s} = {\cal E}_{U_{\rm s}} (\proj{\psi_{\rm in}})$ is given by the action of the unitary $U_{\rm s}$ and noise acting on $\rho_{\rm in}^{\rm s}$, as described in Sec.~\ref{sec:operatorsum}. Observe that, when the quantum channel is described in terms of the process matrix of Eq.~\eqref{eq:generalmapintro}, the state fidelity is lower bounded by $F=\lambda_{00}$. In the following we omit the label ``$\rm s$'' for clarity.

\subsection{Quantum computing implementations}
\label{sec:quantum_implem}
Here, we review the different implementations considered for performing the quantum computations. We elaborate on the noise model associated with each of them. 

\subsubsection{Gate-based quantum computation (GB-QC)}
\label{sec:gb_quantum_comp_intro}
Quantum gates are unitary operations acting as building blocks for quantum circuits. They carry out arbitrary computations on a set of input qubits and are at the basis of GB-QC \cite{nielsen_chuang_2010}. To date, however, experimental apparatuses suffer from decoherence \cite{Bharti2022}, such that the desired effect of a given quantum gate is never perfectly matched in practice. Therefore, an accurate characterization of our system (in this case a quantum computer) is often out of reach, since the microscopic source of the decoherence is inaccessible.

\textbf{Noise model.---} In standard gate-based (GB) quantum computation we consider an error model consisting of an ideal application of each quantum gate followed by noise acting on each of the qubits involved. In other words, given a circuit made of $k$ quantum gates, each of them implementing a unitary operation $U_{i}$ with $i=\{1,\dots,k\}$, the noisy implementation of the computation $U = \prod_{i=1}^k U_i$ is given by
\begin{equation}\label{eq:gen_noise_gate}
    {\cal E}_U \left( {\rho_{\rm in}} \right) =  \bigcirc_{i=1}^k \left[ {\cal E}_{U_{i}}  \left( {\rho_{\rm in}}  \right) \right],
\end{equation}
where $\bigcirc$ indicates concatenation of the maps therein and ${\cal E}_{U_{i}} (\rho) = \sum_j K_j \big( U_{i} \rho U^{\dagger}_{i} \big) K_j^{\dagger}$ is the map associated with the imperfect implementation of $U_{i}$, as in Eq.~\eqref{eq:Krausrepresentation}. 

Alternatively, in our analyses, we also model the circuit noise as the ideal implementation of the whole circuit followed by the noise. Observe that both approaches are equivalent and can be mapped to each other by finding the corresponding relations between the Kraus operators.

While our protocols work with arbitrary noise of the forms in Eqs.~\eqref{eq:Krausrepresentation} and \eqref{eq:generalmapintro}, for the results presented in Secs.~\ref{sec:standardgateperformance}, \ref{sec:mbqcperformance} and \ref{sec:resultsinterfero} we focus on dephasing and depolarizing channels ${\cal E}_U$. For these, in the case of single qubit unitaries $U$ (with $m=1$) the Kraus operators are given by
\begin{subequations}
\label{eq:considered_noises}
\begin{align}
    & \text{Dephasing: } K_0=\sqrt{p_0} \id,  K_1=\sqrt{1-p_0} {Z}
    ,   \label{eq:dephasingnoise}\\
    & \text{Depolarizing: }  K_0=\sqrt{p_0} \id,  K_{i}=\sqrt{\frac{1-p_0}{3}} {\sigma_i},  \label{eq:depolarizingnoise}
\end{align}
\end{subequations}
where $\sigma_{i}$ is the Pauli operator ${X}$, ${Y}$, or ${Z}$ for $i=1,2,3$, respectively, and $p_0$ is the probability of not having an error. The corresponding channels when considering more qubits $m \geq 2$ can then be constructed by taking all tensor products of the possible permutations of $m$ operators $K_i$ (one per qubit) in Eqs.~\eqref{eq:considered_noises}. For instance, with $m=2$ and considering the dephasing channel, one can construct the map with the Kraus operators $K_0 = \sqrt{p_0^2} \id \otimes \id$, $K_1 = \sqrt{p_0 (1-p_0)} \id \otimes Z$, $K_2 = \sqrt{(1-p_0)p_0} Z\otimes\id$, $K_3 = \sqrt{(1-p_0)^2} Z\otimes Z$. We remark that, for a given value of $m$, the probability $p_{\rm ne}$ not having an error becomes $p_0^m$.

\subsubsection{Measurement-based quantum computation (MB-QC)}
\label{sec:MB-QC}
Measurement based quantum computation (MB-QC) is an alternative model for quantum computing that relies on entanglement as a resource for carrying out the computation. Its best-known implementation is the ``one-way quantum computer'', which relies on single-qubit measurements \cite{Briegel2001,Oneway2005,Briegel2009} for modifying a given input state. 

MB-QC makes use of graph states \cite{hein2006entanglement} as a resource. Graph states are multi-qubit quantum states that are stabilized \cite{Gottesman1997} by the elements of the Pauli group, i.e. they are eigenstates with $+ 1$ eigenvalues of the operators ${S}_n={X}_n\prod_{k \in \mathcal{N}_{n}}{Z}_k$, where $\mathcal{N}_{n}$ represents the neighborhood of qubit $n$. Here, subscript ``$n$'' (``$k$'') indicates that the corresponding Pauli operator acts on the $n$-th ($k$-th) qubit. Graph states exhibit correlations corresponding to classical graphs \cite{hein2006entanglement}, and can be represented by a graph $G = (V, E)$, where $V$ is the set of vertices and $E$ to the set of edges of the graph. 

The so-called $2$D-cluster state $\ket{G}$ \cite{Briegel2d}, on which one-way quantum computers are based, is a highly entangled graph state that allows for universal computations \cite{Briegel2001}. A way to visualize this state is given by placing $N$ qubits on a 2D lattice, individually initializing them in $\ket{+} = (\ket{0} + \ket{1})/\sqrt{2}$, and applying $\text{CZ}_{ij}= \proj{0}_{i} \otimes \id_{j} + \proj{1}_{i} \otimes Z_{j}$ to all pairs $i,j$ that are connected by an edge, i.e.,
\begin{equation}\label{eq:resource_state_MBQC}
    \ket{G}= \prod_{\{i,j\} \in E}  \text{CZ}_{ij} \ket{+}^{\otimes N}.
\end{equation}
In practice, there are several approaches for building cluster states that do not rely on the application of $\text{CZ}_{ij}$ gates. Photonic ones, for instance, can make use of parametric down conversion to produce several thousands of entangled photons  \cite{Schwartz2016,Larsen2019}.

To carry out a desired computation, input qubits are encoded into the leftmost qubits of the cluster state.  Quantum gates are implemented by performing local measurements on ancilla qubits \footnote{Throughout this paper, we reserve the word ``ancilla'' for the qubits in the resource states that are measured to perform the MB-QC.}, either in the eigenbasis of the Pauli operators ${X}$, ${Y}$, ${Z}$, or in the rotated basis $R(\theta) \equiv \{(\ket{0} \pm e^{i\theta}\ket{1}) / \sqrt{2}\}$. Depending on the measurement outcomes, the system is probabilistically projected onto different states. Byproduct operators and adaptive measurements are in general required in order to make the computation deterministic. It is noteworthy that in MB-QC all non-adaptive measurements (i.e., the Clifford part \cite{GottesmanClifford} of the corresponding circuit, corresponding to all measurements in the ${X}$, ${Y}$, and ${Z}$ bases) can be simultaneously performed at the beginning of the calculation, or even simulated efficiently classically \cite{Aaronson2004}. This often results in fewer steps required to carry out a desired computation, and consequently less time for the system to decohere. We refer to Ref.~\cite{Briegel2002} for further details.

In MB-QC the main source of noise is the imperfect preparation of $2$D cluster states and the imperfect single qubit measurements \cite{Usher2017}. Errors do not affect the computational level directly. They affect the output state in a highly non-trivial way, depending on the size of the resource state, the local measurements performed and their outcomes. Different kinds of noise have vastly different effects on the outcome of the computation. For instance, 1D MB-QC is resilient against bit-flips that leave the output state unaffected \cite{Usher2017}. 

An important remark is that, given any MB computation, one can always describe the noise with the map in Eq.~\eqref{eq:Krausrepresentation} acting on the input qubit. However, even when it is possible to describe the noise affecting the preparation and measurement of the MB resource state in terms of dephasing or depolarizing errors, the Kraus operators in Eq.~\eqref{eq:Krausrepresentation} become highly non-trivial. They generally depend on the properties of the considered computation, such as the size, the measurements performed on the cluster state and their outcomes.

\textbf{Noise model.---} Here we describe the noise model employed in our work when considering MB-QC. After the resource state is built but before any measurement is performed, we assume that each qubit is affected by local noise, i.e.,
\begin{equation}
    \rho_{G}= \prod_{j=1}^N  {\cal E}_j (\proj{G}),
\end{equation}
where $\ket{G}$ is the $N$-qubit cluster state defined in Eq.~\eqref{eq:resource_state_MBQC} and the map ${\cal E}_j$ describes an arbitrary noisy channel $ {\cal E}_j=\sum_i K_i \rho^j K_i^{\dagger}$ acting on the $j$-th qubit, see Eq.~\eqref{eq:Krausrepresentation}. As in GB-QC, the $K_i$ affecting the resource state in MB-QC depend on the specific noise. While our protocols work for all kinds of noise, below we mainly focus on dephasing and depolarizing channels, as in Eqs.~\eqref{eq:dephasingnoise} and \eqref{eq:depolarizingnoise}, respectively. Our formalism is suitable for describing noise contributions arising from both imperfect state preparation and imperfect measurements. This is done by redefining the Kraus operators.

In the purified version treatment of an MB-QC process based on the Stinespring theorem (see Sec.~\ref{sec:operatorsum} and App.~\ref{sec:appendixMB-QC111}), which we denote as the environmental formalism, we assign an environmental subsystem to each cluster state, such that we can analyze the process based on the unitary evolution defined by the global Kraus operators of the form $K_{i}= \bigotimes_{j} K^{(j)}_{i}$, with $i \in \{0,...,r^N \}$ where $r$ is the number of Kraus operators affecting each qubit and $N$ the number of cluster qubits.

\subsubsection{Interferometric-based quantum computing (IB-QC)}
\label{sec:interferometric_intro}
Besides GB- and MB-QC, we also consider an alternative ``interferometric'' approach (IB-QC) where a computation is applied to a system of interest depending on an extra physical degree of freedom \cite{Friis14}. While this degree of freedom is arbitrary, a simple way to think at this interferometric picture is via superposed branches, i.e., a system is routed towards different trajectories simultaneously (in a quantum superposition fashion), and then recombined afterward. Within each of these paths an identical computation occurs, which can be implemented in either GB or MB fashion. Similar approaches have been investigated, both theoretically and experimentally, in recent works. These include superposition of either orders of gates \cite{Ebler2018,Guerin2019} or trajectories \cite{Chiribella2019,Kristjnsson_2020,Abbott2020}.

What makes IB-QC interesting is that when the state is sent into several branches and then recombined, it interferes with the vacuum, rather than an auxiliary subsystem. As discussed in more details in Sec.~\ref{sec:vacuum} and Ref.~\cite{papercomm}, this allows for distinguishing different errors that have occurred, and consequently correcting for them. Importantly, besides the physical degree of freedom encoding the ``which path'' information, the IB-QC implementation of our SQEM protocols does not require auxiliary states (as in the GB and MB versions --- see Secs.~\ref{sec:standardgatebased} and \ref{sec:enhancedMBQC}). To experimentally achieve IB-QC, moreover, there are several possible approaches including photonic and ions \cite{Friis14} as well as superconducting qubits \cite{Friis2015}. 

\textbf{Noise model.---} The noise model considered in IB-QC depends on how the computation is practically carried out in each branch. For concreteness, we assume to use GB-QC, such that the noise in each branch is modeled as in Sec.~\ref{sec:gb_quantum_comp_intro}. Specifically, a perfect implementation of each gate is applied, followed by a noise channel characterized by Kraus operators $\{K_i\}$. In the purified analysis of this strategy based on the Stinespring theorem (see Sec.~\ref{sec:operatorsum}), we assign an environmental system to each branch of the superposition; in this case, the initial state of the environmental system has a physical meaning on account of the interference with vacuum. For more information, see App.~\ref{sec:MB-QCappendix2}.

\subsection{Relation to previous work}
SQEM protocols for standard GB- and MB-QC rely on implementing a desired computation $U$ via a superposition of different branches, where a branch is defined as each of the orthogonal elements or combinations in the superposition. This is accomplished by exploiting the interference between errors occurring on the input and some auxiliary states that are first mixed together and then separated again after the computation. To achieve this, we develop the basic tools that we previously introduced in Refs.~\cite{MiguelRamiro2021, MR2021} and introduce novel ones. With the details described in the following sections, here we briefly summarize relevant works in the literature that share the idea of employing more resources (e.g. number of qubits) to enhance the fidelity of a desired operation.

In contrast to other error mitigation approaches in Refs. \cite{Temme2017,Huggins2021,Koczor2021}, SQEM is not restricted to the estimation of expectation values. Instead, it provided an output state with enhanced fidelity  by exploiting noise interference and keeping the state. Moreover, SQEM only requires a single copy of the input state and is resilient against noise affecting the additional operations required for its operation.

Superposition has been analyzed and exploited in different quantum communication and computing contexts. Applied to causal orders, it has been proved to be advantageous in multiple scenarios, including superposition of orders of channels, also known as 
the quantum switch \cite{Ebler2018,Guerin2019,Caleffi2020,Guo2020,Chiribella2021}, or orders of quantum gates \cite{Procopio2015}. Superposition of paths and trajectories have been analyzed from theoretical \cite{Chiribella2019,Kristjnsson_2020,Abbott2020} and experimental \cite{Rubino2021} points of view, where noise mitigation has been shown. One of our protocols, namely the interferometric-based one (see Sec.~\ref{sec:vacuum}), is related to these last works. Here, however, it is employed for a different task, i.e., enhancing the fidelity of quantum computation.
Moreover, we try to provide some insights into the fundamental physical understanding of the problem. 

\section{Problem setting and general idea}
\label{sec:Settings}
Current quantum computers suffer from noise and imperfections \cite{Preskill2018}. Our protocols, introduced below, exploit available imperfect resources to enhance the fidelity of a computation. At the cost of resorting to more (noisy) qubits and performing operations multiple times, we do not require error correction or tomography techniques. Furthermore, our strategies are effective regardless of the type or the form of the noise.

\begin{figure}
    \centering
    \includegraphics[width=0.93\columnwidth]{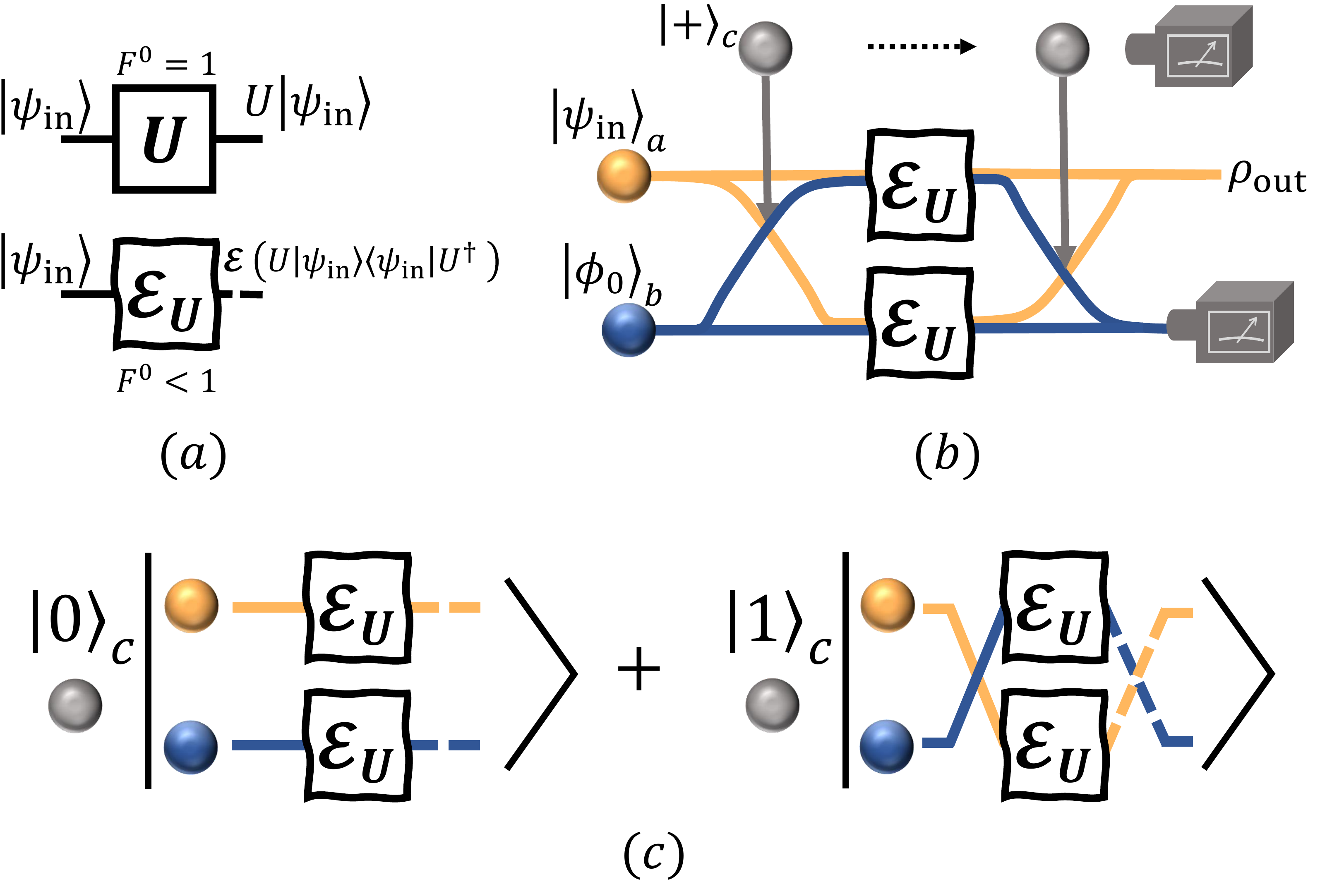}
    \caption{Illustration of our SQEM method applied to an arbitrary computation $U$ acting on $m$ input qubits. (a) The fidelity $F$ is calculated with respect to the output of the perfect computation. In the noiseless case, the absence of error guarantees unit fidelity $F=1$. When considering realistic settings, noise acting on the system lowers the fidelity $F<1$. In panels (b) and (c) we schematically represent the underlying idea behind our protocols. By creating a superposition of two (or more) identical computations, it is possible to enhance the fidelity of the computation.}
    \label{fig:general idea}
\end{figure}

Although the working principles at the basis of our protocols depend on the specific implementation (see below), the underlying idea is the same. Consider a $m$-qubit input state $\ket{\psi_{\rm in}}$ that undergoes some noisy quantum computation described by ${\cal E}_U$, such that the outcome is given by $\rho={\cal E}_U (\proj{\psi_{\rm in}})$ [see Eq.~\eqref{eq:Krausrepresentation}]. Indicating with $U$ the desired (noiseless) unitary operation associated with the computation, we investigate the detrimental effects of decoherence via Eq.~\eqref{eq:fid}, which can be rewritten as
\begin{equation}
    F^{0}=\tr \left( \rho_{\rm out} U \proj{\psi_{\rm in}}{  U^{\dagger}} \right).
\end{equation}
This equation allows determining the expected fidelity $F^{0}$ of the considered computation $U$ with a given input state $\ket{\psi_{\rm in}}$ and noise ${\cal E}_U$. In this manuscript, we refer to $F^{0}$ as the incoherent fidelity. As schematically represented in Fig.~\ref{fig:general idea}, our goal is to enhance $F^{0}$ by exploiting coherent interference between two or more states undergoing the same (noisy) computation ${\cal E}_U$. 

The special case where $U=\id$ can be seen as a particular implementation of a quantum memory \cite{Simon2010,Heshami2016}, where quantum information is stored and our protocols allow to mitigate the noise during the storing process. We refer to the implementation of our protocols for this case as coherent quantum memories.

As discussed in Sec.~\ref{sec:Background}, ${\cal E}_U$ can be modeled by applying arbitrary noisy channels either to the state after implementing the gates within the computation (GB-QC), or to all qubits after the creation of the resource state (MB-QC). We remark that, while these noise models are general, our formalism can be extended to other noisy models, resulting in similar qualitative results --- see  Sec.~\ref{sec:standardgatebased} and Sec.~\ref{sec:enhancedMBQC} for more details.

To explain the underlying idea of our protocols, let us consider the simplest case in Fig.~\ref{fig:general idea} (b) first, which shall be generalized later. An auxiliary degree of freedom, which is initialized in the state $\ket{+}_{\rm c}$, constitutes a control register. As depicted in Fig.~\ref{fig:general idea} (b), we apply the noisy computation ${\cal E}_U$ to $\ket{\psi_{\rm in}}$ multiple times, each controlled by the state of a control system. 

As a result of the coherent superposition, see Fig.~\ref{fig:general idea} (c), the noise acting on the maps ${\cal E}_U$ interfere with each other, resulting in partial cancellation of the errors. A final measurement of the control register in the $X$ basis leads to two possible states (corresponding to each outcome) with fidelities $F_{1}$ and $F_{2}$, each found with probabilities $p_{1}$ and $p_{2}$, respectively. We show that, following our protocols, one finds probabilistic improvement $\max \{ F_{1}, F_{2}\} > F^{0}$ and deterministic improvement $(p_{1} F_{1} + p_{2} F_{2}) > F^{0}$. While the latter usually requires unitary corrections to be applied to the output state, the first does not (but can benefit from them). The fidelity enhancement is significant in all settings we have investigated. 

The above example for the two degrees of freedom represented by the control register $\ket{+}_{\rm c}$ can be generalized. Superposition of more than two elements can be achieved by preparing the control register as a qudit in $\ket{+_{d}}_{\rm c} = \frac{1}{\sqrt{d}}\sum_{i=0}^{d-1} \ket{i}$. As described in Sec.~\ref{sec:standardgatebased}, with an overhead of resources that is constant in the dimension $d$, it is possible to further improve the fidelity and asymptotically reach unit fidelity for a variety of settings.

\subsection{Figure of merit}
\label{sec:figuremerit}
The state fidelity of a noisy computation highly depends on the input state. To better quantify the advantages of our protocols for all input states, we introduce in this section a figure of merit based on Eq.~\eqref{eq:fid}.

The Choi-Jamiołkowski (CJ) isomorphism \cite{Jamiokowski1972} is a practical way to describe the effects of noise on quantum operations via the tools offered by quantum states. In this context, we use the isomorphism to analyze how close a noisy quantum operation is with respect to its ideal implementation.
Consider a maximally entangled state
\begin{equation}
 \ket{\Phi^+}_{\rm t,r} = \frac{\ket{0}_{\rm t}\ket{0}_{\rm r}+\ket{1}_{\rm t}\ket{1}_{\rm r}}{\sqrt{2}}, \label{eq:bellstate}
\end{equation}
where subscripts ``t'' and ``r'' stand for test and result, respectively, and indicate two (entangled) subsystems. The ``result'' subsystem undergoes a computation --- either a standard one as in panel (a) of Fig.~\ref{fig:general idea} or a parallel one as in (b--c) --- which is generally noisy. The CJ fidelity is then defined as
\begin{equation}
\label{eq:CJfid}
  \text{CJ fid.}: F_{\text{CJ}}= \bra{\Phi^{+}}_{\rm t,r} (\id \otimes U^{\dagger}) \rho_{\rm out}^{\rm t,r}   (\id \otimes U)\ket{\Phi^{+}}_{\rm t,r}, 
\end{equation}
where $\rho_{\rm out}^{\rm t,r}$ is the state of the composite target and result systems after the computation is applied to the latter, and $\id \otimes U$ is an operator acting $U$ on the result subsystem. 

For an $m$-qubit input state we require $2m$ qubits for calculating the CJ fidelity, which generalizes to 
\begin{subequations}
\label{eq:CJfidgeneralized}
\begin{align}
    & F_{\text{CJ}} = \bra{ \Phi_{m}^{+}}_{{\rm t},{\rm r}} ( \id_{\rm t} \otimes U^{\dagger}) \rho_{\rm out}^{\rm t,r}   ( \id_{\rm t} \otimes U)   \ket{\Phi_{m}^{+}}_{{\rm t},{\rm r}}
    , \label{eq:CJfidgeneralized_form}\\
    & \ket{\Phi_{m}^{+}}_{{\rm t},{\rm r}} = \left(
    \frac{
    \ket{0}_{\rm t}\ket{0}_{\rm r}+\ket{1}_{\rm t}\ket{1}_{\rm r}
    }
    {\sqrt{2}}
    \right)^{\otimes m}
    , \label{eq:bellstate_m}
\end{align}
\end{subequations}
where $\id_{\rm t}$ and $U$ act on the $m$ qubits constituting the test and the result states, respectively, and $\ket{\Phi_{1}^{+}}_{{\rm t},{\rm r}} = \ket{\Phi^{+}}_{{\rm t},{\rm r}}$.

Given an arbitrary input state, $F_{\text{CJ}}$ provides a lower bound on the fidelity of the computation \cite{Dur2005}.  In fact, if the result subsystem of the maximally entangled state $\ket{\Phi_{m}^{+}}$ in Eq.~\eqref{eq:bellstate_m} decoheres, the entanglement with the test subsystem is reduced and consequently $F_{\rm CJ}$. By using half the qubits in $\ket{\Phi_{m}^{+}}$ as input for some process, from a teleportation point of view, one can see that the CJ fidelity is a lower bound for any fidelity $F$ obtained with an arbitrary input state, as long as the protocol is run under the same settings.

\section{Gate-based SQEM approach} \label{sec:standardgatebased}
As mentioned above, there are different ways to carry out a quantum computation: GB- and MB-QC being the better known. In the remainder of this work, we introduce three different SQEM protocols to coherently enhance quantum computations. The first, explained in this section, is applicable to GB-QC; the second (Sec.~\ref{sec:enhancedMBQC}) to MB-QC, while the third employs interferometric principles (Sec.~\ref{sec:vacuum}).
Here, we consider GB-QC with decoherence arising from imperfect quantum gates. As already explained in Sec.~\ref{sec:Noise_Fidelity}, noise is modeled via the application of quantum channels to qubits after the desired unitary evolution. 

In this section, we first introduce the relevant notation and explain the protocol in Sec.~\ref{sec:gate_protocol}. In Sec.~\ref{sec:gate_prob_prot} we then describe the implementation called ``probabilistic'' where, depending on the outcomes of the measurements on the auxiliary systems, the output state is either kept or discarded. Later in Sec.~\ref{sec:gate_det_prot}, we introduce the other variant named ``quasi-deterministic'', in which the user, at the cost of performing additional correcting operations, can enhance the probability of keeping the state, possibly making the protocol fully deterministic. Finally, in Sec.~\ref{sec:realistic}, we explain how to generalize both these variants in a ``nested'' fashion, which allows for overcoming the limits of the first two in some parameter regimes and further enhance the fidelity of the resulting state.

\subsection{Protocol}
\label{sec:gate_protocol}
\begin{figure}
    \centering
    \includegraphics[width=\columnwidth]{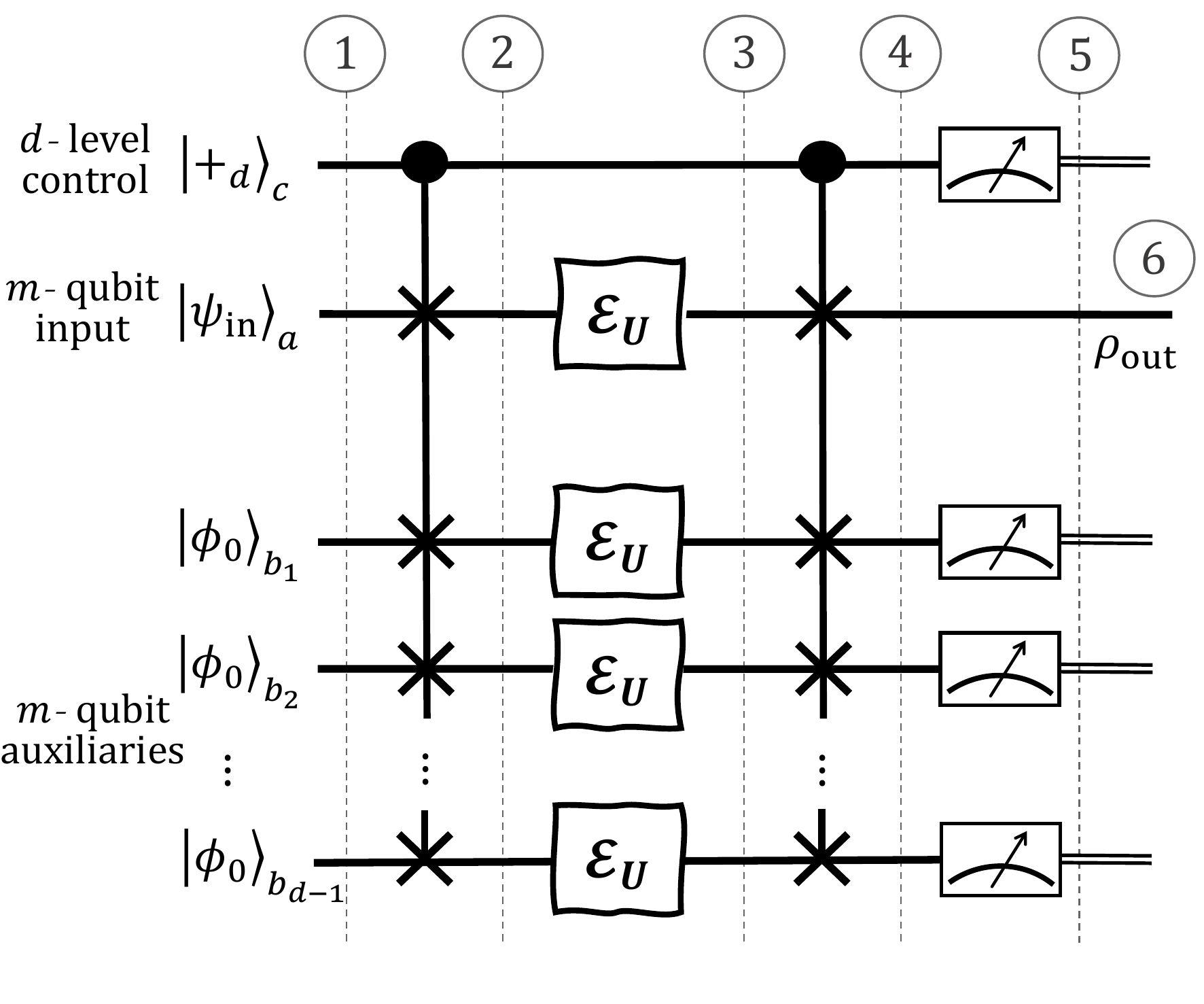}
    \caption{Schematic representation of Protocol \ref{table:GBstandard}. Vertical dashed lines, marked with numbers from one to six, identify the system's state after the corresponding steps of the protocol, as explained in the main text. We explicitly include the initial $\ket{\psi_{\rm in}}_{\rm a}$ and auxiliary $\ket{\phi_{0}}_{{\rm b}_{i}}$ ($i=1,\dots,d-1$) states, as well as the noisy computations ${\cal E}_U$. The multi-qubit gates between lines $1$ and $2$, as well as $3$ and $4$, are cSWAP operations defined in Eq.~\eqref{eq:GB-QCcswap}.}
    \label{fig:compstandard1}
\end{figure}

The aim is to apply a circuit corresponding to the unitary $U$ to an unknown $m$-qubit input state $\ket{\psi_{\rm in}}$. In general, if the gates within $U$ are noisy the effective action of the circuit can be described via the map ${\cal E}_U$ in Eq.~\eqref{eq:Krausrepresentation} such that 
\begin{equation}\label{eq:incoherentmap}
    \rho_0 \equiv {\cal E}_U \left(\proj{\psi_{\rm in}} \right) = \sum_{i} K_{i} U \proj{\psi_{\rm in}} U^{\dagger}  K_{i}^{\dagger}, 
\end{equation}
with $K_{i}$ being Kraus operators, see Sec.~\ref{sec:Noise_Fidelity}. In the following, we refer to $\rho_0$ as the ``incoherent'' output, as opposed to the coherent one, $\rho_{\rm out}$, resulting from the application of our strategies outlined below. From Eq.~\eqref{eq:fid} it is possible to determine the fidelity of $\rho_0$ with respect to the desired output state $U\proj{\psi_{\rm in}}U^{\dagger}$, i.e.
\begin{equation}\label{eq:incoherentfid}
    F^{0}= \bra{\psi_{\rm in}} U^{\dagger} \rho_0 U \ket{\psi_{\rm in}}, 
\end{equation}
which we label as incoherent fidelity. 

As shown in Fig.~\ref{fig:compstandard1} and thoroughly described below, our SQEM protocol relies on the idea of performing the computation in superposition, such that the desired computation $U$ is independently applied to the input and auxiliary registers. Depending on the state of a control register, the input state is spread throughout all branches and correlated with chosen auxiliary states. In Fig.~\ref{fig:compstandard1}, $d$ branches are depicted, the first (``${\rm a}$'') corresponding to the input and the remaining $d-1$ (``${\rm b}_i$'', $i=1,\dots,d-1$) to the auxiliary registers.

The superposition is achieved with controlled-SWAP (cSWAP) gates applied to the input and auxiliary states in a coherent fashion \cite{MiguelRamiro2021}. Identical noisy operations are then implemented in each system, followed by another round of cSWAPs for reassembling the desired output state. Under the assumption that the noise affecting each register is independent, measurements of the auxiliary states project the system state vector such that noise is mitigated for the output state. 

\begin{table}
{\LinesNumberedHidden
    \begin{algorithm}[H]
        \SetKwInOut{Input}{Input}
        \SetKwInOut{Output}{Output}
        \SetAlgorithmName{Protocol}{}
   \justifying  \textit{Input}: An initial state $ \ket{\psi_{\rm in}}$ and a noisy computation ${\cal E}_U$ implementing the unitary $U$ with a fidelity $F^{0}$.
       \begin{enumerate}
           \item Prepare a control qudit in the state $\ket{+_{d}}_{\rm c}$ as in Eq.~\eqref{eq:controldlevel}, and auxiliary qubits in $\bigotimes_{i=0}^{d-1} \ket{\phi_0}_{b_{i
           }}$. \label{prot:gate1}
        
           \item Apply a cSWAP operation, Eq.~\eqref{eq:GB-QCcswap}, coherently swapping the main and auxiliary register states. A superposition is generated.
           \label{prot:gate2}
           
           \item Implement the noisy computation on the main and auxiliary registers independently.
           \label{prot:gate3}
            
            \item Apply again the cSWAP gate for reassembling.
            \label{prot:gate4}
            
            \item Measure the control register in the generalized $X$ basis in Eq.~\eqref{eq:generalizedX} and the auxiliary qubits in suitable bases.
            \label{prot:gate5}
            
            \item Post-select the state depending on the measurement outcomes of the control and the auxiliaries, unless the protocol is fully deterministic. If desired, apply correcting unitaries depending on the measurement outcomes.
            \label{prot:gate6}
        \end{enumerate}
        
    \justifying \textit{Output}: State $\rho_{\rm out}$ characterized by a fidelity $F > F^{0}$, in both the probabilistic and (on average) the deterministic protocols. 
\caption{SQEM for a GB-QC implementation} \label{table:GBstandard}
\end{algorithm}}

\end{table}
Our goal is to design a procedure that, at the expense of more resources in terms of qubits and gates, produces an output state $\rho_{\rm out}$ whose fidelity $F$ is higher than $F^{0}$. This procedure is summarized in Table~\ref{table:GBstandard}. Here, we present an accurate description that step-by-step provides insight on the protocol and the mathematical tools that will be used in the following.

\textbf{Step~\ref{prot:gate1}. ---} In the first step, a $d-$level qudit system, playing the role of control register (hence the subscript ``c''), is prepared in the state
\begin{equation}
  \ket{+_{d}}_{\rm c} = \frac{1}{\sqrt{d}}\sum_{i=0}^{d-1} \ket{i}_{\rm c}.
  \label{eq:controldlevel}
\end{equation}
The dimension $d$ indicates the number of branches (see also Fig.~\ref{fig:compstandard1}) into which the input state will be distributed to create the noise interference at the basis of the fidelity enhancement. The role of the control register is to keep track of the states in all branches, in order to reconstruct the desired output at step~\ref{prot:gate4} of our protocol. We remark that, in practice, it is always possible to realize a $d$-level qudit control system by embedding $n\geq \log_2d$ qubits each in the $\ket{+}$ state.

Additionally, we prepare $d-1$ $m$-qubit auxiliary systems in the $ \bigotimes_{i=1}^{d-1} \ket{\phi_0}_{{\rm b}_i}$ state. The choice of $\ket{\phi_0}$ is not unique and, together with the measurement basis of the auxiliary registers (see step~\ref{prot:gate5} below), determines the amount of advantage achieved
by the protocol.

The state of the whole system at step~\ref{prot:gate1} is therefore characterized by the equation
\begin{equation}\label{eq:gate_state_step1}
    \text{step~\ref{prot:gate1}: } \ket{+_d}_{\rm c} \ket{\psi_{\rm in}}_{\rm a}  \bigotimes_{i=1}^{d-1} \ket{\phi_{0}}_{{\rm b}_{i}},
\end{equation}
where we use subscripts ``${\rm a}$'' and ``${\rm b}_i$'' ($i=1,\dots,d-1$) to indicate the registers associated with the input $\ket{\psi_{\rm in}}$ and auxiliary $\ket{\phi_0}$ states, respectively.

\textbf{Step~\ref{prot:gate2}. ---} In the second step, a controlled-SWAP (cSWAP) operation, which is a generalization of the Fredkin gate \cite{Fredkin82, Daboul_2003,Patel_2016}, is applied to the state in Eq.~\eqref{eq:gate_state_step1}. The action of the cSWAP is described by the unitary operator
\begin{equation}\label{eq:GB-QCcswap}
\begin{split}
      \text{cSWAP} = & \ket{0} \bra{0}_{\rm c} \otimes  \id  + \sum^{d-1}_{i=1}  \ket{i} \bra{i}_{\rm c} \otimes \text{SWAP}_{{\rm a}, {\rm b}_{i}}.
  \end{split}
\end{equation}
Here, the SWAP gate defined as $\text{SWAP}_{n,n'}=\sum_{i,j} \ket{i}\bra{j}_{n}\ket{j}\bra{i}_{n'}$ exchanges states $n$ and $n'$, and we omitted the subscripts in the kets; e.g., $\ket{i}\bra{j}_{n}$ indicates $\ket{i}_{n}\bra{j}_{n}$. We remark that in the case of multi-qubit input and auxiliary states, the SWAP can be constructed by exchanging, qubit by qubit, the state vectors of the two registers to which it is applied.   Note that a cSWAP gate can be decomposed into two cNOT and one Toffoli gate \cite{Heese2022}, for which high fidelity implementations have been achieved \cite{Kim2022}.

The $\text{cSWAP}$ in Eq.~\eqref{eq:GB-QCcswap} distributes the input $\ket{\psi_{\rm in}}_{\rm a}$ among all branches. Specifically, if the control is in $\ket{i}_{\rm c}$, the register ${\rm b}_i$ with its $m$-qubit auxiliary state is swapped with the input $\ket{\psi_{\rm in}}_{\rm a}$ initially stored in the register ${\rm a}$. Therefore, at the end of step~\ref{prot:gate2}, the system state is given by the coherent superposition
\begin{equation}
\label{eq:step2_gb_prot}
\begin{split}
    \text{step~\ref{prot:gate2}: }  \frac{1}{\sqrt{d}} & \left( \ket{0}_{\rm c} \ket{\psi_{\rm in}}_{\rm a}  \bigotimes_{j=1}^{d-1} \ket{\phi_{0}}_{{\rm b}_{j}} \right. \\  & \left. + \sum_{i=1}^{d-1}  \ket{i}_{\rm c}  \ket{\phi_{0}}_{\rm a} \ket{\psi_{\rm in}}_{{\rm b}_{i}} \bigotimes_{j \neq i}^{d-1} \ket{\phi_{0}}_{{\rm b}_{j}} \right).
\end{split}
\end{equation}

\textbf{Step~\ref{prot:gate3}. ---} In the third step, the desired computation $U$ is carried out in each register ${\rm a}$, ${\rm b}_i, i\in \lbrace 1,\dots,d-1 \rbrace$. Realistic implementations of $U$ are associated with noise arising from the imperfect application of the quantum gates within $U$ (see Sec.~\ref{sec:Noise_Fidelity}), yielding the maps ${\cal E}_U$ (see Eq.~\eqref{eq:incoherentmap} and Fig.~\ref{fig:compstandard1}). 

Here, we make two assumptions. First, noise acting on different registers must be uncorrelated \footnote{We remark that our protocol work even if, \textit{within} each branch, noise is correlated.}. This is required to ensure that the noise interference that is built at step~\ref{prot:gate4} of our protocol cancels (some of) the errors affecting the resulting output state once the control and all auxiliary systems are measured. Second, we assume that each register is subjected to the same decoherence. This is motivated by the fact that all computations are carried out identically. However, we remark that device imperfections may introduce different decoherence effects in each branch. Our protocols, particularly the quasi-deterministic version, are resilient against differences in the Kraus operators acting on separate registers, and tolerate weak correlations between them.

The system state vectors at the end of steps~\ref{prot:gate3} and \ref{prot:gate4} are lengthy and the same physical insight is obtained after the collapse of the state following the measurements in step~\ref{prot:gate5}. Therefore, we omit the mathematical expression here and point at App.~\ref{sec:Appendix_density} for further information.

\textbf{Step~\ref{prot:gate4}. ---} In the fourth step a second cSWAP of the form of Eq.~\eqref{eq:GB-QCcswap} is applied, with the purpose of reassembling the desired state and building the interference at the basis of noise cancellation. In the absence of decoherence, the composite system state becomes 
\begin{equation}\label{eq:gate_no_noise}
    \ket{+_d}_{\rm c} U \ket{\psi_{\rm in}}_{\rm a}  \bigotimes_{j=1}^{d-1} U \ket{\phi_{0}}_{{\rm b}_{j}},
\end{equation}
i.e., the cSWAPs in steps~\ref{prot:gate2} and \ref{prot:gate4} do not contribute to the  dynamics of the system. 

When noise is present, on the other hand, the state at this point of the protocol is \textit{not} separable into its branches' state vectors, but is entangled. As will become clear in the next step, this entanglement is the essential ingredient for partially or completely correcting the errors affecting the input state. In fact, since the first cSWAP distributed $\ket{\psi_{\rm in}}_{\rm a}$ among all branches, by measuring the auxiliary system we can either post-select the outcome that is ensured to yield minimal error (see Sec.~\ref{sec:gate_prob_prot}), or correct via unitary operations the output state to boost its fidelity (see Sec.~\ref{sec:gate_det_prot}).

The cSWAP operations are central to this protocol, and an important question to be addressed is how noise resulting from their application affects the protocol output. This is investigated in Sec.~\ref{sec:standardgateperformance}, where we show that for common experimental settings decoherence from the cSWAPs is tolerated.

\textbf{Step~\ref{prot:gate5}. ---} In this step, control and auxiliary subsystems are all measured. The state vector collapses, resulting --- either probabilistically or deterministically (see next step) --- in an output state characterized by a higher fidelity compared to the incoherent case in Eqs.~\eqref{eq:incoherentmap} and \eqref{eq:incoherentfid}. We need first to specify the bases used for the measurements. 

The control is projected onto elements of the generalized $d$-level $X$ basis, i.e.,
\begin{equation}\label{eq:generalizedX}
    \left\lbrace  \frac{1}{\sqrt{d}}\sum_{k=0}^{d-1} e^{\frac{2 \pi i k l}{d}}  \ket{k}_{\rm c} \right\rbrace_{l=0}^{d-1}.
\end{equation}
This choice is motivated since, as outlined above, prior to this step and in the absence of noise we expect to recover the state in Eq.~\eqref{eq:gate_no_noise}. Therefore, we want to be able to discern when the control register is in the $\ket{+_d}_{\rm c}$ state. In principle, the other elements in the control measurement basis in Eq.~\eqref{eq:generalizedX} are arbitrary, and one could even optimize their choice to further enhance the protocol \footnote{Notice that this is equivalent to applying unitary operations on the control register at the beginning of the protocol, in step~\ref{prot:gate1}.}. We selected the generalized $d$-level $X$ basis for concreteness. In fact, when the control register is a collection of qubits (see step~\ref{prot:gate1}), with the basis in Eq.~\eqref{eq:generalizedX} all the measurements can be implemented locally by performing a Pauli $X$-measurement on each  qubit.

On the other hand, the basis chosen for the auxiliary subsystem generally depends on the characteristics of the unitary $U$, the state $\ket{\phi_{ 0}}$, and the noise. As discussed below in Sec.~\ref{sec:gate_det_prot}, its choice can even be optimized. Ideally, following a similar argument as for the control basis, it is desirable to project all auxiliary subsystems onto $U \ket{\phi_0}$ [see Eq.~\eqref{eq:gate_no_noise}]. This naturally occurs when $\ket{\phi_{ 0}}$ is an eigenstate of the unitary $U$. However, depending on $U$ and $\ket{\phi_{ 0}}$, this may be impractical. Different choices of the measurement basis for the auxiliary registers lead to different fidelity enhancements.
A more detailed discussion on this topic is given in Sec.~\ref{sec:gate_prob_prot}. 

\textbf{Step~\ref{prot:gate6}. ---} The last step regards post-processing of the result, which is different depending on whether the SQEM protocol is run in its probabilistic or quasi-deterministic version. We summarize the main features here, and expand in the following Secs.~\ref{sec:gate_prob_prot} and \ref{sec:gate_det_prot}.

As the name suggests, the probabilistic scheme involves post-selection of the resulting state depending on the measurement outcomes of the control and auxiliary subsystems. If all measurements at step~\ref{prot:gate5} yield the desired result, $\rho_{\rm out}$ is kept; otherwise it is discarded. The most appealing aspect of this variant is that it does not require any information on the input state, the unitary or the noise. It is a plug-and-play protocol that is readily implemented and ensures, under broad assumptions, that the resulting state is characterized by a better fidelity than the incoherent case.

While the probabilistic SQEM protocol does not require any information on the hardware, one may ask whether and how accessing extra knowledge (e.g., the noise characteristics) can help improve the resulting state fidelity.  Furthermore, the likelihood to obtain all the desired measurement outcomes in step~\ref{prot:gate5} decreases exponentially with the number of branches $d$, making the probabilistic protocol unfeasible for large values of $d$.  These two observations motivate the quasi-deterministic version of our protocol, where one can both improve the output fidelity and enhance the post-selection probability to a user-specified threshold. The most appealing aspect of this scheme is that it employs a black-box optimization that guarantees the best possible performances given the unknown noise characteristics and the chosen threshold. Notice that, if required, the protocol can work in a completely deterministic way, i.e., without requiring post-selection.

\begin{figure}
    \includegraphics[width=\columnwidth]{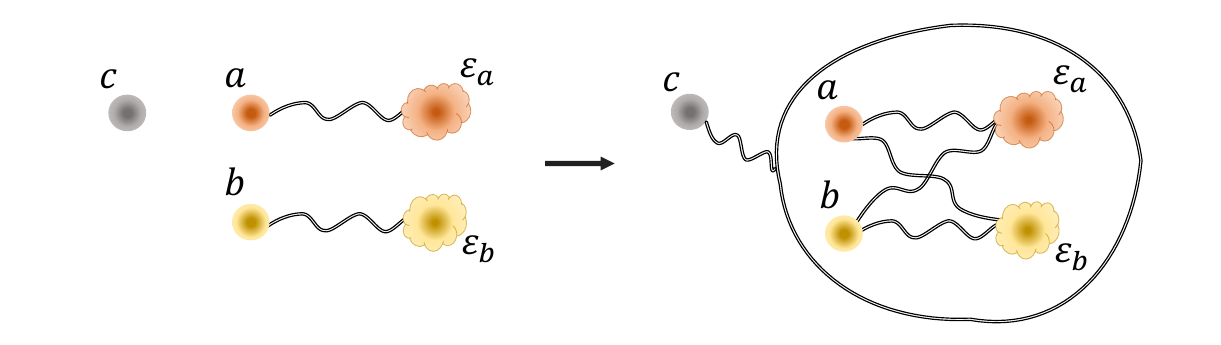}
    \caption{Schematic representation of the evolution of correlations between the systems and environments during the process.} \label{fig:compstandard2} 
\end{figure}

A direct intuition about how the protocol works can also be found analyzing the correlations of the environment spaces where information is leaked out because of noise. Thanks to the coherent superposition, the subspaces corresponding to the environments associated with the input and auxiliary subsystems get correlated with each other and with the system's state vector before the measurements at step~\ref{prot:gate5} (see Fig.~\ref{fig:compstandard2}). At the end of the process, the measurement of the auxiliary and control qubits (partially) reveals to us the nature of the interactions between the system and the environments, effectively suppressing the errors. We refer the reader to App.~\ref{sec:Appendixstandard1} for details. 

In the following subsections, we formally introduce the probabilistic and the quasi-deterministic versions of our protocol. Bounds and asymptotic limits of e.g. the resulting fidelities and the success probabilities will be derived and applied to relevant examples. 

\subsection{Protocol analysis}
We provide now an  analysis of the protocol. For clarity, in this section, we assume projecting \textit{all} auxiliary subsystems onto the same state, which we indicate with $\ket{\phi_{\rm f}}$. With the control and all auxiliary subsystems projected onto $\ket{+_d}_{\rm c}$ and $\ket{\phi_{\rm f}}$, respectively, the resulting unnormalized density matrix $\rho_{\rm out}$ of the system ``${\rm a}$'' reads (after tracing out the measured systems)
\begin{widetext}
\begin{subequations}
\label{eq:gate_out}
\begin{align}
    \rho_{\rm out} = &  \frac{\mathcal{A}_d}{d} 
        \left[ \sum_i K_i U \proj{\psi_{\rm in}} U^{\dagger} K_i^{\dagger}
        + (d-1)
        \sum_{i,j} 
        \left(
        \frac{
        \bra{\phi_{\rm f}} K_j U \ket{\phi_{ 0}}
        \bra{\phi_{ 0}} U^{\dagger} K_i^{\dagger} \ket{\phi_{\rm f}}
        }{
       \mathcal{A}_2
        }
        \right)
        K_i U \proj{\psi_{\rm in}} U^{\dagger} K_j^{\dagger} \right],
        \label{eq:gate_out_main}
    \\
    \mathcal{A}_d = & \left( \sum_{i} \left\lvert 
    \bra{\phi_{\rm f}} K_i U \ket{\phi_{ 0}}
    \right\rvert^2 \right)^{d-1}.
    \label{eq:gate_out_cnst}
\end{align}
\end{subequations}
\end{widetext}
The trace $\tr \left( \rho_{\rm out} \right)$ corresponds to the probability of measuring the chosen state $\ket{+_d}_{\rm c} \bigotimes_{j=1}^{d-1} \ket{\phi_{\rm f}}_{{\rm b}_j}$ for the control and auxiliary subsystems.

From Eq.~\eqref{eq:gate_out_main} it is possible to understand how the protocol works. The first term in the square brackets describes the input state $\ket{\psi_{\rm in}}$ always remaining in register ``${\rm a}$''. Noise interference from the other registers ${\rm b}_i$ is absent, and thus this case resembles, up to an overall constant, the incoherent one in Eq.~\eqref{eq:incoherentmap}. The interesting term of Eq.~\eqref{eq:gate_out_main} is the second one: it is enhanced by a factor $d-1$ and thus becomes dominant for large $d$. This contribution to $\rho_{\rm out}$ contains all cases where the input has been distributed within all other registers ${\rm b}_{i}$ ($i=1,\dots,d-1$) and later on brought back into its original register. Noise interference is evident from the products 
\begin{equation}\label{eq:gate_interf}
    \bra{\phi_{\rm f}} K_j U \ket{\phi_{ 0}}
\bra{\phi_{ 0}} U^{\dagger} K_i^{\dagger} \ket{\phi_{\rm f}},
\end{equation}
which suppress decoherence the more the state $U \ket{\phi_{ 0}}$ is ``sensitive'' to the Kraus operators and the less $\ket{\phi_{\rm f}}$ is orthogonal to $U \ket{\phi_{ 0}}$. In our terminology, $\ket{\psi}$ is fully sensitive to $K_j$ if $\bra{\psi} K_j \ket{\psi} = 0$. By comparison, $\ket{\psi}$ is called insensitive to $K_j$ if it is one of its (nonzero) eigenstates, i.e., $K_j \ket{\psi} \propto \ket{\psi}$.

Eqs.~\eqref{eq:gate_out_main} and \eqref{eq:gate_interf} tell us that the less $\ket{\phi_{\rm f}}$ is orthogonal to $U \ket{\phi_{ 0}}$ and the more the state $U\ket{\phi_{ 0}}$ is affected by the noise, the better our protocol works. This rather counter-intuitive fact is better understood by thinking in terms of the noise. After step~\ref{prot:gate2}, the input and the auxiliary states are in a coherent superposition, and thus subjected to the same noise. 
By measuring the control and auxiliary subsystems in step~\ref{prot:gate5}, we can learn the kind of noise that has been applied to both $U\ket{\phi_{ 0}}$ and $U\ket{\psi_{\rm in}}$, but \textit{only if} $U\ket{\phi_{ 0}}$ is sensitive to the associated Kraus operators. Moreover, this information is accessible \textit{only if} $\ket{\phi_{\rm f}}$ is not orthogonal to $U \ket{\phi_{ 0}}$. 

In order to quantify the noise mitigation obtained with our protocol, we introduce the parameters
\begin{subequations}\label{eq:omegas}
    \begin{align}
    \omega_1 & =1 - \frac{\sum_{j \geq 1} \left\lvert\bra{\phi_{ 0}} U^{\dagger} K_j U \ket{\phi_{ 0}} \right\rvert^2}{1-p_{\rm ne}} 
    ,\label{eq:omegas_1}\\
    \omega_2 & =|\bra{\phi_{\rm f}} U \ket{\phi_0} |^2
    , \label{eq:omegas_2}
    \end{align}
\end{subequations}
where $p_{\rm ne}$ is the probability not having an error of any kind. From the properties of the Kraus operators (see Sec.~\ref{sec:operatorsum}), it is possible to bound the right-hand side of Eq.~\eqref{eq:omegas_1} between $0$ and $1$, with the former (latter) being achieved if and only if $U \ket{\phi_{ 0}}$ is completely insensitive (sensitive) to all Kraus operators. Therefore, the extreme points $(\omega_1, \omega_2)=(1,1)$ and  $(\omega_1, \omega_2)=(0,0)$, respectively, relate with maximum and minimum mitigation of the error affecting the computation $U$. Any other pair of values of $(\omega_1, \omega_2)$ generally indicates a certain degree of noise mitigation and the corresponding advantage of our protocols. 

\subsubsection{ \texorpdfstring{$(\omega_1, \omega_2)=(1,1)$}{(omega1,omega2)=(1,1)}}
To better understand how our scheme works, let us consider first the best possible scenario, i.e., the chosen auxiliary state $\ket{\phi_0}$ and $\ket{\phi_{\rm f}}$ are such that $U\ket{\phi_{ 0}}$ is fully sensitive to all Kraus operators and $\bra{\phi_{\rm f}} U \ket{\phi_0} = 1$, such that $(\omega_1, \omega_2)=(1,1)$ in Eqs.~\eqref{eq:omegas}. 

For $(\omega_1,\omega_2) = (1,1)$, of all possible products in Eq.~\eqref{eq:gate_interf}, only the one with $i=j=0$ survives, and the resulting state $\rho_{\rm out}$ in Eq.~\eqref{eq:gate_out_main} becomes
\begin{equation}\label{eq:gate_out_noise_suppr}
    \begin{split}
        \rho_{\rm out} = & \frac{p_{\rm ne}^{d-1}}{d} 
        \sum_{i\geq 1} K_i U \proj{\psi_{\rm in}}_{\rm a} U^{\dagger} K_i^{\dagger}
        \\
        &
        + p_{\rm ne}^d
        U \proj{\psi_{\rm in}}_{\rm a} U^{\dagger}
        .
    \end{split}
\end{equation}
For sufficiently many branches $d\gg 1$, we can approximate this last equation (after normalization) by $\rho_{\rm out} \approx U \proj{\psi_{\rm in}}_{\rm a} U^{\dagger}$, recovering the perfect outcome. In other words, $(\omega_1, \omega_2)=(1,1)$ is a sufficient condition to achieve the desired outcome in the limit $d \gg 1$.

In order to obtain $\omega_1 = 1$, a necessary and sufficient requirement is that each term in the sum within Eq.~\eqref{eq:omegas_1} is zero, i.e., the auxiliary state is orthogonal to all the eigenvectors of all Kraus operators different from the identity $\id$. For some kinds of noise (e.g., the depolarizing channel) this is not possible without resorting to larger entangled states, since these eigenvectors form a basis of the Hilbert space within which $\ket{\phi_0}$ resides. However, for rank-2 noise \cite{nielsen_chuang_2010}, $\omega_1=1$ is attainable with non-entangled auxiliary states. For instance, with dephasing we can set each qubit of $\ket{\phi_0}$ to be in an eigenstate of either the Pauli $X$ or $Y$ operator, ensuring $\omega_1 = 1$. Alternatively, one can always ensure complete sensitivity to any Kraus operator by employing $m$ Bell states for each auxiliary $\ket{\phi_0}$.

Another question is how to obtain $\omega_2 = 1$, i.e., post-selecting the measurements at step~\ref{prot:gate5} which are associated with noise suppression and hence fidelity improvement. As mentioned above, a possibility is to apply the reverse computation $U^\dagger$ to the auxiliary states after the second cSWAP, and to post-select $\ket{\phi_{0}}$ (which is known) after the measurement. While this strategy could be a viable option, undoing the unitary $U$ is a noisy process that (while still being advantageous) lowers the resulting fidelity of the output state $\rho_{\rm out}$. It is thus relevant to investigate the scenarios where one can ensure $\omega_2 = 1$ without applying $U^\dagger$ before the measurements of the auxiliary states.

To have $\omega_2 = 1$ it is required to know $U$ and how it acts on $\ket{\phi_{0}}$. The simplest scenario is when $\ket{\phi_{0}}$ is one of its eigenstates, such that $U\ket{\phi_{0}} = \ket{\phi_{0}}$ and it becomes redundant to apply $U^\dagger$. For instance, for $U$ being a rotation on the Bloch sphere, we can pick $\ket{\phi_{ 0}}$ along the rotational axis, e.g., $\ket{0}$ or $\ket{1}$ for $U=T$, where $T$ is the $T=e^{i \pi Z / 4}$ gate. 

When $U$ is a Clifford gate \cite{GottesmanClifford}, it is possible to generalize this approach to a larger class of auxiliary states. In this scenario, we consider $\ket{\phi_{0}}$ to be a stabilizer state \cite{Gottesman1997}, such that it is classically efficient to determine $U \ket{\phi_{0}}$. Specifically, by ensuring that all stabilizers of both $\ket{\phi_{ 0}}$ and $U \ket{\phi_{0}}$ are bitwise commuting, we can prepare $\ket{\phi_{0}}$ \textit{and} measure $U \ket{\phi_{0}}$ by employing local rotations only. We remark that the cNOT gate, which is a limiting factor to scaling up quantum computation \cite{Preskill2018}, is a Clifford gate and as such suited to this choice for the state $\ket{\phi_{0}}$.

We refer to App.~\ref{app:otheranalytics} for further analysis.

\subsubsection{ \texorpdfstring{$(\omega_1,\omega_2) < (1,1)$}{(omega1, omega2)<(1,1)}}
We have identified above different scenarios where maximum noise mitigation can be obtained with our SQEM protocols. However, to  achieve $(\omega_1,\omega_2) = (1,1)$, one may require additional resources that themselves can contribute to the noise affecting the output of our SQEM protocols. This is the case with, e.g., entangled auxiliary states (with the overheads and additional noise associated), or with  $\ket{\phi_{\rm f}}=U\ket{\phi_{ 0}}$ which requires an additional application of $U$.

When $(\omega_1,\omega_2) < (1,1)$, it is possible to avoid these hidden resources. For instance, one may restrict the available correction operations to e.g., only single qubit unitary corrections when implementing a multi-qubit gate, or only Clifford operations when implementing a T gate.  As we demonstrate numerically in Sec.~\ref{sec:standardgateperformance} and theoretically in the following, Sec.~\ref{sec:variants}, the SQEM probabilistic protocol is robust for $(\omega_1,\omega_2) < (1,1)$ in a broad range of relevant settings. Finally, to further enhance the advantage of the SQEM protocols, we introduce the nested variation in Sec.~\ref{sec:gate_nested}, which uses different auxiliary states to lower the noise acting on the resulting state after the measurement at step~\ref{prot:gate5}.

\subsection{Protocol post-processing variants }
\label{sec:variants}
Below, we analyze in detail the possible post-processing alternatives in step~\ref{prot:gate6} of our SQEM protocols, namely the probabilistic and the quasi-deterministic versions.

\subsubsection{Probabilistic SQEM}
\label{sec:gate_prob_prot}
The probabilistic protocol post-selects the resulting state $\rho_{\rm out}$ depending on the measurement outcomes of step~\ref{prot:gate5}. Here, for the sake of clearness, we consider the case of local noise and describe a procedure that ensures 
$(\omega_1,\omega_2)=(1,1)$. In these settings, analytical results are more compact, and the basic principles on which our schemes rely are better understood.  We analyze the protocol under the action of depolarizing noise and we generalize afterward to arbitrary noise. We discuss the practical implementation of our schemes and consider more realistic scenarios in which the auxiliary states can be efficiently prepared in the following Sec.~\ref{sec:realistic}.

\textbf{Depolarizing noise.---} Here, we investigate the case of depolarizing noise affecting each of the $m$ qubits undergoing a desired operation $U$, such that the Kraus operators are the ones in Eq.~\eqref{eq:depolarizingnoise}. We remark that depolarizing noise is widely considered as one of the most difficult to handle \cite{nielsen_chuang_2010}, as with probability $2(1-p_0)/3$ the resulting state is completely mixed, i.e., no information survives. 

Recalling that $p_{0}$ is the probability that each of the $m$ qubits is not subject to an error, we have that $p_{\rm ne} = p_0^m$ and the resulting CJ fidelity $F^{0}$ (Sec.~\ref{sec:figuremerit}) in the incoherent case is
\begin{equation}\label{eq:gate_ex_1qubut_incoh}
    F^{0} = p_{\rm ne} = p_{0}^m.
\end{equation}
Employing our protocol with the auxiliary state and measurement in Eqs.~\eqref{eq:bellstate_m}, we find that
\begin{equation}\label{eq:gate_bell_input_cross}
    \bra{\phi_{ 0}} U^\dagger K_i U \ket{\phi_{ 0}} = \sqrt{p_{0}^m} \delta_{i,0}
\end{equation} 
for all $i=0,\dots,4^m-1$.  From Eq.~\eqref{eq:gate_bell_input_cross} we conclude that $\bra{\phi_{\rm f}} U \ket{\phi_0} = 1$ and the auxiliary state is fully sensitive to all Kraus operators acting on the system, such that $(\omega_1, \omega_2) = (1,1)$ making it is possible to employ Eq.~\eqref{eq:gate_out_noise_suppr} for determining $\rho_{\rm out}$.

Specifically, the post-selection probability ${P} = \tr(\rho_{\rm out})$ and a lower bound of the fidelity $F$ of the state are (see also App.~\ref{sec:Appendix_density})
\begin{subequations}
\label{eq:gate_ex_1qubit_out}
\begin{align}
    {P} & = p_0^{m d} + \frac{p_0^{m d}}{d} \left( 
    \frac{1}{p_0^{m}}-1
    \right),\label{eq:gate_ex_1qubit_out_P}\\
    F & \geq \frac{d p_0^m}{p_{0}^{m}(d-1) +1 }\label{eq:gate_ex_1qubit_out_F}
\end{align}
\end{subequations}
where, for the same reasons for which it is maximally sensitive to the noise, the auxiliary state $\ket{\phi_0}$ in Eq.~\eqref{eq:bellstate_m} saturates the inequality in Eq.~\eqref{eq:gate_ex_1qubit_out_F} (and $F$ becomes the CJ fidelity --- see Sec.~\ref{sec:figuremerit}).
For $d\gg 1$, we thus find ${P} \approx p_0^{m d}$ and $F \approx 1$. 
As formally proven below, increasing the number of channels within our protocol is always beneficial, and when the auxiliary state is maximally sensitive to all Kraus operators we consistently achieve unit fidelity for $d \gg 1$.

\textbf{Arbitrary noise.---} Above, we analyzed the protocol performance for depolarizing noise, and demonstrated through Eqs.~\eqref{eq:gate_ex_1qubit_out} that our scheme is always beneficial and achieves unit fidelity for $d \gg 1$. Here, we consider the general case with arbitrary noise. For practical reasons, it is better to work in the process matrix representation, such that the map ${\cal E}_U$ takes the form in Eq.~\eqref{eq:generalmapintro}. 

With the chosen auxiliary $\ket{\phi_0}$ and projecting onto the state $\ket{\phi_{\rm f}}=U\ket{\phi_{ 0}}$, these assumptions can be again formulated as $(\omega_1,\omega_2) = (1,1)$. Furthermore, as one can see from Eq.~\eqref{eq:generalmapintro}, the coefficient associated with the identity Pauli operator is $\lambda_{00}$, such that we have $p_{\rm ne} = \lambda_{00}$.

Through the same steps that took us to Eq.~\eqref{eq:gate_out}, the application of our protocol yields a state $\rho_{\rm out}$ that can be described with a noisy map as in Eq.~\eqref{eq:generalmapintro}, but with coefficients $\lambda_{ij} \rightarrow \lambda'_{ij}$ defined by
\begin{subequations}\label{eq:generallambdageneral}
\begin{align}
    & \lambda'_{00}= \frac{ \lambda^{d }_{00}}{P}
    , \label{eq:generallambdageneral0}
    \\
    & \lambda'_{ij}= \frac{ \lambda_{00}^{(d-2)} \left[ \lambda_{00} \lambda_{ij} + (d-1) \lambda_{i0} \lambda_{0j}\right]}{d P}
    ,  \label{eq:generallambdageneral1}
    \\
    & P = \frac{1}{d} \left[ \lambda_{00}^{ (d-1)} + (d-1)  \lambda_{00}^{ (d-2)} \left(\sum_{i} \left\lvert \lambda_{i0} \right\rvert^2 \right) \right],
     \label{eq:generallambdageneral2}
\end{align}
\end{subequations}
where the $\lambda'$ are normalized to ensure $\tr(\rho_{\rm out}) = 1$.
Here, $P$ is the success probability, and lower bounds for the fidelities after the application of our protocol and the incoherent one are $F = \lambda'_{00}$ and $F^{0}  = \lambda_{00}$, respectively.
We remark that $\lambda_{ij}=\lambda_{ji}^{*}$, which follows from the hermiticity of the map ${\cal E}_U$ in Eq.~\eqref{eq:generalmapintro}, implies $\lambda_{ij}'=(\lambda_{ji}')^*$. Eqs.~\eqref{eq:generallambdageneral} can be derived from Eqs.~\eqref{eq:gate_out} (and vice versa),  via the change of basis described in Sec.~\ref{sec:operatorsum}, which allows to switch from the Kraus to the Pauli decomposition of the noise. 

As shown in App.~\ref{app:Proof}, it is possible to demonstrate that
\begin{equation}\label{eq:finalresult1}
   \lambda'_{00} > \lambda_{00}
\end{equation}
whenever $\lambda_{00} \in (\frac{1}{2},1)$. Furthermore, larger values of $\lambda'_{00}$ are associated with higher $d$. Therefore, our protocol is always advantageous independently of the noise affecting the desired computation, and the resulting fidelity is always enhanced when increasing the number of branches $d$. 

Observe that Eqs.~\eqref{eq:generallambdageneral} can be used to 
rederive the results in Eqs.~\eqref{eq:gate_ex_1qubit_out} when depolarizing noise affects locally each of the $m$ input qubits in an uncorrelated way. This follows from the fact that any diagonal coefficient can be written as $\lambda_{ii}=\sqrt{(1-p_0/3)^k p_0^l}$ for certain non-negative integers $k$ and $l$ such that $k+l = m$, and $\lambda_{ij} = 0$ for all $i\neq j$. The coefficient $\lambda'_{00}$ then reduces to Eq.~\eqref{eq:gate_ex_1qubit_out_F}.

\subsubsection{quasi-deterministic SQEM}
\label{sec:gate_det_prot}
The probabilistic protocol investigated above has the  advantage of being plug-and-play. Even with both the input state $\ket{\psi_{\rm in}}$ and the noise (i.e., the Kraus operators) completely unknown, one can choose an auxiliary state $\ket{\phi_{ 0}}$ and obtain an improved fidelity $F \geq F^0$. The quasi-deterministic protocol discussed in this section, on the other hand, addresses a complementary situation. Namely, it assumes that the user wants to maximize the success probability of the protocol, possibly achieving deterministic advantage.  

As described in Sec.~\ref{sec:gate_protocol}, the only different step between the probabilistic and the quasi-deterministic versions of our scheme is the \ref{prot:gate6}$^{\rm th}$ one. Indeed, while the probabilistic protocol keeps the resulting $\rho_{\rm out}$ only for a single set of measurement outcomes, the quasi-deterministic one allows for more, possibly all (in this case it is fully deterministic). 

However, now correcting unitaries must be applied to the output state, depending on the measurement outcomes of the control and auxiliary states. Determining the correcting unitaries  may be challenging, particularly with many branches and/or large input and auxiliary states. Furthermore, it is fundamental to ensure that these correcting unitaries are characterized by much smaller errors than the desired computation $U$. While an analytical approach can be pursued, in this work we restrict to a numerical analysis of the process. We introduce in the following an optimization process for evaluating the protocol performance. The process is also capable of determining the best auxiliary state and measurement bases to be employed, and is independent of the computation and the noise. 

The idea is to perform an initial calibration to gain knowledge about the noise affecting the setup. This knowledge is then used to choose the best auxiliary states, correcting unitaries, and measurement bases to be employed, such that when the protocol is run with some state $\ket{\psi_{\rm in}}$ as input, the highest possible fidelity is achieved with a desired post-selection probability. 
The $q=\{1,\dots,r\}$ different output states that are post-selected and corrected are jointly considered by evaluating the weighted average of the associated CJ fidelities $F_{\rm CJ}^{(q)}$
\begin{equation}
\label{eq:CJaverage}
    F_{\rm CJ}= \frac{1}{P}\sum_{q=1}^{r} P^{(q)} F_{\rm CJ}^{(q)}, 
\end{equation}
where $P = \sum_{q=1}^{r} P^{(q)}$ is the probability not to discard the measurement outcomes at step~\ref{prot:gate5}.

Since $F_{\rm CJ}$ is a lower bound on the fidelity for an arbitrary input state, it is well suited as a cost function to be maximized for determining the best auxiliary states, correcting unitaries, and measurement bases to be employed. These represent the knobs \footnote{In principle, one could also optimize the state of the control register. While in the settings considered in this work this is redundant, it would be important in the scenario in which noise affects different branches differently.} that can be modified within the quasi-deterministic protocol. In general, one has some freedom in deciding how to tune these knobs, which can be also limited by experimental constraints. Owing to the presence of noise, we follow a specific route (below) to limit the detrimental effects from various sources of decoherence. However, we remark that optimizing the CJ fidelity can be done in different ways, and in principle, one may even want to use different cost functions, e.g., the Haar fidelity \cite{Nielsen2002}, that may be more suited to the specific setup in consideration. 

In general, varying both the auxiliary state and the measurement basis may seem redundant, particularly if one wants to ensure that $U\ket{\phi_{ 0}}$ is one of the elements that can be discerned by the measurement. 
However, since in practical implementations there can be inhomogeneities that are not captured by our analytic calculations, we keep the measurement bases' optimization independent to compensate for those. For instance, if the branches were characterized by different (possibly correlated) decoherence processes, the extra knob could allow for better performances of the protocol. One of the appealing aspects of the quasi-deterministic scheme is that, being based on a black-box optimization, it is quite insensitive to all sorts of inhomogeneities, even unforeseen ones, as long as it has enough many knobs to act on. Specific performance results are shown in Sec.~\ref{sec:standardgateperformance}.

Given the features of the probabilistic and the  quasi-deterministic protocols, the former approach could be  particularly useful for increasing the fidelity of whole computations, while the latter could be advantageous when one wants to optimize one or a few gates that are
repeated within a larger circuit. This scenario is encountered in most quantum computations, where a set of universal gates is used for carrying out any desired unitary transformation \cite{nielsen_chuang_2010}. Our quasi-deterministic protocol can then be used to enhance the fidelity of each of those gates (or the ones most susceptible to errors), while keeping a large overall post-selection probability. 

\subsection{Protocol extensions} \label{sec:realistic}
In the sections above, we have introduced and analyzed the probabilistic and quasi-deterministic SQEM protocols. The first can be employed without any knowledge of the noise acting on the system, and always yields an improved fidelity compared to the incoherent one. The latter optimizes the choice of several parameters for further enhancing the process success probability. 

As it is possible to infer from Eq.~\eqref{eq:gate_out_main}, however, when the auxiliary state is (partially) insensitive to the noise, even in the asymptotic case $d \gg 1$ unit fidelity cannot be achieved.
Furthermore, in several experimental scenarios, the decoherence is such that it is not possible to choose auxiliary states that are fully sensitive to the noise --- consider, for instance, the depolarizing channel.

In the following, we introduce a nested extension of our protocol that employs an iterated application of either the probabilistic or the quasi-deterministic scheme. It is capable of correcting for most errors and greatly enhances the fidelity of the output state $\rho_{\rm out}$.

\subsubsection{Nested SQEM protocol}
\label{sec:gate_nested}
\begin{figure}
    \includegraphics[width=\columnwidth]{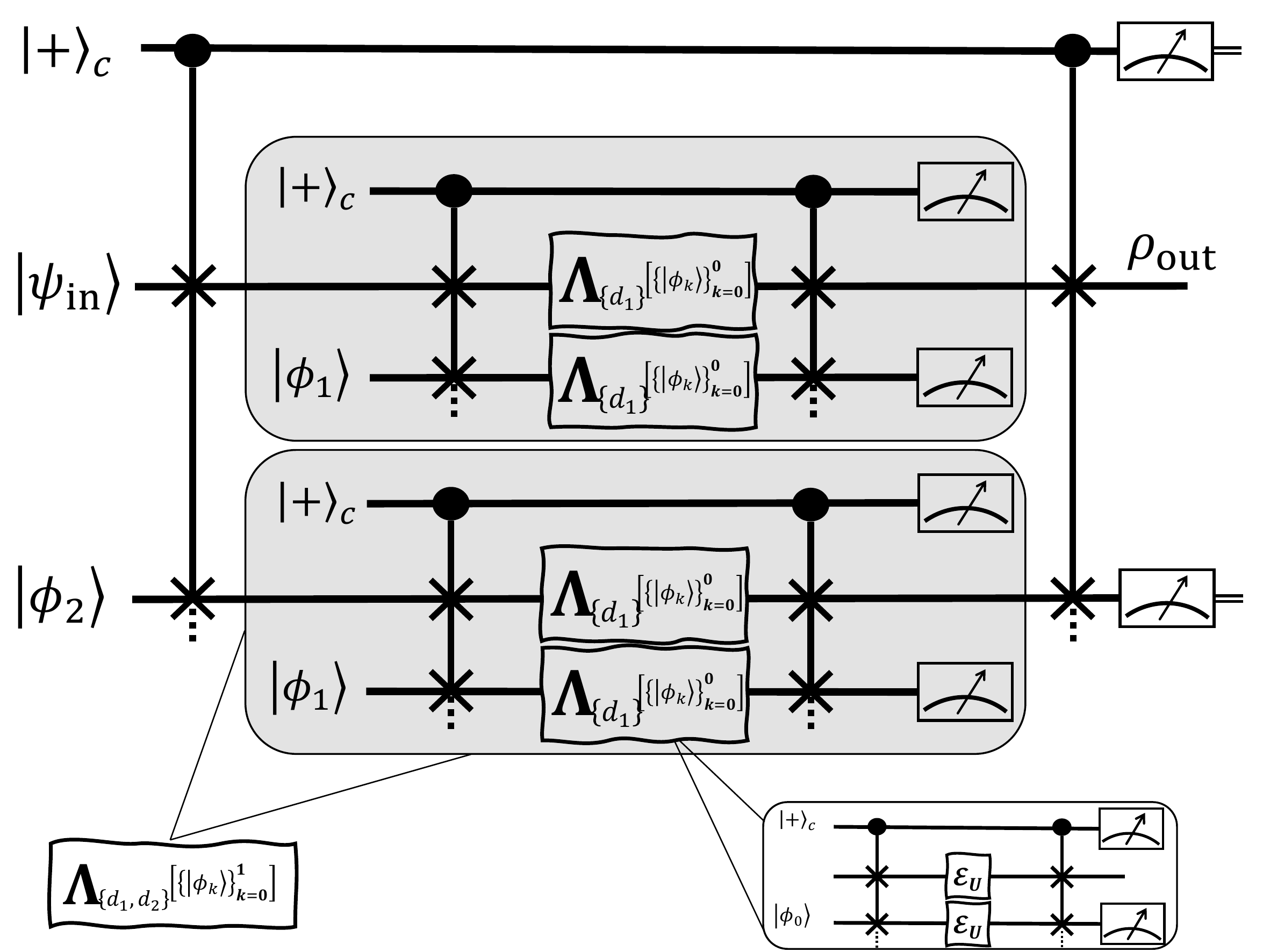}
    \caption{Example of $3$ concatenations of the nested approach that allows one to effectively implement our protocol with a superposition of $n$ channels by using only the basic cSWAP operations, Eq.~\eqref{eq:GB-QCcswap}, for $d=2$.} \label{fig:GB_scalable} 
\end{figure}

In this section we introduce an extension of our SQEM protocol for, first, scaling up the setup without the requirement of additional higher dimensional systems and operations, and second, maximizing error mitigation when using unentangled auxiliary states. As schematically represented in Fig.~\ref{fig:GB_scalable}, this extension consists in nesting basic runs of the protocol into each other, such that each run is an effective building block of the next one. Based on this observation, it is then possible to understand and analytically describe the effective action of the whole nested extension by studying each building block separately.

The fidelity enhancement from the nested extension comes from using different auxiliary states such that, if one is (partially) insensitive to a kind of noise characterized by one or more Kraus operators, another will compensate for that. Different auxiliary states $\ket{\phi_0}$, $\ket{\phi_1}$, ..., $\ket{\phi_{n-1}}$ work together at each of the different $n$ layers, each correcting certain types of decoherence, such that the resulting state approximates well the desired one. 

We remark that it is in general not possible to directly employ different auxiliary states in either the probabilistic or the quasi-deterministic protocol. By dedicating a fraction of the $d$ branches to $\ket{\phi_0}$, another one to $\ket{\phi_1}$ and so on, we find cross terms such as $
\bra{\phi_k} U^{\dagger} K_j U \ket{\phi_k}
\bra{\phi_l} U^{\dagger} K_i^{\dagger} U \ket{\phi_l}
$ in $\rho_{\rm out}$ that, for different $k$ and $l$, retain errors in the computation. The nested extension cancels in consecutive iterations \textit{each} kind of noise contribution. This is the reason behind its iterative nature, which is evident from Fig.~\ref{fig:GB_scalable} and explained in more detail below.

As one can see in the figure, the fundamental idea is that each application of either the probabilistic or the quasi-deterministic SQEM protocol within the nested scheme can be viewed as a noisy map that depends on all previous applications. Therefore, the whole process is described by $\Lambda_{\vec{d}} \left[ \lbrace \ket{\phi_k} \rbrace_{k=0}^{n-1} \right]$, where $n$ and $\vec{d} = \lbrace d_0,\dots d_{n-1} \rbrace$ refer to the numbers of iterations and their associated branches, respectively. For completeness, we also indicated the auxiliary states $\lbrace \ket{\phi_k} \rbrace_{k=0}^{n-1}$ used at each application. 

To better understand how the nested extension works, we first give the map $\Lambda_{\vec{d}} \left[ \lbrace \ket{\phi_k} \rbrace_{k=0}^{n-1} \right]$ for a specific example, and later on discuss its form in more general settings. For making the following clearer, in the remainder of this section we consider the scenario in which we apply the nested extension to the probabilistic protocol with $\omega_2 = 1$, i.e., we assume that at step~\ref{prot:gate5} (see Sec.~\ref{sec:gate_protocol}) of the $k$-th iteration we always project control and auxiliary registers onto
\begin{equation}\label{eq:projected_prob_right}
    \ket{+_{d_{k}}}_{\rm c} \bigotimes_{j=1}^{d_k-1}  U\ket{\phi_{k}}_{{\rm b}_j},
\end{equation}
for all $k=0,\dots,n-1$. We remark, however, that with the quasi-deterministic version there is no qualitative difference. The maps $\Lambda_{\vec{d}} \left[ \lbrace \ket{\phi_k} \rbrace_{k=0}^{n-1} \right]$ and the probability with which we post-select the outcomes would generally vary, but the physics behind remains unchanged. 

\textbf{Same auxiliary state.---} An illustrative way to explain how the nested extension works is to analyze its characterizing map $\Lambda_{\vec{d}} \left[ \lbrace \ket{\phi_0} \rbrace_{k=0}^{n-1} \right]$ with $\omega_1 = 1$ and all auxiliary states being the same $\ket{\phi_0}$ for all $k = 0,\dots,n-1$. As previously done for deriving Eq.~\eqref{eq:gate_out_noise_suppr}, we assume $\bra{\phi_{0}} U^\dagger K_i U \ket{\phi_{ 0}} = \sqrt{p_{\rm ne}} \delta_{i,0}$, i.e., the auxiliary and $\ket{\phi_{\rm f}} = U \ket{\phi_0}$ are chosen to be fully sensitive to all Kraus operators. Furthermore, for simplicity, we assume here that at each iteration of the protocol we employ the same number of branches, i.e., $d_k = d$ for all $k=0,\dots,n-1$.

For $n=1$ the nested scheme is the same as the probabilistic protocol. Therefore, the map $\Lambda_{\vec{d}} \left[ \lbrace \ket{\phi_0} \rbrace_{k=0}^{0} \right]$  applied to $\proj{\psi_{\rm in}}$ yields the output state in Eq.~\eqref{eq:gate_out_noise_suppr}.
From this equation, we conclude that the first ($n=1$) application of the protocol can be described with a noisy map with new Kraus operators
\begin{equation}
\label{eq:mat_prot_first_out}
\begin{split}
    \rho_{\rm out} = & 
    \Lambda_{d} \left[ \lbrace \ket{\phi_0} \rbrace_{k=0}^{0} \right] 
    \left( \proj{\psi_{\rm in}} \right)
    \\ = &
    \sum_i K_{i}^{(1)} U \proj{\psi_{\rm in}} U^\dagger  \left(K_{i}^{(1)}\right)^{\dagger},
\end{split}
\end{equation}
where $K_i^{(1)}$ takes different forms depending on $d$. Specifically, we have that
\begin{subequations}
\label{eq:new_krauses_mat_simple}
\begin{align}
    K_{0}^{(1)} & 
    = \sqrt{\frac{d}{(d-1) p_{\rm ne} + 1}} K_0 
    ,\label{eq:new_krauses_mat_simple_id}\\
    K_{i}^{(1)} & 
    = \sqrt{\frac{1}{(d-1) p_{\rm ne} + 1}} K_i
    \text{ for $i \geq 1$}
    ,\label{eq:new_krauses_mat_simple_others}
\end{align}
\end{subequations}
with $K_i$ being the original Kraus operators describing the noise in the incoherent case.

Once it is understood how the map $\Lambda_{\vec{d}} \left[ \lbrace \ket{\phi_0} \rbrace_{k=0}^{0} \right]$ looks for $n=1$, it is possible to investigate what happens when more iterations are performed ($n \geq 2$): the noise affecting the input state within the $k^{\rm th}$ application is described by the Kraus operators found in the $(k-1)^{\rm th}$ one. For instance, in the considered example and for $n=2$ the map $\Lambda_{\vec{d}} \left[ \lbrace \ket{\phi_0} \rbrace_{k=0}^{1} \right]$ is found following the same steps that took us to Eqs.~\eqref{eq:mat_prot_first_out} and \eqref{eq:new_krauses_mat_simple} with the substitution $K_i \rightarrow K_{i}^{(1)}$. Therefore, after $n$ applications we find 
\begin{subequations}
\label{eq:new_krauses_mat_gen}
\begin{align}
\begin{split}
    \rho_{\rm out} = & 
    \Lambda_{\vec{d}} \left[ \lbrace \ket{\phi_0} \rbrace_{k=0}^{n-1} \right] 
    \left( \proj{\psi_{\rm in}} \right)
    \\ = &
    \sum_i K_{i}^{(n)} U \proj{\psi_{\rm in}} U^\dagger  \left( K_{i}^{(n)} \right)^{\dagger},
\end{split}
    \\
    K_{0}^{(n)} = & 
    \sqrt{\frac{\beta_{d,n}}{(\beta_{d,n} - 1) p_{\rm ne} + 1}} K_0 
    ,\label{eq:new_krauses_mat_gen_id}\\
    K_{i}^{(n)} = & 
    \sqrt{\frac{1}{(\beta_{d,n} - 1) p_{\rm ne} + 1}} K_i
    \text{ for $i \geq 1$}
    ,\label{eq:new_krauses_mat_gen_others}
\end{align}
\end{subequations}
where $\beta_{d,n} = d^{2^{n-1}}$ is the total number of registers employed in the whole protocol.

From these last equations, it is finally possible to obtain a lower bound for the fidelity $F[d,n]$ of the whole nested protocol when $n$ iterations each with $d$ branches are employed,
\begin{equation}\label{eq:fidelity_mat_same_aux}
     F[d,n] \geq 
     1 - \frac{1 - p_{\rm ne}}{1+\left( \beta_{d,n} -1 \right) p_{\rm ne}}.
\end{equation}
The lower bound is derived as in Sec.~\ref{sec:gate_prob_prot}, namely, by assuming that the decoherence affecting the system is maximally detrimental to $U\ket{\psi_{\rm in}}$, resulting in an incoherent fidelity of $F^{0} = p_{\rm ne}$. As one can see from the equation, $F[d,n] \rightarrow 1$ when either the parameter $d$ or $n$ is much bigger than one. Specifically, we can set $d=2$ and $n \gg 1$ to get unit fidelity.

Before considering the more interesting case in which we use different auxiliary states at each iteration, let us make an important remark. The fidelity resulting from the application of the nested extension in Eq.~\eqref{eq:fidelity_mat_same_aux} is the same as found for the probabilistic protocol in Eq.~\eqref{eq:gate_ex_1qubit_out_F}, provided we substitute $\beta_{d,n} \leftrightarrow d$. This is unsurprising, considering that we always employ the same auxiliary state $\ket{\phi_0}$, which is maximally sensitive to the noise. However, this does not mean that the practical implementations of these two protocols are equally demanding. Compared to the probabilistic protocol, the nested scheme with $d=2$ requires simpler cSWAP operations to generate the superposition [see Eq.~\eqref{eq:GB-QCcswap}]. In particular, having two parameters ($d$ and $n$) to tune can be useful when we analyze the influence of the noise resulting from the application of different numbers of cSWAPs with varying $d$.

\textbf{Different auxiliary states.---} Above, we have investigated how the nested extension works when employing, at each application, the same auxiliary state $\ket{\phi_{0}}$ that is maximally sensitive to the noise. Here, we examine the more interesting case in which at each iteration we apply a different $\ket{\phi_{k}}$, $k=0,\dots,n-1$. We also consider general settings; namely, that there are kinds of noise to which each auxiliary state is (partially) insensitive. Therefore, within a given $k$-th application of the nested scheme, Eq.~\eqref{eq:gate_out_noise_suppr} is not valid anymore. Instead, we must fall back on Eq.~\eqref{eq:gate_out_main} with $\ket{\phi_{\rm f}}$ as in Eq.~\eqref{eq:projected_prob_right}.

Following the same steps as above, we find recursive relations (in $n$) for the maps $\Lambda_{\vec{d}} \left[ \lbrace \ket{\phi_k} \rbrace_{k=0}^{n-1} \right]$. Unfortunately, when the auxiliary states are not fully sensitive to the noise, Eqs.~\eqref{eq:new_krauses_mat_gen} also cease to be valid. Because of the cross terms on the right-hand side of Eq.~\eqref{eq:gate_out_main}, the representation of $\Lambda_{\vec{d}} \left[ \lbrace \ket{\phi_k} \rbrace_{k=0}^{n-1} \right]$ in terms of Kraus operators cannot be simply derived. To avoid this problem, we make use of the process matrix representation  in Eq.~\eqref{eq:generalmapintro}. Rather than the Kraus operators, we then base our analysis on the coefficients $\lambda_{ij}$ and investigate how they are updated at each application of the nested protocol. 

By plugging Eqs.~\eqref{eq:krausinpauli} and \eqref{eq:def_lambda_coef} into Eqs.~\eqref{eq:gate_out}, we express 
\begin{widetext}
\begin{subequations}
\label{eq:matryoshka_general}
\begin{align}
    \rho_{\rm out} = & \Lambda_{\vec{d}} \left[ \lbrace \ket{\phi_k} \rbrace_{k=0}^{n-1} \right]\left( \proj{\psi_{\rm in}} \right) 
    = 
    \sum_{i,j}\lambda_{ij}^{(n-1)}\sigma_i U \proj{\psi_{\rm in}} U^{\dagger}\sigma_{j}^{\dagger}
    ,\label{eq:matryoshka_general_out}
    \\
    \lambda_{ij}^{(k)} = & \frac{\mathcal{A}_{d_{k}}^{(k)}}{d_{k}}\left[
    \lambda_{ij}^{(k-1)} + \frac{d_{k}-1}{\mathcal{A}_{2}^{(k)}}
    \sum_{n,m} \lambda_{mj}^{(k-1)}\lambda_{in}^{(k-1)}
    \bra{\phi_{k}}U^{\dagger}\sigma_{m}U\ket{\phi_{k}}
    \bra{\phi_{k}}U^{\dagger}\sigma_{n}^{\dagger}U\ket{\phi_{k}}
    \right]
    ,\label{eq:matryoshka_general_coeffs}
    \\
    \mathcal{A}_{d}^{(k)} = & \left( \sum_{i,j} \lambda_{ij} \bra{\phi_{k}}U^{\dagger}\sigma_{i}U\ket{\phi_{k}}
    \bra{\phi_{k}}U^{\dagger}\sigma_{j}^{\dagger}U\ket{\phi_{k}} \right)^{d-1},
    \label{eq:matryoshka_general_Ad}
\end{align}
\end{subequations}
\end{widetext}
where $k=0,\dots,n-1$ and in Eq.~\eqref{eq:matryoshka_general_coeffs} for $k=0$ the coefficients $\lambda_{ij}^{(-1)} = \lambda_{ij}$ are those characterizing the channel in the incoherent case. We remark that, as in Eqs.~\eqref{eq:gate_out}, the resulting density matrix $\rho_{\rm out}$ is not normalized. $\tr \left( \rho_{\rm out} \right)$ is the post-selection probability associated with all measurement outcomes at step~\ref{prot:gate5} of each iteration yielding the desired result [see Eq.~\eqref{eq:projected_prob_right}].

While cumbersome to evaluate, Eqs.~\eqref{eq:matryoshka_general} can be qualitatively investigated to help understand several characteristics of the nested protocol. Specifically, there are two fundamental observations on which we base our following strategy in terms of auxiliary states $\ket{\phi_k}$ and $\vec{d}$ to be employed. The first is that the Pauli operator $\sigma_0 = \id$ is ``special'', as $\bra{\phi_{k}}U^{\dagger}\sigma_{0}U\ket{\phi_{k}} = 1$ for all $U$ and $\ket{\phi_{k}}$. On the one hand, since $\sigma_0 = \id$ is associated with the absence of noise, this is the reason for which our protocol works, as the corresponding term $\lambda_{00}^{(k)}$ in Eq.~\eqref{eq:matryoshka_general_coeffs} is generally enhanced. On the other hand, this also limits the maximum achievable noise suppression. In fact, when $\omega_1 < 1$ [see Eq.~\eqref{eq:omegas_1}] for each iteration $k=0,\dots,n-1$, the auxiliaries are (partially) insensitive to the noise and there are cross terms contributing to $\lambda_{0i}^{(k)}$ and $\lambda_{i0}^{(k)}$ for $i \neq 0$ in Eq.~\eqref{eq:matryoshka_general_coeffs}. These cross terms cannot be suppressed at the next iterations by using another auxiliary state that is insensitive to the associated noise, as one of the corresponding Paulis is the identity and as such will always survive. How to limit their detrimental effect is related to the second observation below.

Even though the highest achievable fidelity is ultimately limited by the coefficients $\lambda_{i0}^{(n)}$ ($i \neq 0$), 
it is possible to suppress them via interference between subsequent iterations. For simplicity, let us consider the $k$-th and $(k+1)$-th applications of the nested protocol, and assume that $\lambda_{0i}^{(k-1)} = \lambda_{i0}^{(k-1)} = 0$ for all $i \neq 0$.
The idea is that if the state $U\ket{\phi_k}$ is insensitive to a specific kind of noise ---say $\sigma_i$ ($i \neq 0$)--- then after the $k$-th iteration there will be nonzero contributions to $\lambda_{i0}^{(k)}$ and $\lambda_{0i}^{(k)}$ [see Eq.~\eqref{eq:matryoshka_general_coeffs}]. These contributions will have the sign of $ \bra{\phi_{k}}U^{\dagger}\sigma_{i}^{\dagger}U\ket{\phi_{k}}$, which depends on the auxiliary $\ket{\phi_k}$. Therefore if, at the next iteration $k+1$, we choose an auxiliary $\ket{\phi_{k+1}}$ such that $\bra{\phi_{k+1}}U^{\dagger}\sigma_{i}^{\dagger}U\ket{\phi_{k+1}}$ has the opposite sign, we can reduce the magnitude of both $\lambda_{i0}^{(k+1)}$ and $\lambda_{0i}^{(k+1)}$ (compared to the previous iteration) and thus enhance the fidelity of the output. 

We remark that this trick does not set $\lambda_{i0}^{(k+1)}$ and $\lambda_{0i}^{(k+1)}$ to zero, as their relative change depends on the other values $\lambda_{ij}^{(k)}$, which have been modified after the $k$-th iteration. However, as long as $d_{k}$ and $d_{k+1}$ are both small, cancellation of $\lambda_{i0}^{(k+1)}$ and $\lambda_{0i}^{(k+1)}$ can be substantial and leads to improvements of several orders of magnitude in the resulting fidelity (see Sec.~\ref{sec:standardgateperformance}). The reasons for which $d_{k}$ and $d_{k+1}$ must be small is that if (say) the second is negligible compared to the first, then we suppress too much the coefficients $\lambda_{ij}^{(k)}$ that at the following $(k+1)$-th iteration contribute to the cancellation of $\lambda_{i0}^{(k+1)}$ and $\lambda_{0i}^{(k+1)}$.

Based on these two observations we outline our strategy, in terms of $\vec{d}$ and $\ket{\phi_{k}}$ ($k=0,\dots n-1$), when applying the nested protocol. Specifically, on the one hand, the procedure should employ different auxiliary states such that for each kind of noise, there is at least one $\ket{\phi_{k}}$ that is sensitive to it. On the other hand, the chosen $\ket{\phi_{k}}$ must ensure as much suppression as possible of the elements $\lambda_{i0}^{(k+1)}$ and $\lambda_{0i}^{(k+1)}$ ($i \neq 0$).
For the reasons explained above, we choose the number of branches $d_k$ at each iteration to be $d_k = 2$ for all $k$. This ensures that at each step $k$ the relative change in the magnitude of $\lambda_{ij}^{(k)}$ is small enough to be able, at the next iteration, to compensate for the detrimental contributions to $\lambda_{i0}^{(k+1)}$ and $\lambda_{0i}^{(k+1)}$ ($i\neq 0$). This compensation is then secured by choosing auxiliary states $\ket{\phi_k}$ such that the signs of $ \bra{\phi_{k}}U^{\dagger}\sigma_{i}^{\dagger}U\ket{\phi_{k}}$ are, for all $i > 0$ and in subsequent iterations, as different as possible. A simple way for doing this is to set 
\begin{equation}\label{eq:aux_mat}
    \begin{split}
        \left\lbrace \ket{\phi_k} \right\rbrace_{k=0}^{n-1} = \lbrace & \ket{1}^{\otimes m}, \ket{0}^{\otimes m}, \ket{+}^{\otimes m}, \ket{-}^{\otimes m},
        \\
        & 
        \ket{R}^{\otimes m}, \ket{L}^{\otimes m}, \dots \rbrace,
    \end{split}
\end{equation}
where dots indicate repeating the pattern and $\ket{R}$, $\ket{L}$ are the eigenstates of the Pauli $Y$ with eigenvalues $\pm 1$, respectively. 

We highlight that our strategy, with $d_k = 2$ for all $k=0,\dots,n-1$ and the auxiliaries $\left\lbrace \ket{\phi_k} \right\rbrace_{k}$ as in Eq.~\eqref{eq:aux_mat}, may not be optimal. Specifically, different approaches may yield better results depending on the input state and the noise. Also, in higher dimensions $m>1$, it is possible to devise alternatives based on other stabilizer states. However, as demonstrated in Sec.~\ref{sec:standardgateperformance}, this strategy consistently and considerably outperforms the probabilistic protocol, is noise and input state independent, and works well even when both $\omega_1$ and $\omega_2$ are much smaller than one.

\subsubsection{Coherent memories}

To conclude this section, we consider a specific case of the nested extension, namely when the applied unitary $U$ is the identity $U=\id$. Rather than a computation this is a quantum memory \cite{Simon2010, Heshami2016}, where information is stored, and the goal is to mitigate the decoherence affecting the input state.

In this case, independently of the noise, we can make use of depolarization techniques \cite{Dur2005} to ensure that the noise affecting any run of the protocol is described by an effective depolarizing channel (see Sec.~\ref{sec:gb_quantum_comp_intro}). This has two important advantages. First, with the Kraus operators known (up to a multiplying constant), we can determine a priori the best possible auxiliary state(s) to be employed. Specifically, Bell states are characterized by $(\omega_1,\omega_2) = (1,1)$ while any non-entangled state, thanks to the symmetry of the depolarizing channel, are characterized by the same $\omega_1 < 1$. Second, by means of depolarization techniques \cite{Dur2005}, we are guaranteed to suppress all terms $\lambda_{i0}$ and $\lambda_{0i}$ in Eq.~\eqref{eq:generalmapintro} that are the main limiting factor to the maximum achievable fidelity (see Sec.~\ref{sec:gate_prob_prot}).

We remark that, generally, it is not possible to apply depolarization techniques when both $U$ and the noise are arbitrary \cite{Dur2005}, and therefore generalization of this simplified approach is only possible for some selected $U$ and kinds of noise.

\subsection{Protocol performance. Numerical analysis} \label{sec:standardgateperformance}

In this section, we study the performance of the probabilistic, quasi-deterministic and nested SQEM strategies under different assumptions and settings. As a figure of merit, we consider the infidelity ratio $\mathcal{R}$ between the incoherent and coherent results:
\begin{equation}\label{eq:cj_fid_ratio}
    \mathcal{R} = \frac{1-F_{\rm CJ}^{0}}{1-F_{\rm CJ}},
\end{equation}
where $F_{\rm CJ}$ ($F_{\rm CJ}^{0}$) is the CJ fidelity when our protocol is (not) employed. Note that $F_{\rm CJ}^{0}=p_{\rm ne}$ --- see Sec.~\ref{sec:figuremerit} for more details.
To characterize our protocols, we consider first the ideal case in which the control register and the cSWAP operations are perfect. In real-world scenarios, however, cSWAP gates are noisy and will contribute to the infidelity of the resulting state $\rho_{\rm out}$. Therefore we also include the noise affecting the cSWAP and control registers, and numerically demonstrate that it sets an upper bound on the maximum achievable fidelity. This suggests that most advantage is obtained with the nested extension (with $d_k=2$ for each $k=0,\dots,n-1$, see Sec.~\ref{sec:gate_nested}), and when the considered $U$ comprises several gates whose total noise contribution is dominant over that of the cSWAP.  We consider Pauli rank-2 (e.g., dephasing) and depolarizing noises associated with the imperfect implementations of $U$'s.

In our analysis we study two of the building blocks for arbitrary computations. We analyze the T and the cNOT gates for systems of one and two input qubits, respectively. In the former case, Clifford operations in, e.g., the preparation of ancilla qubits or correcting operations are assumed noiseless by analogy with magic state distillation \cite{Campbell2017}. In the cNOT case, single-qubit states and operations are assumed noiseless by analogy with standard multi-qubit computation formalisms \cite{Preskill2018}. We remark, however, that these assumptions are not required for the SQEM protocols to work and are only meant to make the results clearer. We then extend the study to concatenation of gates.
\begin{figure}
    \includegraphics[width=\columnwidth]{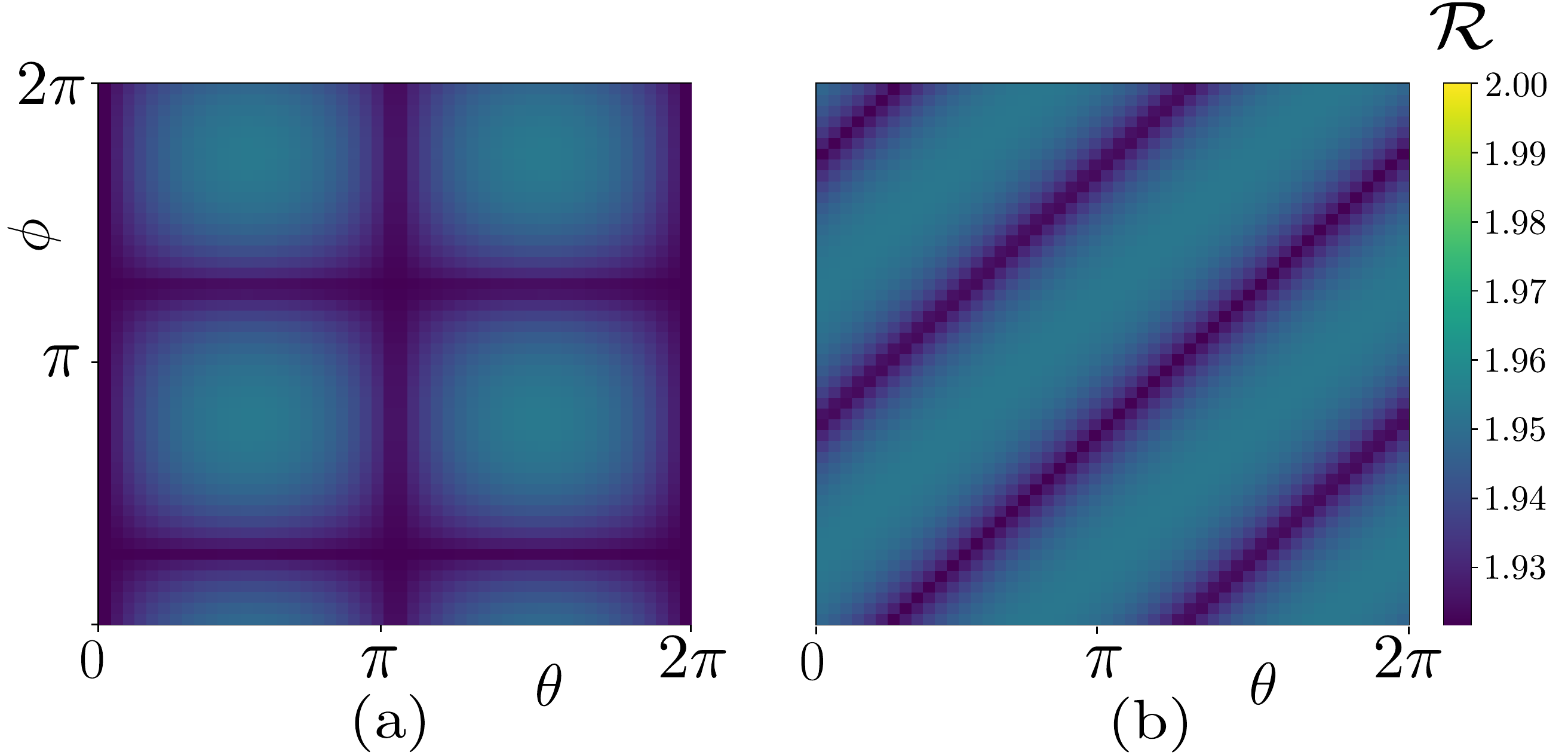}
    \caption{Infidelity ratio $\mathcal{R}$ [see Eq.~\eqref{eq:cj_fid_ratio}] of the probabilistic protocol for a $U=T$ gate with depolarizing noise, $d=2$ and $p_{0}=0.97$ [see Eq.~\eqref{eq:depolarizingnoise}]. In (a), the auxiliary state is set to $\ket{\phi_{0}} = \cos{(\theta/2)}\ket{0} + \sin{(\theta/2)} e^{i \phi} \ket{1}$, while the measurement basis is the $X$ basis. In (b), the auxiliary state is $\ket{\phi_{0}} = (\ket{0} + e^{i \theta} \ket{1})/\sqrt{2}$ and the measurement basis contains the element $(\ket{0} + e^{i \phi} \ket{1})/\sqrt{2}$. In both panels, the chosen state $\ket{\phi_{\rm f}}$ for the post-selection at step~\ref{prot:gate5} of the protocol is the one characterized by the largest probability.}
    \label{fig:differentaux} 
\end{figure}
\begin{figure*}
    \includegraphics[scale=0.33]{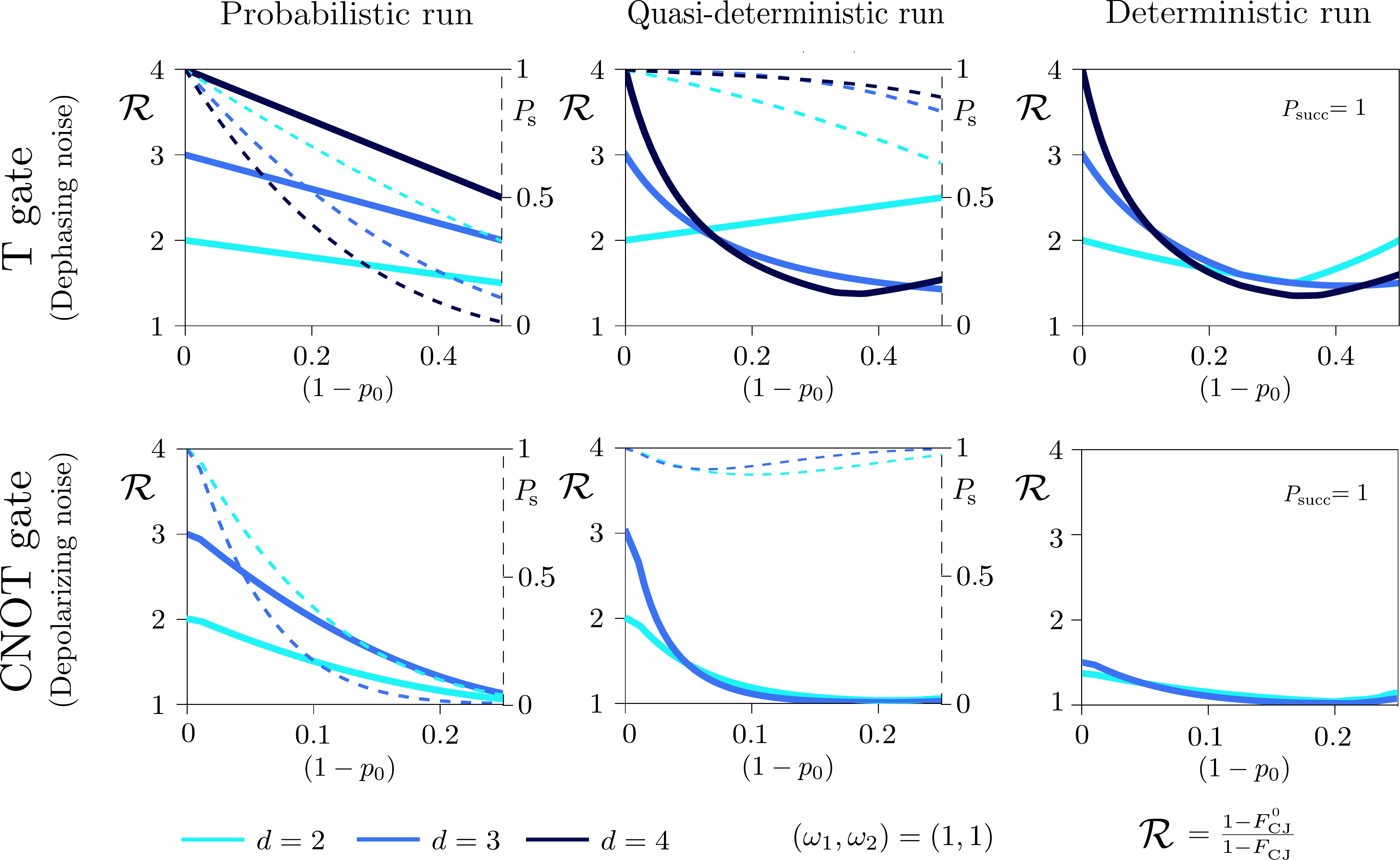}
    \caption{Performance of the probabilistic (first column), quasi-deterministic (second column) and completely deterministic (third column) protocols applied to a T gate (first row, $m=1$) and a cNOT gate (second row, $m=2$). For the T (cNOT) gate we consider dephasing (depolarizing) noise, with single qubit no-error probability $p_0$ [see Sec.~\ref{sec:gb_quantum_comp_intro}]. Colours are used to indicate different values of $d$, as shown in the legend, and we set $(\omega_1,\omega_2) = (1,1)$. For the quasi-deterministic and completely deterministic protocols, only single-qubit Clifford operations are considered as correcting unitaries [see Sec.~\ref{sec:gate_det_prot}].
    }\label{fig:GB_T_depha} 
\end{figure*}
\subsubsection{General protocol performance}
We start our analysis by showing how the performance of our protocols is not jeopardized when $(\omega_1, \omega_2) < (1,1)$ in Eq.~\eqref{eq:omegas}.
In Fig.~\ref{fig:differentaux} we set $d=2$ and present the advantage of the probabilistic protocol when using different auxiliary states and measurement bases. Specifically,  $\ket{\phi_{\rm f}}$ is generally different from $U\ket{\phi_{0}}$, which in turn is not maximally sensitive to the noise. We consider the $U=T$ gate with depolarizing noise, and the auxiliary states are chosen as single-qubit states around the Bloch sphere. As one can see from the figure, $\mathcal{R}$ varies depending on both $\ket{\phi_{\rm f}}$ and $\ket{\phi_0}$. Yet, even for the smallest possible $\omega_2$, $\mathcal{R}$ is consistently above $1$. The advantage comes from the fact that when the probability of not having an error $p_{\rm ne}$ is larger than $0.5$, it is always more likely to post-selected the outcome that is associated with no errors.

The performance of all our protocols for imperfect T and cNOT gates affected by either dephasing or depolarizing noise is reported in Fig.~\ref{fig:GB_T_depha} for different values of $d$. As discussed in Sec.~\ref{sec:gate_prob_prot}, when increasing the number of branches $d$ in the probabilistic protocol, $\mathcal{R}$ is always enhanced. Specifically, for $(\omega_1,\omega_2) = (1,1)$ we have that $\mathcal{R} = p_{\rm ne}(d-1) + 1$, where $p_{\rm ne}$ equals $p_0$ and $p_0^2$ for the T and the cNOT imperfect gates, respectively. Notice that in the limit $p_{\rm ne} \rightarrow 1$ we find $\mathcal{R} = d$, in agreement with the figure. 

While the probabilistic protocol consistently achieves sizable improvements, the associated success probability (see right-hand side axes) is also the lowest, and is reduced exponentially when increasing $d$. To overcome this limitation, we apply the quasi-deterministic protocol introduced in Sec.~\ref{sec:gate_det_prot}. For the sake of convenience, instead of setting a fixed lower limit for the post-selection probability, here we discard only the outputs $q$ with the worst fidelity $F_{\rm CJ}^{(q)}$ in Eq.~\eqref{eq:CJaverage} at step~\ref{prot:gate6} of the protocol. As one can see in the central panels of Fig.~\ref{fig:GB_T_depha}, the quasi-deterministic protocol has similar performances as the probabilistic one for large no-error probabilities $p_{\rm ne}$. In particular, it achieves the same limit $\mathcal{R} \rightarrow d$ for $p_{\rm ne} = 1$. Yet, to ensure high post-selection probabilities, it must include outcomes that are generally associated with low fidelities. Hence, when increasing $d$ the quasi-deterministic scheme is ensured to yield higher $\mathcal{R}$ only for large values of $p_{\rm ne}$. However, there are instances in which some of the outcomes that were discarded by the probabilistic protocol can have a higher (on average) fidelity. This is the case of the $d=2$ quasi-deterministic advantage for the T gate, which is consistently higher than the corresponding advantage of the probabilistic scheme. 

Similar conclusions are drawn for the deterministic protocol, whose results are reported in the rightmost panels of Fig.~\ref{fig:GB_T_depha}. Without any post-selection, all outcomes are kept, including the ones characterized by the smallest fidelity. Therefore, the obtained values of $\mathcal{R}$ are always lower compared to the quasi-deterministic protocol. Yet, even in this case $\mathcal{R} > 1$ for a wide range of the no-error probability $p_{\rm ne}$, showing consistent advantage against the incoherent case.
\begin{figure}
    \includegraphics[width=\columnwidth]{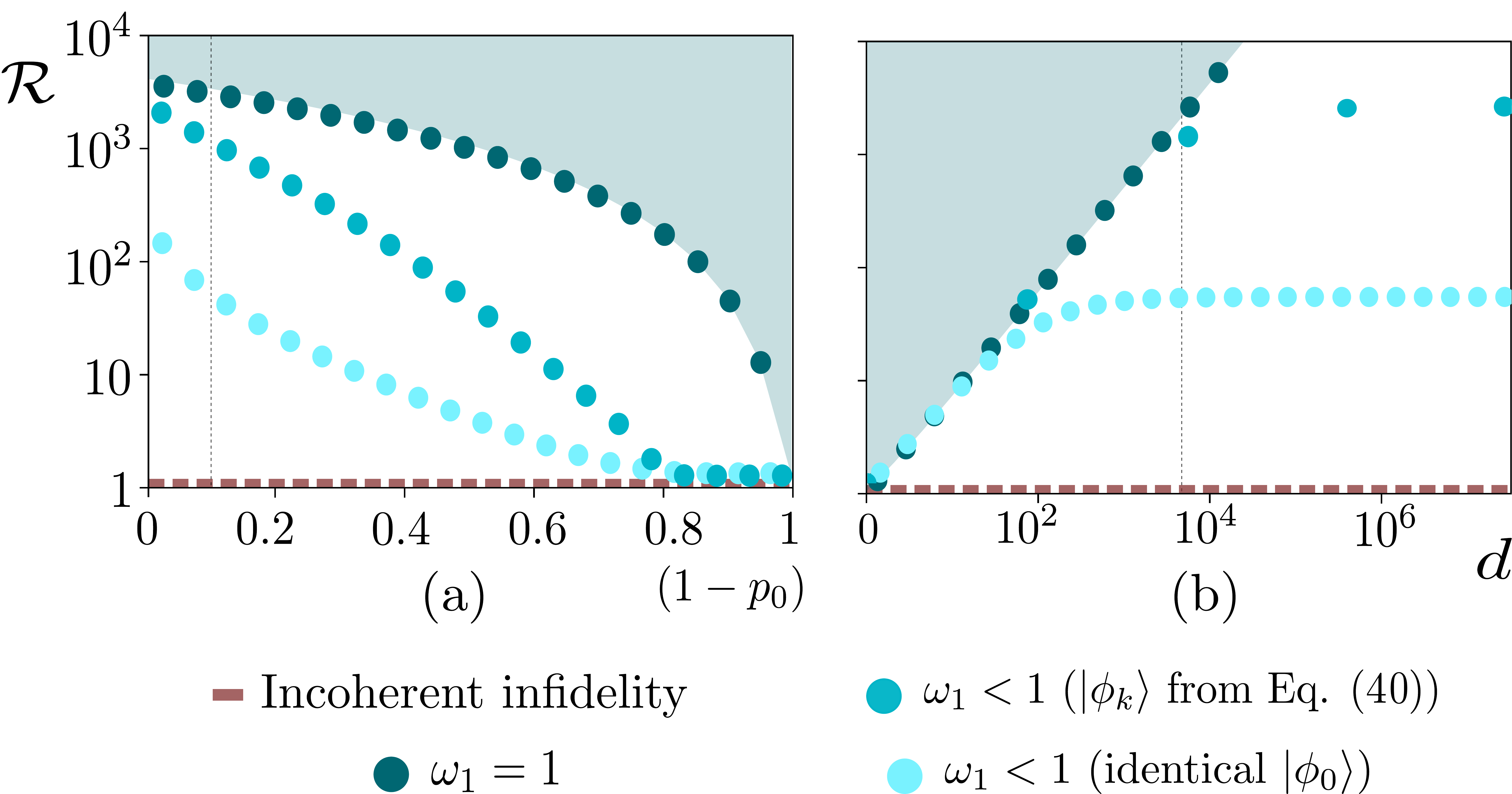}
    \caption{
    Performance of the nested protocol applied to a cNOT ($m=2$) with depolarizing noise. In panel (a) we vary the single qubit no-error probability $p_0$ [see Sec.~\ref{sec:gb_quantum_comp_intro}], while in (b) we vary the number of iterations $n$ and the associated total number of auxiliary states (equivalent branches) employed $d_{\rm tot} = \sum_{k=0}^{n-1} d_k$. Dark blue dots are obtained with $\omega_1 = 1$, and represent the upper bound on the ratio $\mathcal{R}$ that can be achieved by our protocols (hence the shadowed area). Light blue dots are derived for $\ket{\phi_{k}} = \ket{++}$ for all $k = 0,\dots,n-1$, while blue dots are found from the auxiliary states in Eq.~\eqref{eq:aux_mat}. 
    Vertical lines indicate the values of $p_0 = 0.9$ and $n=12$ (i.e., $d_{\rm tot} = 4096$) that are employed in the other panel. We set $\omega_2 = 1$ for clarity, and numerical results are obtained by simulating Eqs.~\eqref{eq:matryoshka_general}.
    } \label{fig:nestedperformance} 
\end{figure}
\begin{figure}
    \includegraphics[width=0.9\columnwidth]{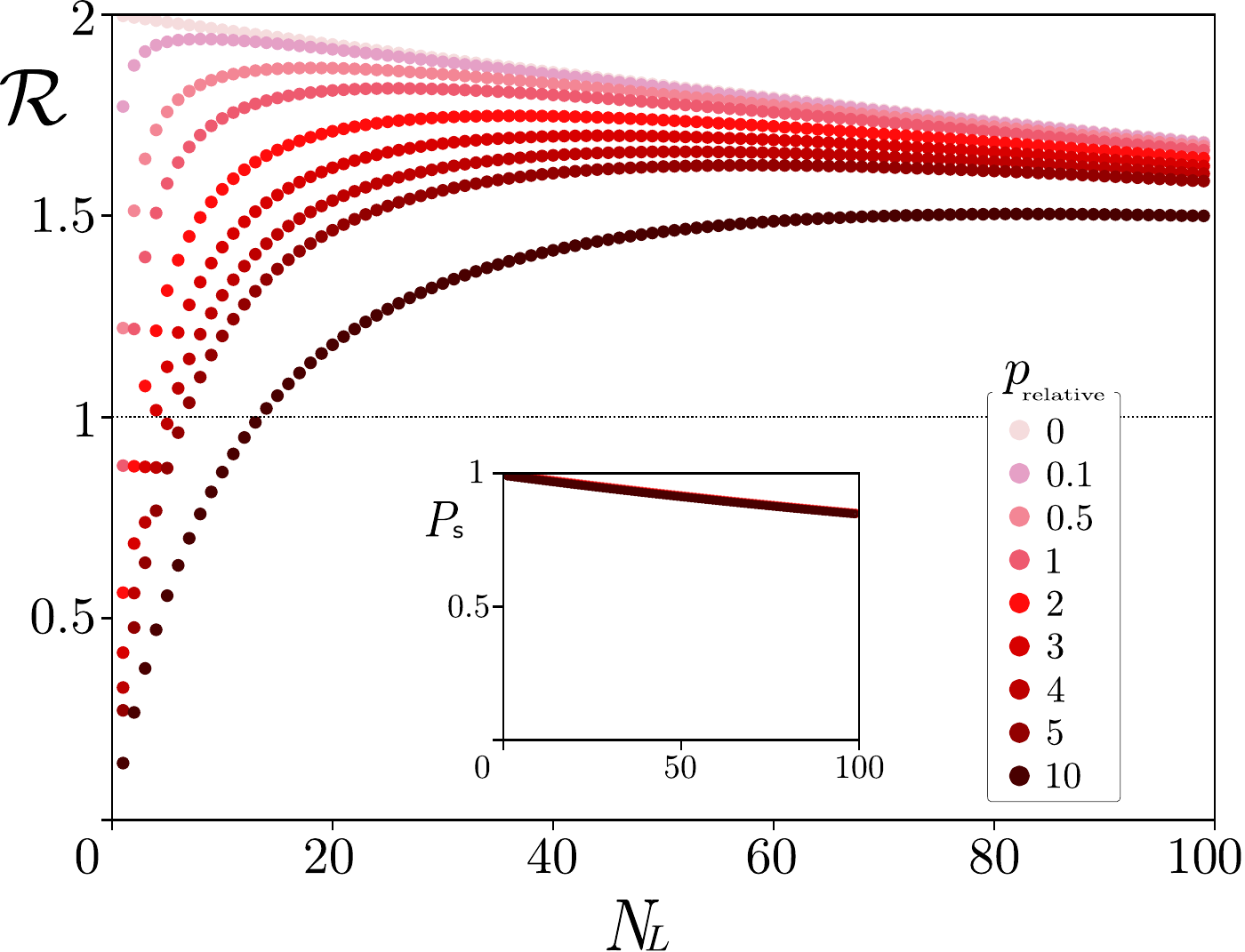}
    \caption{ Performance of the quasi-deterministic protocol ($d=2$) for  circuits U of different depth. The number of layers $N_{\rm L}$ of cNOT and T gates (corresponding to the circuit depth) is plotted against $\mathcal{R}$. Success probability is given in the inset. A single layer comprises a cNOT followed by two T gates applied to each of the $m=2$ qubits. The auxiliary is chosen such that $\omega_2 = 1$ and the noise is modeled by depolarizing channels after each gate, with $p_0 = 1-3\times 10^{-4}$. As reported in the legend, colored dotted lines are associated with different relative error probabilities, i.e., $p_{\rm relative}=(1-p_{\rm cswap})/(1-p_0)$ characterizing the depolarizing channels acting after each of the two Fredkin gates that a cSWAP is decomposed into.}
    \label{fig:GB_concat_CNOTTT_depol} 
\end{figure}
\subsubsection{Nested protocol}
The advantage of the nested extension can be found in  Fig.~\ref{fig:nestedperformance}. The shadow blue area indicates the limits of the protocol, and we set $\omega_2 = 1$ for clarity. The probabilistic protocol (at parity of resources, i.e., total number of auxiliary states) is equivalent to the nested one when either $\omega_1 = 1$ (dark blue dots) or the same auxiliary state with $\omega_1 < 1$ is employed (light blue dots). We also plot the result of employing the strategy outlined in Sec.~\ref{sec:gate_nested}, i.e. choosing different auxiliary states with $\omega_1 < 1$ at successive iterations (blue dots). In any case, we achieve remarkable advantage with respect to the incoherent case (red dotted line) when increasing the number of auxiliary states.

As one can see from the two panels in Fig.~\ref{fig:nestedperformance}, when varying either the total number of branches (equivalently, auxiliary states) $d_{\rm tot} = \sum_{k=0}^{n-1} d_k$ or the no-error probability $p_{\rm ne} = p_0^2$, there is ample margin in which, for fixed $\omega_1 < 1$, the nested extension yields results that are considerably better than the probabilistic protocol. With the additional advantage that the cSWAP operations are simpler, the nested protocol represents a promising route in scaling up our protocols.

\subsubsection{Noisy cSWAP}
Finally, we investigate the scenario in which both the cSWAPs and the computation $U$ are noisy. Specifically, here $U$ comprises $N_{\rm L}$ layers acting on $m=2$ qubits; each layer contains one cNOT gate followed by one T gate on each qubit, with depolarizing channels acting after each operation. In Fig.~\ref{fig:GB_concat_CNOTTT_depol}, we consider the quasi-deterministic SQEM protocol and present the ratio $\mathcal{R}$ against $N_{\rm L}$ for different cSWAP error probabilities. Each cSWAP can be decomposed into two Fredkin gates; the relative error probability of each cSWAP is $p_{\rm relative}=(1-p_{\rm cswap})/(1-p_0)$, where $p_{\rm cswap}$ is the no-error probability associated with the depolarizing channels acting after each Fredkin gate on its control and both of its targets.

For the case in which the cSWAPs are noiseless, the results are in agreement with Fig.~\ref{fig:GB_T_depha} and the associated $\mathcal{R}$ is constantly above $1$. However, with the cSWAPs becoming noisier, our protocol ceases to be advantageous for sufficiently small values of $N_{\rm L}$.  This suggests that the most advantage is achieved for large computations $U$ or in situations in which the cSWAP is particularly stable (e.g., \cite{Monz09,Reed2012,Levine2019,Gu2021} and Sec.~\ref{sec:vacuum}).  As a last observation, we remark that the post-selection probability is not significantly changed by the presence of noise affecting the cSWAPs (see the inset in Fig.~\ref{fig:GB_concat_CNOTTT_depol}).

\section{Measurement-based  SQEM approach}
\label{sec:enhancedMBQC}
As in GB-QC, one of the obstacles to large-scale MB-QC is the presence of decoherence, which severely limits the size and complexity of achievable computations.
Although noise arises from different sources compared to GB-QB, our SQEM protocols outlined in Sec.~\ref{sec:standardgatebased} 
can be adapted to MB-QC. In Sec.~\ref{sec:MBQC_prot}, we first explain how to generalize the probabilistic, quasi-deterministic and deterministic schemes to MB-QC. Then, in Sec.~\ref{sec:mbqcperformance}, we consider the main error sources of MB-QC and study the performance of our protocols. 

\subsection{Protocol}
\label{sec:MBQC_prot}
In this section, we describe the protocol for enhancing the fidelity of a noisy MB-QC, where the computations are carried out in the standard MB-QC fashion \cite{Briegel2001,Oneway2005,Briegel2009}. Aside from this and the specific form of the Kraus operators, the process is conceptually identical to the standard GB procedure presented in Sec.~\ref{sec:standardgatebased}. 

To allow a better comparison with the interferometric-based (IB-QC) implementation in Sec.~\ref{sec:vacuum}, here we employ the environmental formalism for deriving the main results. As explained in more detail in Sec.~\ref{sec:Background}, the action of a noisy channel with Kraus operators $K_i$ acting on a state $\rho_{\rm in}=|\psi_{\rm in}\rangle \langle\psi_{\rm in}|$ can be also described as a unitary evolution in a larger Hilbert space including the environment [see Eq.~\eqref{eq:Stinespring0}], into which information leaks in the decoherence process. By tracing the environment out, from Eq.~\eqref{eq:Stinespring0} one recovers the standard action of a channel in Eq.~\eqref{eq:Krausrepresentation}, showing the equivalence of the two approaches. The advantage of the representation based on the Stinespring theorem is that the state of the system is always pure. Detailed derivations of the results presented in this section are given in App.~\ref{sec:appendixMB-QC111}.

\begin{figure}
    \includegraphics[width=\columnwidth]{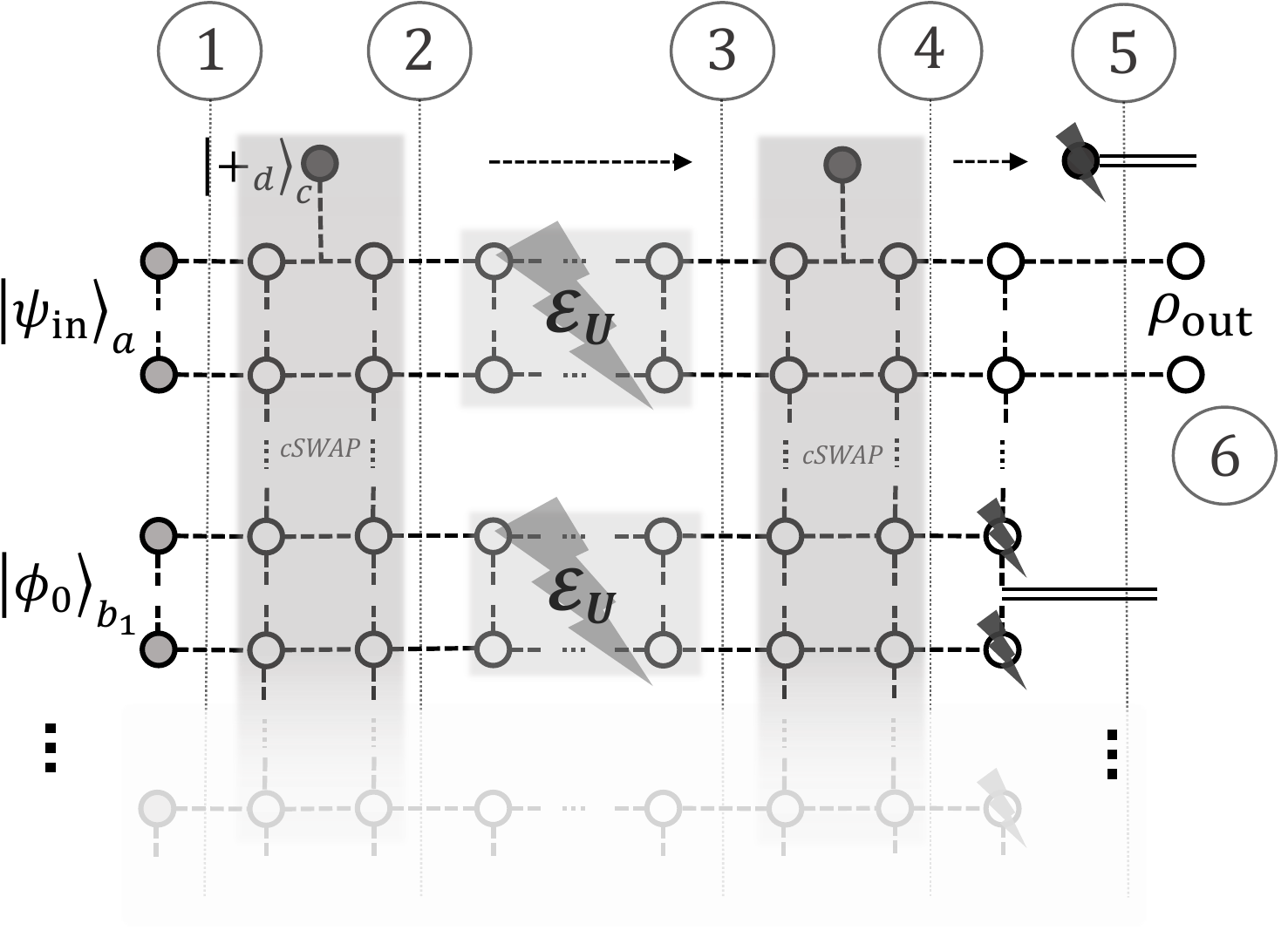}
    \caption{
    Schematic representation of the SQEM process for an MB-QC implementation, which enhances the fidelity of any noisy computation $\cal{E}_{\rm U}$. A control register of dimension $d$ generates the superposition by swapping in a controlled way the $m$-qubit input state with $d$ $m$-qubit auxiliaries. All the operations can be performed in a measurement-based fashion, where noise arises from the imperfect preparation of the resource state.
    }
    \label{fig:mbqcbasic} 
\end{figure}

\begin{table}
{\LinesNumberedHidden
    \begin{algorithm}[H]
        \SetKwInOut{Input}{Input}
        \SetKwInOut{Output}{Output}
        \SetAlgorithmName{Protocol}{}
   \justifying  \textit{Input}: An initial state $ \ket{\psi_{\rm in}}$ and a noisy cluster state implementing the map ${\cal E}_U$, characterized by a fidelity $F^{0}$ with respect to the desired output state $U \ket{\psi_{\rm in}}$.
       \begin{enumerate}
            \item Prepare a cluster state that is sufficiently large for implementing the cSWAP gates in steps~\ref{prot:mb3} and    \ref{prot:mb5}, and the desired computation $U$ in step~\ref{prot:mb4}. Initialize the input qubits of the cluster state in $\ket{+_d}_{\rm c}\ket{\psi_{\rm in}}_{\rm a}  \bigotimes_{i=1}^{d-1}\ket{\phi_0}_{{\rm b}_i}$.
            \label{prot:mb1} 
            \item Apply the cSWAP gate in Eq.~\eqref{eq:GB-QCcswap} (in an MB-QC fashion) for distributing the input and auxiliary states into all $d$ branches.
            \label{prot:mb2}
            \item Implement the computation $U$ by measuring the dedicated ancilla qubits in the cluster \cite{Briegel2001,Oneway2005,Briegel2009}.
            \label{prot:mb3} 
            \item Apply again the cSWAP gate for reassembling.
            \label{prot:mb4} 
            \item Fix the output state by means of byproduct operators (see Sec.~\ref{sec:MB-QC} and Refs.~\cite{Briegel2001,Oneway2005,Briegel2009}).
            \label{prot:mb5} 
            \item Proceed as in Protocol~\ref{table:GBstandard} for steps from \ref{prot:gate5} to \ref{prot:gate6}, i.e., measuring the control and auxiliary systems and running the probabilistic or deterministic protocol. 
            \label{prot:mb6} 
        \end{enumerate}
        
    \justifying \textit{Output}: State $\rho_{\rm out}$ characterized by a fidelity $F > F^{0}$, in both the probabilistic and (on average) the deterministic protocols. 
\caption{SQEM for an MB-QC implementation} \label{table:MBstandard}
\end{algorithm}}
\end{table}
As with the GB-QC, we consider an $m$-qubit input state $\ket{\psi_{\rm in}}$ upon which a given computation $U$ acts. As in standard MB-QC, this computation is performed via a $2$D cluster state that is, however, noisy. 
For reducing the impact of decoherence affecting the computation, it is possible to follow the procedure presented in Protocol~\ref{table:MBstandard}. Below, we present a more detailed description (see also Fig.~\ref{fig:mbqcbasic}) with the main results.

\textbf{Step~\ref{prot:mb1}. ---} The scope is to prepare the cluster state required for the protocol and initialize its input qubits to $\ket{+_d}_{\rm c}\ket{\psi_{\rm in}}_{\rm a}  \bigotimes_{i=1}^{d-1}\ket{\phi_0}_{{\rm b}_i}$. The control qudit system $\ket{+_{d}}_{\rm c}$ of dimension $d$ in Eq.~\eqref{eq:controldlevel} is prepared by embedding more qubits initialized into $\ket{+}$, as explained in Sec.~\ref{sec:gate_protocol}. If we explicitly consider the noise, the system is characterized by
\begin{equation}
\left|+ _{ d}\right\rangle_{\rm c} \sum_{q_0} K_{q_0} \left|G_{\psi_{\rm in}}\right\rangle_{a}   \left|q_0\right\rangle_{\epsilon_{a}}  \bigotimes_{i=1}^{d-1} \left( \sum_{q_i} K_{q_i}  \left|G_{\phi_{0}}\right\rangle_{b_i}  \left|q_i\right\rangle_{\epsilon_{_{b_i}}} \right),
\end{equation}
which, aside from the cluster states explicitly included, corresponds to Eq.~\eqref{eq:gate_state_step1}. Observe that, unlike the GB case, noise already affects the systems at this stage.
We associate an environmental system with  each cluster state and $\left|G_{\psi_{\rm in}}\right\rangle$ and $\left|G_{\phi_{0}}\right\rangle$ represent cluster states with the first qubits prepared in the states $\ket{\psi_{\rm in}}$ and  $\ket{\phi_{0}}$ respectively (see App.~\ref{sec:MB-QCappendix2} for details). 

\textbf{Step~\ref{prot:mb2}. ---} To generate a superposition, we apply a cSWAP operation, which can be implemented in an MB fashion (for details, see App.~\ref{sec:cswap_patterm}). Ancilla qubits in the cluster are measured to effectively implement Eq.~\eqref{eq:GB-QCcswap}, see Fig.~\ref{fig:mbqcbasic}. The state of the system becomes 
\begin{equation}
\label{eq:mb_prot_sup}
\begin{split}
 &\frac{1}{\sqrt{d}}\ket{0}_{c} \sum_{q} K_{q} \left|G_{\psi_{\rm in}}\right\rangle_{a}   \left|q\right\rangle_{\epsilon_{a}} \bigotimes_{i=1}^{d-1} \sum_{q_i} K_{q_i}  \left|G_{\phi_{0}}\right\rangle_{b_i}  \left|q_i\right\rangle_{\epsilon_{b_i}}+ \\
  +& \frac{1}{\sqrt{d}} \sum_{j=1}^{d-1}  \ket{j}_{c}\sum_{q_0} K_{q_0} \left|G_{\phi_{0}}\right\rangle_{a}   \left|q_0\right\rangle_{\epsilon_{a}}    \sum_{q_j} K_{q_j}  \left|G_{\psi_{\rm in}}\right\rangle_{b_j}  \left|q_j\right\rangle_{\epsilon_{b_j}} \\
  & \qquad \qquad   \qquad \qquad \qquad \qquad  \bigotimes_{i \neq j} \sum_{q_i} K_{q_i}  \left|G_{\phi_{0}}\right\rangle_{b_i}  \left|q_i\right\rangle_{\epsilon_{b_i}},
\end{split}
\end{equation}
by analogy with Eq.~\eqref{eq:step2_gb_prot}, where the cluster states  $\left|G_{\psi_{\rm in}}\right\rangle_{\rm a}$ and $\left|G_{\phi_{0}}\right\rangle_{\rm b}$ represent remaining clusters once the corresponding qubits have been measured in the application of the cSWAP operation.

\textbf{Steps~\ref{prot:mb3} to \ref{prot:mb6}. ---} From step~\ref{prot:mb3} to \ref{prot:mb6}, the computation $U$ followed by the second cSWAP is implemented via the measurements of ancilla qubits. In the environmental formalism, this equation reads
\begin{equation}
\begin{split}
\label{eq:mb_prot_sup_2}
&\frac{1}{\sqrt{d}}\ket{0}_{c} \sum_{q} L_{q} U\left|{\psi_{\rm in}}\right\rangle_{a}   \left|q\right\rangle_{\epsilon_{a}} \bigotimes_{i =0}^{d-1} \sum_{q_i} L_{q_i}  U\left|{\phi_{0}}\right\rangle_{b_i}  \left|q_i\right\rangle_{\epsilon_{b_i}} \\
  +& \frac{1}{\sqrt{d}} \sum_{j=1}^{d-1}  \ket{j}_{c}\sum_{q_0} L_{q_0} U\left|{\phi_{0}}\right\rangle_{b_j}   \left|q_0\right\rangle_{\epsilon_{a}}    \sum_{q_j} L_{q_j}  U\left|{\psi_{\rm in}}\right\rangle_{a}  \left|q_j\right\rangle_{\epsilon_{b_j}} \\
 & \qquad \qquad   \qquad \qquad \qquad \qquad \bigotimes_{i \neq j} \sum_{q_i} L_{q_i}  U\left|{\phi_{0}}\right\rangle_{b_i}  \left|q_i\right\rangle_{\epsilon_{b_i}},
\end{split}
\end{equation}
where we recall that, compared to the GB version, here the Kraus operators $L_i$ depend \cite{Usher2017} on the noise acting locally on each of the qubit of the cluster state prepared at step~\ref{prot:mb1} and given by Kraus operators $K_i$.

In the final step~\ref{prot:mb6}, control and auxiliary subsystems are measured in bases modified to take into account the byproduct operators, so as to obtain the desired output $\rho_{\rm out}$ of our protocol. Assuming the measurement outcomes are $\ket{+_d}_{\rm c}$ and $\ket{\phi_{\rm f}}$ respectively, one recovers Eq.~\eqref{eq:gate_out} by simply tracing out the environment systems. For the details of the probabilistic implementation, see App.~\ref{sec:MB-QCappendix2}.
 
The total number of logical qubits involved in the protocol is $dm + \log_{2} d$, taking into account input, auxiliary and control registers. Therefore, denoting the depths of the incoherent computation and the cSWAP by $K$ and $S$ respectively, the whole Protocol~\ref{table:MBstandard} requires a cluster state of size $\mathcal{O} [(dm + \log_{2} d) \times (K+2S)]$. 

\subsection{Protocol performance. Numerical analysis}
\label{sec:mbqcperformance}
\begin{figure*}
   \includegraphics[width=0.87\textwidth]{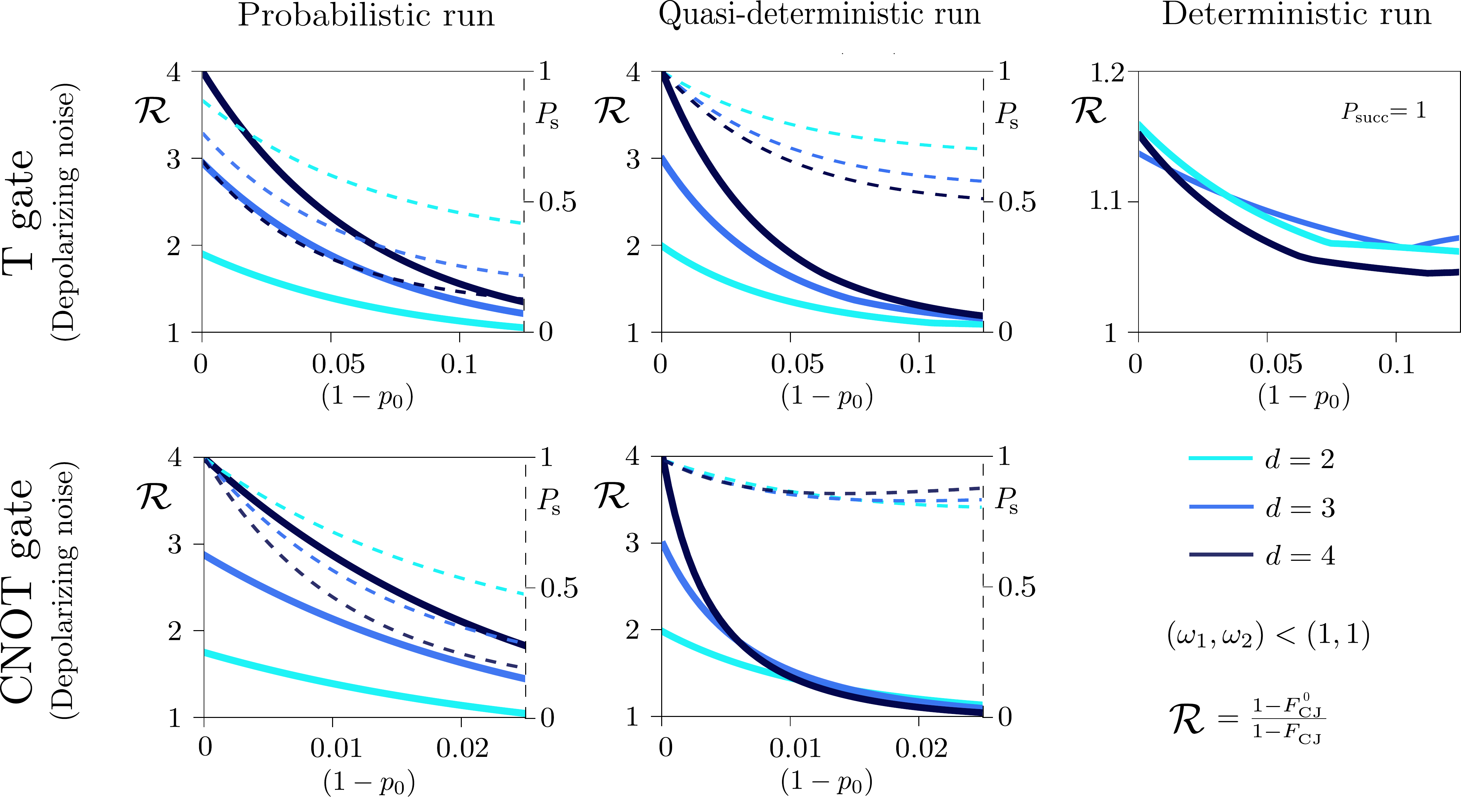}
    \caption{Performance of the protocol 2 for mitigating the noise of a T gate and a cNOT gate in an MB setting, whose imperfect implementation is modeled by depolarizing noise with no-error probability $p_0$ acting on \textit{every} qubit of the resource state.  In this case, the auxiliary state, measurement bases and correcting operations are all chosen from the Clifford group, such that  $\omega_2 < 1$ for the T gate,  $\omega_2 = 1$ for the cNOT gate, and $\omega_1 < 1$.}\label{fig:MB_CNOT_depol} 
\end{figure*}

\begin{figure}
    \includegraphics[width=0.9\columnwidth]{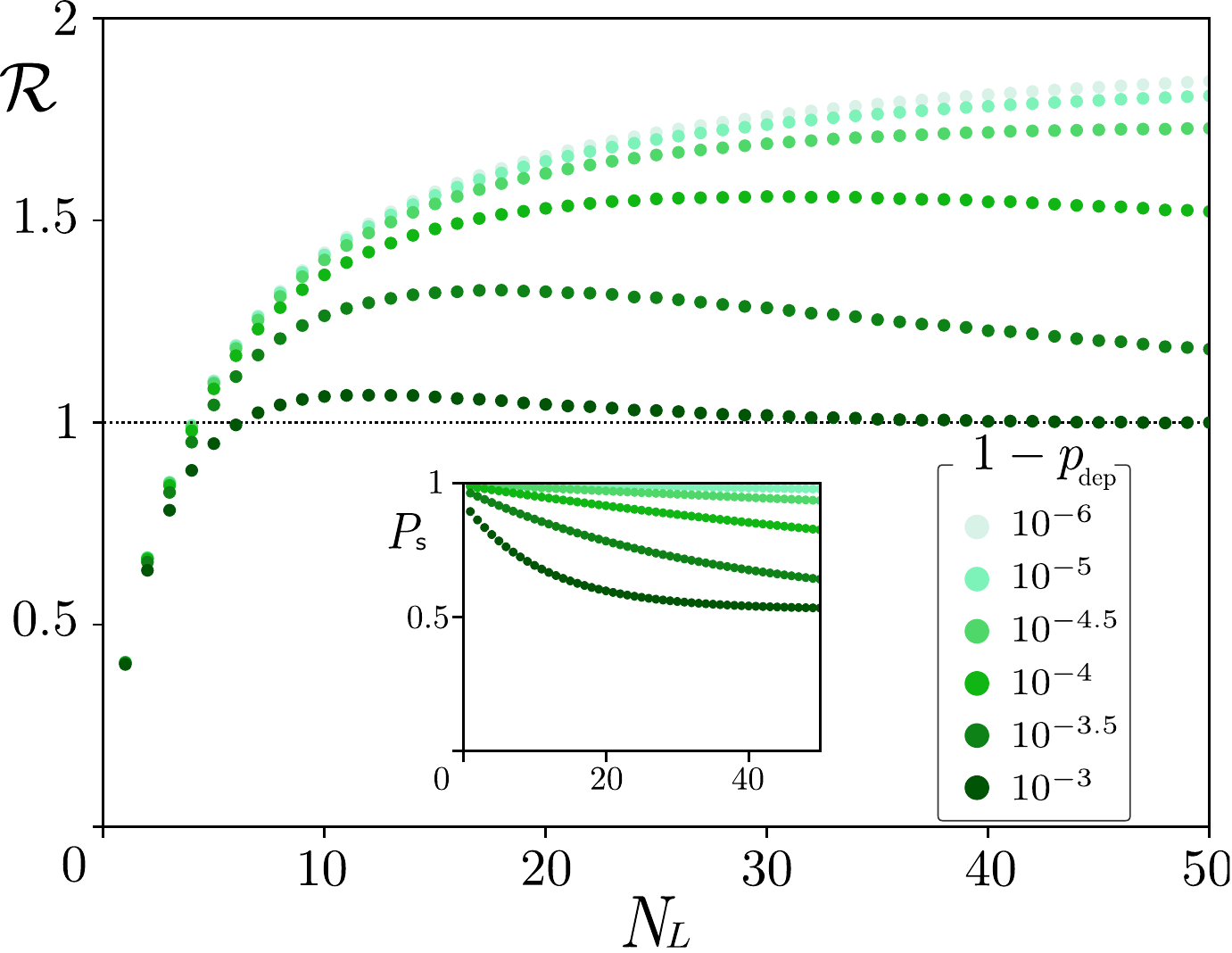}
    \caption{Advantage ratio, Eq.~\eqref{eq:cj_fid_ratio},  of MB-QC with noisy cSWAP gates, where each cSWAP operation is implemented in an MB fashion, and each qubit involved is subjected to the same depolarizing noise (with no-error probability $p_0$) as the rest of the resource qubits (i.e. the ones that carry out the computation). As in the GB case, each of the $N_{\rm L}$ layers comprises a cNOT followed by two T gates applied to each of the $m=2$ qubits.}\label{fig:MB_T_concatenation} 
\end{figure}

In this section, we analyze the performance of the probabilistic, quasi-deterministic and deterministic MB SQEM protocols under different assumptions and settings. We consider the figure of merit $\mathcal{R}$ in Eq.~\eqref{eq:cj_fid_ratio}; numerical results are obtained by simulating \textit{all} qubits in the system and the noise affecting them. This includes input, auxiliary, control, and ancilla qubits in the cluster. To do that, our numerical simulator is based on a series of steps such that only the qubits that must be measured and the ones directly connected to them are considered each time. The process is therefore carried out in a concatenated way, a feature allowed for MB implementations \cite{Oneway2001}.
The noise is introduced by applying a channel as in Eq.~\eqref{eq:Krausrepresentation} to all (or a subset of) qubits within the cluster, see Fig.~\ref{fig:mbqcbasic} and Sec.~\ref{sec:MB-QC} for details. Specifically, in Fig.~\ref{fig:MB_CNOT_depol} noise affects only the portions of the cluster that are dedicated to implementing $U$, while in Fig.~\ref{fig:MB_T_concatenation} it affects all ancilla qubits that are required for both cSWAP (see App.~\ref{sec:cswap_patterm}) and the computation $U$.

Despite the differences in the Kraus operators acting on the output state $\rho_{\rm out}$, the performance of the protocol is qualitatively similar to the GB scheme in Sec.~\ref{sec:standardgateperformance}. This can be seen from Fig.~\ref{fig:MB_CNOT_depol}, presenting the results for the T and the cNOT gate, where each qubit of the cluster state is affected by depolarizing noise with parameter $p_0$. We remark that in the probabilistic implementation we always employ $\ket{+}^{\otimes m}$ both as auxiliary state $\ket{\phi_0}$ and for the post-selection in step~\ref{prot:mb6}. As demonstrated by Fig.~\ref{fig:differentaux}, this is a viable choice, albeit it generally implies $\omega_1 < 1$ and (for the T gate) $\omega_2 < 1$ and thus slightly lowers $\mathcal{R}$ and the success probability $P_{\rm s}$. This is the reason for which, in the limit $p_0 \rightarrow 1$, the probabilistic implementation of the T gate in Fig.~\ref{fig:MB_CNOT_depol} is characterized by $P_{\rm s} < 1$.

Fig.~\ref{fig:MB_CNOT_depol} shows that for the same single channel no-error probability $p_0$, the gain $\mathcal{R}$ of the MB protocols is lower compared to the GB case (see Sec.~\ref{sec:standardgateperformance}). This stems from the fact that the probability $1-p_0$ of getting an error refers to many qubits within the cluster, implying that the overall Kraus operators $L_i$ affecting the output $\rho_{\rm out}$ in Eq.~\eqref{eq:mb_prot_sup_2} describe a much more disruptive channel. However, despite the fact that both $\omega_1$ and $\omega_2$ in Eq.~\eqref{eq:omegas} are generally lower than one, the advantage compared to the incoherent case is still consistent and the salient features are unchanged compared to the GB case. 

In Fig.~\ref{fig:MB_T_concatenation} we study the impact to the infidelity ratio $\mathcal{R}$ of the noise affecting the two cSWAPs. Following the example of Fig.~\ref{fig:GB_concat_CNOTTT_depol}, we consider $N_{\rm L}$ layers of cNOT plus T gates and the probabilistic implementation with $\ket{+}$ as auxiliary state and for the post-selection. Here, we investigate the scenario in which every ancilla qubit in the cluster is affected by the same noise with probability $1-p_0$, and different colours are used for the values of $p_0$ reported in the legend. The horizontal, black dotted line serves as a reference to discriminate when our protocol is advantageous. As in the GB case, the noise affecting the cSWAP poses a limitation to the advantage of our protocols, which are beneficial when the computation $U$ is noisier than the two cSWAP gates. From the figure, we conclude that there is an ample window both in the value of $p_0$ and the size of the employed cluster state (proportional to $N_{\rm L}$) in which the MB scheme yields an  advantage over the incoherent case.

\section{Interferometric-based SQEM approach}
\label{sec:vacuum}

Both the GB and MB implementations of our SQEM protocols are limited by the noise affecting the two cSWAP gates that are required for the creation of superposition between the $d$ branches. This is numerically investigated in Figs.~\ref{fig:GB_concat_CNOTTT_depol} and \ref{fig:MB_T_concatenation}, and discussed in the corresponding sections. Qualitatively, our schemes are beneficial as long as the noise affecting the computation $U$ is stronger than the one affecting the cSWAP operations. It is therefore of paramount importance to find ways to create the superposition between different branches with high fidelity. 

In this section, we propose an alternative SQEM approach for an interferometric-based (IB) implementation. Specifically, it employs additional physical degrees of freedom that are naturally available,  instead of the cSWAP, in a way similar to what has been done (with different scopes) in Refs.~\cite{Friis14,Araujo2014,Chiribella2019,Kristjnsson_2020,Abbott2020,Rubino2021,MR2021}. The idea is that, instead of using auxiliary states $\ket{\phi_i}$ that are later measured to gain information about the noise, the input $\ket{\psi_{\rm in}}$ \textit{alone} is coherently distributed into different branches along with the vacuum. The simplest way to understand the working principle is to use one of the possible physical implementations of our IB protocols, namely, a photonic interferometer (see Fig.~\ref{fig:vaccum}). While we base our analysis on the photonic platform, we stress that other setups, such as ions \cite{Friis14} and superconducting qubits \cite{Friis2015}, are also suitable for our IB schemes. 

In Sec.~\ref{sec:IB_protocol} below we describe the protocol, highlighting the main differences with respect to the GB and the MB alternatives in Secs.~\ref{sec:gate_protocol} and \ref{sec:MBQC_prot}, respectively. The working principle of the IB scheme is based on interference with the vacuum. We propose a theoretical model based on the stochastic Hamiltonian formulation \cite{Molmer1992,Gregoratti_2001} in Sec.~\ref{sec:IB_stoqastic}. Finally, we present different possible realizations of the IB schemes in Sec.~\ref{sec:IB_different_implementations}, and provide numerical simulations in Sec.~\ref{sec:resultsinterfero}.

\subsection{Protocol}
\label{sec:IB_protocol}

\begin{figure}
    \centering
    \includegraphics[width=\columnwidth]{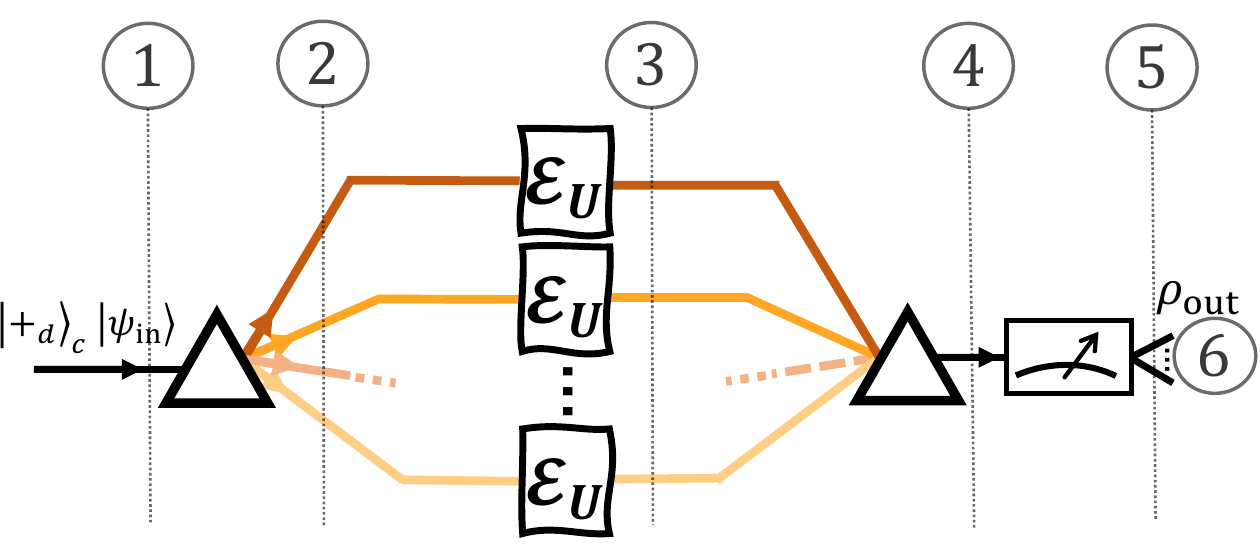}
    \caption{Illustration of the strategy for enhancing the fidelity of a noisy computation in an interferometric-like circuit model. The \textit{noisy} computation is identically applied in all branches of the superposition and no dedicated auxiliary system is needed. The control register can be encoded in some extra degree of freedom of the input qubit(s). System-vacuum correlations (see text) allow us to dilute the effect of the noise and lead to computational fidelity advantage.}
    \label{fig:vaccum}
\end{figure}

Since the IB schemes do not rely on auxiliary states $\ket{\phi_{i}}$, we must use the environmental formalism (outlined in App.~\ref{sec:appendixMB-QC111}), previously employed for describing the MB protocols in Sec.~\ref{sec:MBQC_prot}. The underlying idea of the IB schemes is similar: distributing the input between different branches, it is possible to either post-select the best outcome or (partially) correct the resulting state based on the knowledge gathered from the protocol. Before, these processes were based on the measurement outcomes on the auxiliary states. 
As we shall see, here they result from probing an extra physical degree of freedom such as the path taken within an interferometer. In the following we describe, one by one, the steps in the IB scheme presented in Protocol~\ref{table:GBinterfer} and schematically shown in Fig.~\ref{fig:vaccum}. 

\begin{table}
{\LinesNumberedHidden
    \begin{algorithm}[H]
        \SetKwInOut{Input}{Input}
        \SetKwInOut{Output}{Output}
        \SetAlgorithmName{Protocol}{}
   \justifying  \textit{Input}: An initial state $ \ket{\psi_{\rm in}}$ and a noisy computation ${\cal E}_U$, implementing the unitary $U$ with a fidelity $F^{0}$.
       \begin{enumerate}
            \item Find a suitable physical degree of freedom ---the control system--- and initialize it in $\ket{+_{d}}_{\rm c}$ as in Eq.~\eqref{eq:controldlevel}. 
            \label{prot:ib1} 
            \item Distribute the input state $\ket{\psi_{\rm in}}$ into $d$ branches according to the state of the control system. Specifically, create the superposition of all $d$ states labeled by $i = 0,\dots,d-1$, where $\ket{\psi_{\rm in}}$ is in the $i$-th branch and vacuum is in all other branches.
            \label{prot:ib2} 
            \item Implement the noisy computation ${\cal E}_U$ in every path.
            \label{prot:ib3} 
            \item Recombine the paths such that, in the absence of noise, $U \ket{\psi_{\rm in}}$ is deterministically found in a chosen branch $i=0$, and all others $i=1,\dots d$ are empty. 
            \label{prot:ib4} 
            \item Measure the control register in the generalized $X$ basis and run the probabilistic or deterministic protocol as in step~\ref{prot:gate6} of the GB Protocol~\ref{table:GBstandard}.
            \label{prot:ib5} 
        \end{enumerate}
        
    \justifying \textit{Output}: State $\rho_{\rm out}$ characterized by a fidelity $F > F^{0}$, in both the probabilistic and (on average) the deterministic protocols.
\caption{SQEM for a IB-QC implementation}
\label{table:GBinterfer}
\end{algorithm}}
\end{table}

\textbf{Step~\ref{prot:ib1}. ---} First, a control qudit of dimension $d$ is initialized in the state $\ket{+_{d}}_{\rm c}$ [see Eq.~\eqref{eq:controldlevel}] such that the system can be described by
\begin{equation}
     \text{step~\ref{prot:ib1}: }  \ket{+_d}_{\rm c}  \ket{\psi_{\rm in}} 
 \bigotimes_{i=0}^{d-1} \left|\varepsilon_{i}\right\rangle _{\epsilon_{i}},
\end{equation}
where $\ket{\psi_{\rm in}}$ is the $m$-qubit input.
We explicitly keep track of the environment state $\ket{\varepsilon_{i}} _{\epsilon_{i}}$, which is generally unknown, inaccessible and associated with the input in the $i$-th path (see App.~\ref{sec:appendixMB-QC111}). The physical meaning of the control register is the $d$ possible paths/branches of an interferometer with $d$ inputs and outputs, as shown in Fig.~\ref{fig:vaccum}. Initializing the corresponding state in $\ket{+_d}_{\rm c}$ is equivalent to saying that $\ket{\psi_{\rm in}}$, alongside the vacuum at all other input ports, is equally distributed between all branches with the same phase. This notation is convenient, as it allows a better comparison with the GB and MB protocols, where the control register are actual qudits (or qubits) and are essential to the cSWAP operations.

\textbf{Step~\ref{prot:ib2}. ---} The superposition is generated in an interferometric fashion \cite{Friis14}, such that for each state $\ket{i}_{\rm c}$ within $\ket{+_d}_{\rm c}$, the input state follows a different branch. Therefore, at this point of the protocol we can describe the composite state vector by
\begin{equation}
    \text{step~\ref{prot:ib2}: }  \frac{1}{d} \sum_{j=0}^{d-1} \ket{j}_{\rm c}  \ket{\psi_{\rm in}}^{(j)}
 \bigotimes_{i=0}^{d-1} \left|\varepsilon_{i}\right\rangle _{\epsilon_{i}},
\end{equation}
where $\ket{\psi_{\rm in}}^{(j)}$ indicates the input state in the $j$ path and we omit the vacuum for clarity. We point out that in the corresponding GB and MB situations in Eqs.~\eqref{eq:step2_gb_prot} and \eqref{eq:mb_prot_sup} the superposition is created with known auxiliary states and a cSWAP gate. Here, on the other hand, it resembles the action of a generalized beam splitter with $d$ inputs and outputs, see Fig.~\ref{fig:vaccum}.

\textbf{Steps~\ref{prot:ib3} to \ref{prot:ib5}. ---} As shown in Fig.~\ref{fig:vaccum}, the same noisy computation ${\cal E}_U$ is applied in each branch to the input state $\ket{\psi_{\rm in}}$. The action of the noise is given by the environmental formalism (see also App.~\ref{sec:appendixMB-QC111}). The state after tracing out the environmental systems reads
\begin{equation}
\begin{split}
&\frac{1}{d} \sum_{i} \left|i\right\rangle_{\rm c} \langle {i}| \sum_{r} K_{r} U \rho_{\rm in} U^{\dagger} K_{r}^{\dagger} \\
+&\frac{1}{d} \sum_{i \neq j} \left|i\right\rangle_{\rm c} \langle {j}|    \left( \sum_{r} \langle \varepsilon_{i}|r \rangle K_{r}  \right)  U \rho_{\rm in} U^{\dagger}  \left( \sum_{s} \langle s|  \varepsilon_{j} \rangle K^{\dagger}_{s}  \right),
\label{eq:rhointerf3}
\end{split}
\end{equation}
where $K_i$ are the Kraus operators affecting the $m$-qubit input.

Afterwards, we apply the inverse of the transformation that distributed $\ket{\psi_{\rm in}}$ into all branches at step~\ref{prot:ib2}. Practically, this is done by measuring the control register in the $X$ basis. In the absence of noise this measurement yields the outcome $\ket{+_d}_{\rm c}$ deterministically, i.e., $U\ket{\psi_{\rm in}}$ is found at the desired output branch of the interferometer, which for convenience is indicated with the same index ``$0$'' as the input one. 

When noise is present, on the other hand, there is a finite probability to get any of the $d$ outcomes and the corresponding state, not only the desired $0$-th one. A given combination of errors corresponds to a particular probability distribution of finding $\rho_{\rm out}$ at different outputs. It is then possible to acquire knowledge about the noise acting on the system, and from this we can choose the best unitary operation to be applied to correct $\rho_{\rm out}$, or post-select $\rho_{\rm out}$ if $\ket{+_d}_{\rm c}$ is obtained.

In practice, this corresponds to running the probabilistic, quasi-deterministic or the fully deterministic versions of the IB SQEM protocol. In the probabilistic version, only the states $\rho_{\rm out}$ that are found at the $0$-th output branch are kept. Indeed, they indicate noise suppression without the requirement of correcting unitaries. For the other two versions, one can proceed with the same steps as in the GB or MB quasi-deterministic (deterministic) schemes, except that there are no auxiliary qubits to be measured, and the correction depends only on the measurement outcome of the control register, i.e. the specific branch $\rho_{\rm out}$ found at the output.
Besides this difference, the optimization for enhancing the fidelity is done as described in Sec.~\ref{sec:gate_det_prot}.

For clarity, let us here consider the probabilistic scheme. The state of the system after the projection onto $\ket{+_d}_{\rm c}$ is
\begin{equation}
\label{eq:interfer_out}
\begin{split}
&\frac{{\mathcal{A}_d}}{d} \sum_{r} K_{r} U \rho_{\rm in} U^{\dagger} K_{r}^{\dagger} \\
+&\frac{{\mathcal{A}_d}}{d} \sum_{i\neq j} \left( \sum_{r} \langle \varepsilon_{i}|r \rangle K_{r}  \right)  U \rho_{\rm in} U^{\dagger}  \left( \sum_{s} \langle s|  \varepsilon_{j} \rangle K^{\dagger}_{s}  \right),
\end{split}
\end{equation}
which is compared to Eqs.~\eqref{eq:gate_out} and \eqref{eq:mb_prot_sup_2} for the GB and MB protocols, respectively. From this comparison, we see the main difference between the IB and the other schemes: the unitary operation $U$ in Eq.~\eqref{eq:interfer_out} acts \textit{exclusively} on the input $\ket{\psi_{\rm in}}$, with the vacuum being unaffected. 

However, the vacuum does play an important role in determining how advantageous the IB protocols are. This can be understood from the operators 
\begin{equation}
\label{eq:IB_vacuum_operators}
    \sum_r \langle \varepsilon_{i}|r \rangle K_r,
\end{equation}
which are known by the different names of ``transformation matrices'' \cite{Abbott2020} and ``vacuum interference operators'' \cite{Kristjnsson_2020}. As we explain in more details in Ref.~\cite{papercomm}, these terms follow from the linearity of quantum mechanics and are physically understood in terms of relative phases between the environment states, see Appendix~\ref{sec:Appendixvacuum1}. 

In the incoherent case, the GB and the MB protocols, $\ket{\psi_{\rm in}}$ is not distributed along with the vacuum between the branches and one can redefine the vacuum/environment state $\ket{\epsilon_{0}}_\epsilon$ in Eq.~\eqref{eq:Stinespring0} by a global phase, leaving the dynamics unaffected. However, in the IB protocol, the relative differences between these phases acting on \textit{each} of the $d$ branches modify the interference at step~\ref{prot:ib4} of the protocol, and hence the probability distribution of finding $\rho_{\rm out}$ at different outputs. Ultimately, this changes the efficiency of the IB protocols and thus the resulting fidelity of the output $\rho_{\rm out}$. Furthermore, determining and (possibly) controlling these phases can be challenging in realistic scenarios. While a detailed discussion on these aspects is given in Ref.~\cite{papercomm}, in Sec.~\ref{sec:IB_stoqastic} below we consider a model that allows calculating the vacuum phases and therefore investigating the efficiency of the IB protocols under realistic experimental conditions.

\subsubsection{Classical noise and stochastic Hamiltonian formalism}
\label{sec:IB_stoqastic}

As highlighted above, the performance of the IB SQEM protocols depends on the vacuum interference operators in Eq.~\eqref{eq:IB_vacuum_operators}. However, as the microscopic details of the environment are difficult to access, their determination is generally infeasible. In practice, it is convenient to represent the environment by \textit{classical} random variables. For instance, in the ion setting, stray magnetic fields are a common source of qubit decoherence, and are easily modeled as classical random variables fluctuating with time. In this section, we show how to determine the vacuum interference operators corresponding to such classical noises. Specifically, we employ the stochastic Hamiltonian formalism, while a more thorough analysis is provided in Ref.~\cite{papercomm}.

Let us assume the noise takes the form of a number of classical random variables $\mathbf{r}$ that parametrize the system Hamiltonian $H(\mathbf{r})$. The noisy computation is then described by the set of corresponding unitary evolution operators of the system, $V_{\mathbf{r}}=\mathcal{T}\exp [-i\int dt H(\mathbf{r})/\hbar]$, where $\mathcal{T}$ is the time-ordering operator and we have suppressed the possible time dependence of $H(\mathbf{r})$. $V_{\mathbf{r}}$ is generally different from the unitary $U$ characterizing the noiseless computation. If we denote the probability distribution of the random variables as $p_{\mathbf{r}}$, the output of the incoherent computation is simply the ensemble average over all possible noise realizations,
\begin{equation}
\rho_{\rm in} \to \sum_{\mathbf{r}} p_{\mathbf{r}} V_{\mathbf{r}} \rho_{\rm in} V_{\mathbf{r}}^{\dagger};
\label{eq:ib_output_incoh}
\end{equation}
the Kraus operators of the stochastic Hamiltonian are thus identified as
\begin{equation}
K_{\mathbf{r}}=\sqrt{p_{\mathbf{r}}}V_{\mathbf{r}}U^{\dagger}.
\label{eq:ib_kraus}
\end{equation}

In an interferometric-like system comprising $d$ branches, the time evolution is governed by the total Hamiltonian $H_{\rm total} = \sum_{i=0}^{d-1} H_{i} (\mathbf{r}_{i})$; here $\mathbf{r}_{i}$ labels the noise variables in branch $i$. Since all $H_{i}$ commute with each other, the unitary evolution operator is the product of evolution operators in individual branches, $V_{\rm total}=\prod_{i=0}^{d-1} V_{i \mathbf{r}_i}$. To apply the IB SQEM scheme in Protocol~\ref{table:GBinterfer}, we also make the following two standard assumptions: the noise in different branches is independent, so that the averaging over all noise variables factorizes into branches $\prod_{i=0}^{d-1} (\sum_{\mathbf{r}_{i}} p_{\mathbf{r}_i})$; the vacuum state in each branch is nondegenerate, so that the unique vacuum state in branch $i$ is an eigenstate of $V_{i \mathbf{r}_i}$ with eigenvalue $\nu_{i \mathbf{r}_i}$.

When we implement the noisy computation by applying $V_{\rm total}$, we need to average over all noise variables. As in Eq.~\eqref{eq:rhointerf3}, the output density matrix has both diagonal and off-diagonal terms in the Hilbert space of the control system. In each diagonal term $\left|i\right\rangle_{\rm c} \langle {i}|$, the evolution of any branch other than $i$ gives a trivial identity factor which remains trivial upon noise averaging, and we recover the incoherent form in Eq.~\eqref{eq:ib_output_incoh}. Meanwhile, each term proportional to $\left|i\right\rangle_{\rm c} \langle {j}|$ contains two nontrivial noise averages.

The state of the system after projecting the control system onto $\ket{+_d}_{\rm c}$ is
\begin{equation}
\label{eq:rhointerf2}
\begin{split}
&\frac{{\mathcal{A}_d}}{d} \sum_{i} \sum_{\mathbf{r}_{i}} p_{i\mathbf{r}_{i}} V_{i\mathbf{r}_{i}} \rho_{\rm in} V_{i\mathbf{r}_{i}}^{\dagger} \\
+&\frac{{\mathcal{A}_d}}{d} \sum_{i\neq j} \left( \sum_{\mathbf{r}_{i}} p_{i\mathbf{r}_{i}} V_{i\mathbf{r}_{i}} \nu^{*}_{i \mathbf{r}_i} \right)  \rho_{\rm in} \left( \sum_{\mathbf{r}_{j}} p_{j\mathbf{r}_{j}} V_{j\mathbf{r}_{j}}^{\dagger} \nu_{j \mathbf{r}_j} \right),
\end{split}
\end{equation}
where we are able to move $\nu^{*}_{i \mathbf{r}_i}$ to the left of $\rho_{\rm in}$ and $\nu_{j \mathbf{r}_j}$ to the right because they are pure phases rather than operators. Comparing Eq.~\eqref{eq:rhointerf2} with Eq.~\eqref{eq:interfer_out}, we recognize the objects
\begin{equation}
    \sum_{\mathbf{r}_{i}} p_{i\mathbf{r}_{i}} V_{i\mathbf{r}_{i}} \nu^{*}_{i \mathbf{r}_i}U^{\dagger}
    \label{eq:ib_stoc_vio}
\end{equation}
as the vacuum interference operator of branch $i$ in Eq.~\eqref{eq:IB_vacuum_operators}, with the difference that now $p_{i\mathbf{r}_{i}}$ and $V_{i\mathbf{r}_{i}}$ can be easily determined from a stochastic Hamiltonian that contains much less intractable microscopic detail on the environment.

As a simple example, let us consider the aforementioned stray magnetic fields acting on a trapped-ion qubit. Depending on the direction of the stray fields, the noise channel can be dephasing or depolarizing. If the fields are confined in the $z$ direction, the qubit Hamiltonian is written as
\begin{equation}
    H(\mu)=\frac{\hbar\mu}{2\sqrt{\Delta t}} Z;
    \label{eq:stoc_ham_dephase_h}
\end{equation}
here $\mu$ is a time-dependent classical random variable proportional to the field strength, and $\Delta t$ is the time scale on which $\mu$ can be approximated as a constant \footnote{In the continuum limit $\Delta t\to0$, the normalization factor $\sqrt{\Delta t}$ ensures that $\Gamma$ is well-defined.}. For simplicity, we assume $\mu$ follows a Gaussian distribution with variance $2\Gamma$, i.e.
\begin{equation}
    \sum_{\mathbf{r}} p_{\mathbf{r}}\to \int \frac{d\mu}{\sqrt{4\pi\Gamma}}e^{-\frac{\mu^{2}}{4\Gamma}}.
    \label{eq:stoc_ham_field_dist}
\end{equation}

The diagonal elements of the qubit density matrix are unaffected by the noise, which is itself diagonal in the computational basis. On the other hand, the noise-averaged evolution of the off-diagonal elements of the density matrix during the short time interval $\Delta t$ has the following form:

\begin{align}
\ket{1}\bra{0} &\to \int \frac{d\mu}{\sqrt{4\pi\Gamma}}e^{-\frac{\mu^{2}}{4\Gamma}}e^{-\frac{i}{\hbar}H\Delta t}\ket{1}\bra{0}e^{\frac{i}{\hbar} H\Delta t} \nonumber \\
&=e^{-\Gamma \Delta t}\ket{1}\bra{0}.
\end{align}
Thus, over a finite amount of time $t$, the off-diagonal density matrix elements decay exponentially as $e^{-\Gamma t}$. This corresponds to a dephasing noise with a no-error probability
\begin{equation}
    p_0=\frac{1}{2}(1+e^{-\Gamma t}). \label{eq:stoc_ham_dephase_ne_prob}
\end{equation}

It is also possible to calculate the vacuum interference operator during $\Delta t$,

\begin{equation}
    \sum_{\mathbf{r}} p_{\mathbf{r}} V_{\mathbf{r}} \to \int \frac{d\mu}{\sqrt{4\pi\Gamma}}e^{-\frac{\mu^{2}}{4\Gamma}}e^{-\frac{i}{\hbar}H\Delta t}=e^{-\frac{1}{4}\Gamma \Delta t}\mathbb{1}.
\end{equation}
Over a finite amount of time $t$, this becomes
\begin{equation}
    e^{-\frac{1}{4}\Gamma t}\mathbb{1} = (2p_0-1)^{\frac{1}{4}}\mathbb{1}.
    \label{eq:stoc_ham_dephase_vio}
\end{equation}
It is interesting to note that the vacuum interference operator Eq.~\eqref{eq:stoc_ham_dephase_vio} vanishes for the completely dephasing noise $p_0=1/2$; in other words, the IB scheme ceases to be effective in the limit $t\to\infty$.

On the other hand, a qubit in spatially isotropic stray fields is described by the Hamiltonian
\begin{equation}
    H(\mu_x, \mu_y, \mu_z)=\frac{\hbar}{2\sqrt{\Delta t}} (\mu_x X+\mu_y Y+\mu_z Z),
    \label{eq:stoc_ham_depol_h}
\end{equation}
where $\mu_x, \mu_y$ and $\mu_z$ are assumed to independently follow the distribution Eq.~\eqref{eq:stoc_ham_field_dist}. A similar calculation shows that the noise is depolarizing, with a no-error probability
\begin{equation}
    p_0=\frac{1}{4}(1+3e^{-2\Gamma t}) \label{eq:stoc_ham_depol_ne_prob}
\end{equation}
and a vacuum interference operator
\begin{equation}
    e^{-\frac{3}{4}\Gamma t}\mathbb{1}=\left(\frac{4p_0-1}{3}\right)^{\frac{3}{8}} .
    \label{eq:stoc_ham_depol_vio}
\end{equation}
Again, the vacuum interference operator Eq.~\eqref{eq:stoc_ham_depol_vio} vanishes for the completely depolarizing noise $p_0=1/4$.

\subsubsection{Generating the superposition with different physical realizations}
\label{sec:IB_different_implementations}

In this section, we briefly discuss possible physical implementations for the IB schemes. In the protocol description above, we employed a control register to analytically derive the main results. Here, we explain different ways to split the input state and at the same time perform in each branch the desired unitary computation.

The most illustrative implementation is inspired by photonic interferometry \cite{Dell2018,Rubino2021}, where time- or path-encodings can be employed, as well as other recently investigated degrees of freedom such as the orbital angular momentum \cite{ZahidyLiu2022,cozzolino2019orb}. A most promising feature of these implementations is that high controllability has been demonstrated in photonic setups \cite{cozzolino2019orb,Cozzolino:19}. In contrast to, e.g., path-encodings demonstrated for communication \cite{Rubino2021}, we do not fall into impractical assumptions such as noiseless control registers. This suggests that the creation of the superposition at the basis of our protocols can be achieved with limited losses that do not jeopardize the outcome fidelity (see Figs.~\ref{fig:GB_concat_CNOTTT_depol}, \ref{fig:MB_T_concatenation}, and related discussions). On the other hand, multi-photon interactions are challenging to obtain and superpositions of large cluster states have not been directly investigated. Our work shall perhaps suggest this last route for future research. 

Photonic setups are not the only ones in which the IB protocols can be implemented. To distribute $\ket{\psi_{\rm in}}$ into different branches, it is also possible to employ auxiliary levels that are naturally available in, e.g., ions \cite{Friis14} or superconducting qubits \cite{Friis2015}. The mathematical description is then analogous to the one presented above, with the different branches represented by different (pairs of) levels in the setup. In this case, the most promising aspect is that one could use systems that have already demonstrated quantum computing capabilities \cite{Preskill2018}. On the other hand, the number $d$ of branches is here limited by the accessible stable levels. Furthermore, the number of controls required (e.g., lasers or microwave pulses) also scale up linearly in $d$. Recent progresses in manipulating qudit systems in ion-based setups \cite{Ringbauer2022, Low2020} show that this route is indeed possible for near-term applications.

\begin{figure}
    \includegraphics[width=\columnwidth]{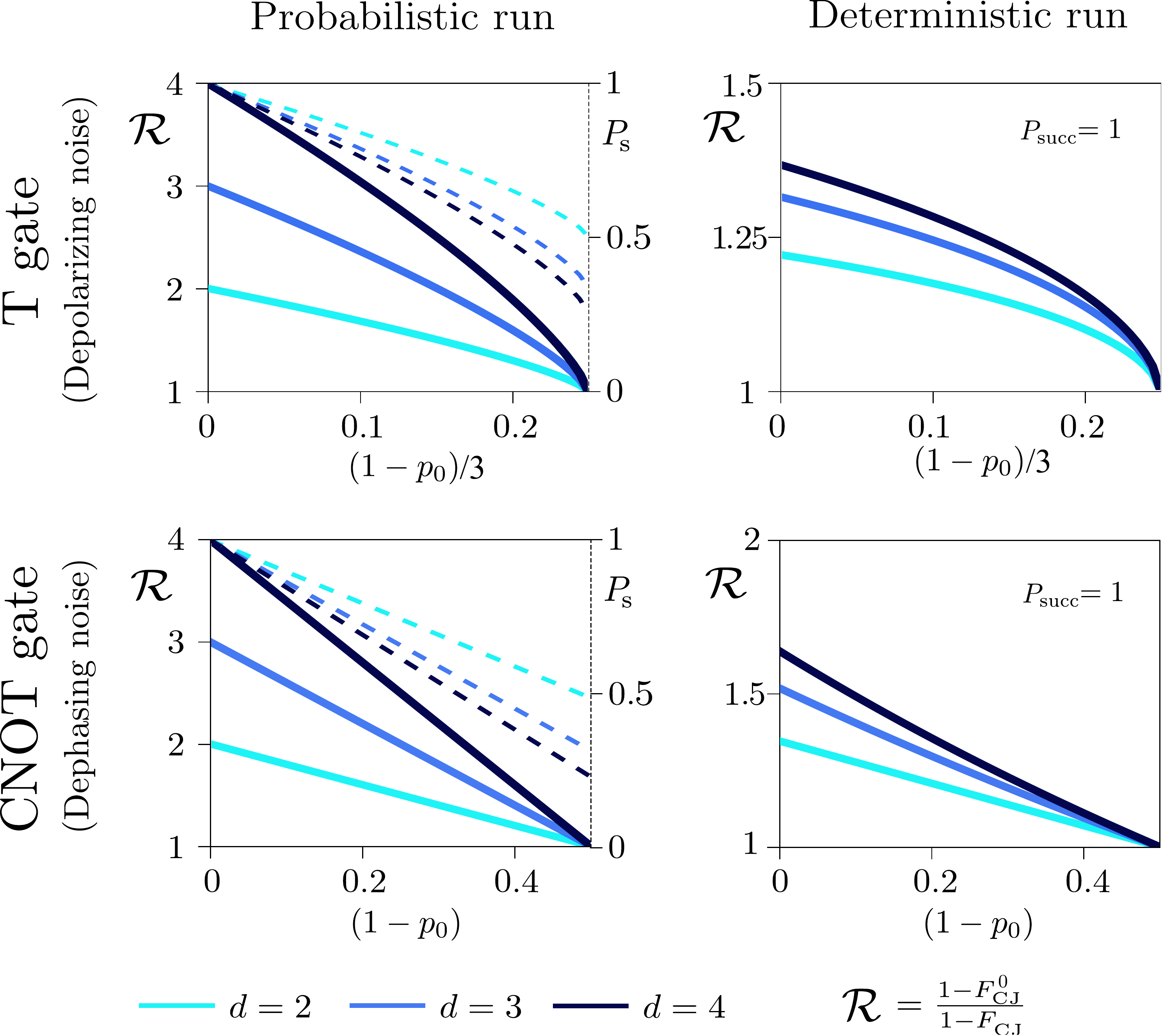}
    \caption{
    Performance of IB SQEM protocol applied to  a T gate and a cNOT gate, affected by depolarizing and dephasing noises, respectively. The performance in our coherent protocol is based on the stochastic magnetic field models Eqs.~\eqref{eq:stoc_ham_dephase_h} and \eqref{eq:stoc_ham_depol_h} with each field component following the Gaussian distribution \eqref{eq:stoc_ham_field_dist}; the vacuum interference operators of these models are given by Eqs.~\eqref{eq:stoc_ham_dephase_vio} and \eqref{eq:stoc_ham_depol_vio} respectively.}
    \label{fig:SH_T_depol} 
\end{figure}
\begin{table*}
\centering
\caption{Comparison table between implementations}
\label{tab:tablecomparision}
\begin{tabular}{|c||c|c|c|}
\hline
 & Gate-based & Measurement-based & Interferometric-based \\
\hline
\textit{Computational setting} & Quantum circuits & Measurement patterns on entangled state & Any \\
\textit{Superposition generation} & cSWAP gates & cSWAP gates & Interferometers \\
\textit{How?} & Artificial noise interference & Artificial noise interference &  Interference with the vacuum\\
\textit{Ancillas required?} & Yes & Yes & No \\
\textit{Free tunability?} & Yes & Yes & No \\
\hline
\end{tabular}
\end{table*}

\subsection{Protocol performance. Numerical analysis}
\label{sec:resultsinterfero}
In this section, we provide a numerical analysis of the performance of the IB protocols. All the numerical calculations are computed using the stochastic Hamiltonian formalism introduced in Sec.~\ref{sec:IB_stoqastic}, based on either the dephasing channel Eqs.~\eqref{eq:stoc_ham_dephase_ne_prob} and \eqref{eq:stoc_ham_dephase_vio} or the depolarizing channel Eqs.~\eqref{eq:stoc_ham_depol_ne_prob} and \eqref{eq:stoc_ham_depol_vio}, with error probability $1-p_0$ each time $U$ is implemented (see Sec.~\ref{sec:gb_quantum_comp_intro}). In Fig.~\ref{fig:SH_T_depol} we present the results for a T gate and a cNOT gate for different values of $d=2,3,4$. Here we only show the probabilistic and deterministic runs, because aside from the ideal output with the index ``0'', all output states yield the same CJ fidelity $F_{\rm CJ}^{(q)}$ in Eq.~\eqref{eq:CJaverage}, meaning discarding the worst outcome (as we have done previously for the quasi-deterministic protocol) is the same as the probabilistic protocol.

As one can see, the IB schemes present similar features as the GB-QC and MB-QC cases. For large $p_0$ we observe increasing values of the ratio $\mathcal{R}$ with respect to the value $d$, both for the probabilistic and deterministic runs of the protocol. Specifically, $\mathcal{R} \rightarrow d$ for $p_0$ approaching one in the probabilistic implementations. Since the noise acts as in standard GB-QC, it is meaningful to compare the T gate results here presented with the ones in Fig.~\ref{fig:GB_T_depha} (first row). The main difference is that the GB protocol achieves higher values of $\mathcal{R}$ for the same $p_0$. The reason for this lies in the different abilities of distinguishing errors. For the GB protocols the auxiliary states were chosen such that $(\omega_1,\omega_2) = (1,1)$, resulting in the maximum advantage $\mathcal{R}$. Here, the vacuum states and their interaction with the experimental setup are set by the considered physical model, which in this case is outlined in Sec.~\ref{sec:gb_quantum_comp_intro}. Therefore, we have less control over the probabilities and the evolution operators in Eq.~\eqref{eq:ib_stoc_vio} [or more generally the phases and weights of the Kraus operators from Eq.~\eqref{eq:IB_vacuum_operators}], which affect the outcome $\rho_{\rm out}$ in Eq.~\eqref{eq:rhointerf2} [or Eq.~\eqref{eq:interfer_out}] and thus the protocol advantage.

\section{Outlook and conclusions}
\label{sec:conclusions}
In this work, we have extended and analyzed several protocols for mitigating the noise associated with any quantum computation, denoted as superposed quantum error mitigation (SQEM)  and introduced in Ref.~\cite{prl_us}. In contrast to error correction and fault-tolerant quantum computing, where errors are corrected actively, our protocols work via interference between the noise acting in different branches.
The key idea behind our strategies consists in performing identical computations in coherent superposition, such that the imperfect quantum gates act either on the desired input or on some auxiliary states (or the vacuum). We demonstrated how this principle leads to a consistent advantage in several parameter regimes. Furthermore, we showed its applicability to different setups based either on GB- or MB-QC, as well as an IB approach introduced here (see also Table \ref{tab:tablecomparision}). 

We have provided analytical and numerical evidence of the advantage of our approaches. We analyzed how the choice of the auxiliary states and measurement bases influence the achievable advantage in the GB and MB settings. Crucially, our protocols can tolerate noise arising from the additional systems employed for generating the desired superposition. The techniques and methods that we introduce are platform-independent and are applicable to different setups and approaches. 

The strategies we have introduced entail a viable near-term approach for mitigating noise in quantum computations. Similar ideas will be explored in the context of self-calibrating quantum networks in a future work~\cite{papercomm}.

After completing this work, we became aware of a similar approach independently put forward in~\cite{Gideon22,Hann_19}.

\section*{Acknowledgments}

This work was supported by the Austrian Science Fund (FWF) through projects No. P36009-N and No. P36010-N. Finanziert von der Europäischen Union - NextGenerationEU. Furthermore, we acknowledge support from the Natural Sciences and Engineering Research
Council of Canada (NSERC), the Canada First Research Excellence Fund (CFREF, Transformative Quantum Technologies), New Frontiers in Research Fund (NFRF), Ontario Early Researcher Award, and the Canadian Institute for Advanced Research (CIFAR). LD acknowledges the EPSRC Quantum Career Development grant EP/W028301/1.

\bibliographystyle{apsrev4-2}
\bibliography{bib}

\clearpage


\onecolumngrid
\renewcommand\appendixname{Appendix}
\appendix

\section{Enhanced standard circuit gate-based computation using auxiliary qubits. Analytic analysis based on the density matrix formalism}\label{sec:Appendix_density}
We consider the protocol introduced in Sec.~\ref{sec:standardgatebased} and detail the mathematical evolution of the input states step by step, making use of the standard density operator formalism. We provide analytical evidence that the probabilistic protocol leads to an advantage without correcting unitaries. We start by analyzing the case with two branches, $d=2$, and analyze the general situation afterward. Moreover, for the sake of a better understanding, we restrict in this section to a simplistic case where $(\omega_1, \omega_2)=(1,1)$ in Eq.~\eqref{eq:omegas} and rank-2 noise. Generalization to arbitrary settings is straightforward and the results of it are stressed in the main text in Sec.~\ref{sec:standardgatebased}. 

Consider first the case with a single-qubit input and a single-qubit auxiliary system prepared in the states $\ket{\psi_{\rm in}}_{\rm a}$ and $\ket{\phi_{0}}_{\rm b}$  respectively. Consider also a certain computation $U$, whose realistic implementation leads to some noisy computation ${\cal E}_U$, see Eq.~\eqref{eq:gen_noise_gate}, modelled by the ideal one followed by a certain noisy channel with Kraus operators $\{K_i\}$, where we assume identical noise in each branch for simplicity. The incoherent effect of the noisy computation acting on the input state simply reads
\begin{equation}
    \rho_{\rm out}=  \sum_{i} K_{i} \left( U \proj{\psi_{\rm in}}_{\rm a} U^{\dagger} \right) K_{i}^{\dagger}, \label{eq:incoherentappendix}
\end{equation}
so that the incoherent fidelity is given by $F^{0}= \bra{\psi_{\rm in}} U^{\dagger} \rho_{\rm out} U \ket{\psi_{\rm in}}$. 

Consider now the SQEM process introduced in Sec.~\ref{sec:standardgatebased}. A control register is also prepared in the state $\ket{+}_{\rm c}$. A controlled-SWAP operation (Eq.~\eqref{eq:GB-QCcswap}) is applied to coherently swap the main and auxiliary registers depending on the state of the control, i.e. 
\begin{equation}
    \ket{+}_{\rm c} \ket{\psi_{\rm in}}_{\rm a} \ket{\phi_{0}}_{\rm b} \overset{(2)} {\rightarrow}  \frac{1}{\sqrt{2}} \ket{0}_{\rm c} \ket{\psi_{\rm in}}_{\rm a} \ket{\phi_{0}}_{\rm b}+ \frac{1}{\sqrt{2}} \ket{1}_{\rm c} \ket{\phi_{0}}_{\rm a} \ket{\psi_{\rm in}}_{\rm b}.
\end{equation}
The noisy computation is then applied in both registers, the main and auxiliary, such that the resulting state (after step~\ref{prot:gate3}) reads
\begin{equation}
\begin{split}
    &\frac{1}{{2}} \proj{0}_{\rm c} \sum_{i,j}\left( L_{i}  \proj{\psi_{\rm in}}_{\rm a} L_{i}^{\dagger} \right) \left(L_{j}  \proj{\phi_{0}}_{\rm b} L_{j}^{\dagger}\right) + \frac{1}{{2}} \proj{1}_{\rm c} \sum_{i,j} \left(L_{i}  \proj{\phi_{0}}_{\rm a} L_{i}^{\dagger}\right)  \left(L_{j}  \proj{\psi_{\rm in}}_{\rm b} L_{j}^{\dagger}\right) + \\
    & \frac{1}{{2}} \ket{0}\bra{1}_{\rm c} \sum_{i,j} \left(L_{i}  \ket{\psi_{\rm in}}\bra{\phi_{0}} _{\rm a} L_{i}^{\dagger}\right)  \left( L_{j} \ket{\phi_{0}}\bra{\psi_{\rm in}} _{\rm a} L_{j}^{\dagger} \right) + \frac{1}{{2}} \ket{1}\bra{0}_{\rm c} \sum_{i,j}\left( L_{i}  \ket{\phi_{0}}\bra{\psi_{\rm in}} _{\rm a} L_{i}^{\dagger}\right)  \left( L_{j} \ket{\psi_{\rm in}}\bra{\phi_{0}} _{\rm b} L_{j}^{\dagger}\right),
\end{split}
\end{equation}
where we have defined $L_i=U K_i$. A second cSWAP gate is subsequently applied (step~\ref{prot:gate4}), i.e.
\begin{equation}
\label{eq:densitymatrixprevious}
\begin{split}
    &\frac{1}{{2}} \proj{0}_{\rm c} \sum_{i,j}\left( L_{i}  \proj{\psi_{\rm in}}_{\rm a} L_{i}^{\dagger} \right) \left(L_{j}  \proj{\phi_{0}}_{\rm b} L_{j}^{\dagger}\right) + \frac{1}{{2}} \proj{1}_{\rm c} \sum_{i,j} \left(L_{i}  \proj{\psi_{\rm in}}_{\rm a} L_{i}^{\dagger}\right)  \left(L_{j}  \proj{\phi_{0}}_{\rm b} L_{j}^{\dagger}\right) + \\
    & \frac{1}{{2}} \ket{0}\bra{1}_{\rm c} \sum_{i,j} \left(L_{i}  \proj{\psi_{\rm in}}_{\rm a} L_{j}^{\dagger}\right)  \left( L_{j} \proj{\phi_{0}}_{\rm b} L_{i}^{\dagger} \right) + \frac{1}{{2}} \ket{1}\bra{0}_{\rm c} \sum_{i,j}\left( L_{j}   \proj{\psi_{\rm in}}_{\rm a} L_{i}^{\dagger}\right)  \left( L_{i} \proj{\phi_{0}}_{\rm b} L_{j}^{\dagger}\right), 
\end{split}
\end{equation}
where we can observe a correlation between the noise (Kraus operators) in the coherence terms. Finally, the control and auxiliary registers are measured in the Pauli $X$ and some suitable basis respectively. The effect of a measurement in the $X$ basis of the control register can be better understood by rewriting Eq.~\eqref{eq:densitymatrixprevious} as 
\begin{equation}
\label{eq:densitymatrixprevious2}
\begin{split}
    &\frac{1}{{2}}  \proj{+}_{\rm c} \left[\sum_{i,j}\left( L_{i}  \proj{\psi_{\rm in}}_{\rm a} L_{i}^{\dagger} \right) \left(L_{j}  \proj{\phi_{0}}_{\rm b} L_{j}^{\dagger}\right) +  \sum_{i,j} \left(L_{i}  \proj{\psi_{\rm in}}_{\rm a} L_{j}^{\dagger}\right)  \left(L_{j}  \proj{\phi_{0}}_{\rm b} L_{i}^{\dagger}\right) \right] + \\
    &\frac{1}{{2}}  \proj{-}_{\rm c} \left[ \sum_{i,j}\left( L_{i}  \proj{\psi_{\rm in}}_{\rm a} L_{i}^{\dagger} \right) \left(L_{j}  \proj{\phi_{0}}_{\rm b} L_{j}^{\dagger}\right) -  \sum_{i,j} \left(L_{i}  \proj{\psi_{\rm in}}_{\rm a} L_{j}^{\dagger}\right)  \left(L_{j}  \proj{\phi_{0}}_{\rm b} L_{i}^{\dagger}\right) \right],
\end{split}
\end{equation}
from which one can see how, in the absence of noise, the probability of measuring the $\proj{+}_{\rm c}$ outcome is 1. 

We focus in this analytical derivation on the branch where desired outcomes are obtained, in order to show how an advantage can be probabilistically always found independently of the input state and the noise, without any correcting operation. See Sec.~\ref{sec:standardgateperformance} for further details and numerical evidence that deterministic enhancement can be also always achieved. 

For a better understanding, we assume rank-2 Pauli noise where the probability that the computation is implemented in a noiseless way is $\geq \frac{1}{2}$, i.e., $K_i = \{K_0, K_1 \}= \{\sqrt{p_{0}} \id, \sqrt{1-p_{0}} K_1\}$ with $p_{0} \geq \frac{1}{2}$. We show afterward how this can be extended for arbitrary noise.

Moreover, we restrict to the case where  $(\omega_1, \omega_2)=(1,1)$ in Eq.~\eqref{eq:omegas}, such that the state $\ket{\phi_{0}}$  is mapped to orthogonal states under $L_0$ and $L_1$, i.e. $\ket{\phi'_{0}} = L_0\ket{\phi_{0}}$ satisfies $\bra{\phi'_{0}} L_1\ket{\phi_{0}}=0$.  Considering that the control register is measured in the $X$ basis and the outcome $\ket{+}_{\rm c}$, while the auxiliary system is measured in the $\{ \proj{\phi'_{0}}, \id - \proj{\phi'_{0}} \}$ basis, and the first outcome is found, the remaining state reads 
\begin{gather}
    \frac{1}{{2}}  \sum_{i}  \left( L_{i}  \proj{\psi_{\rm in}}_{\rm a}  L_{i}^{\dagger} \right)  +   \frac{1}{{2}}  \sum_{i,j}  \left( L_{j}   \proj{\psi_{\rm in}}_{\rm a} L_{i}^{\dagger}\right)  \bra{\phi'_{0}} \left( L_{i} \proj{\phi_{0}}_{\rm b} L_{j}^{\dagger}\right)  \ket{\phi'_{0}}_{\rm b },\label{eq:unaeq}
\end{gather}
up to normalization. Since only terms of the form $\bra{\phi'_{0}} L_0 \ket{\phi'_{0}}$ survive in the right part of the previous expression, one can see from this simplified assumption how the contribution of the noiseless-computation terms (associated with the Kraus $K_0$) gets enhanced.  The fidelity of this state is
\begin{gather}
   F= N \sum_{i} \bra{\psi_{\rm out}} \left( L_{i}  \proj{\psi_{\rm in}}_{\rm a}  L_{i}^{\dagger} \right) \ket{\psi_{\rm out}}  + N   \bra{\psi_{\rm out}} \left( L_{0}   \proj{\psi_{\rm in}}_{\rm a} L_{0}^{\dagger}\right) \ket{\psi_{\rm out}}  \label{eq:unaeq2}
\end{gather}
where $\ket{\psi_{\rm out}}= U \ket{\psi_{\rm in}}$ is the ideal output state and $N$ is a normalization factor. In order to find an advantage in the coherent case, the following expression has to be fulfilled
\begin{gather}
   F= N \sum_{i} \bra{\psi_{\rm out}} \left( L_{i}  \proj{\psi_{\rm in}}_{\rm a}  L_{i}^{\dagger} \right) \ket{\psi_{\rm out}}  + N   \bra{\psi_{\rm out}} \left( L_{0}   \proj{\psi_{\rm in}}_{\rm a} L_{0}^{\dagger}\right) \ket{\psi_{\rm out}}   > F^{0},
   \label{eq:A8}
\end{gather}
where $F^{0}$ is the incoherent fidelity from Eq.~\eqref{eq:incoherentappendix}. For concreteness, we consider the case where the only term that contributes to the fidelity of the  $\ket{\psi_{\rm in}}_{\rm a}$ state is $L_0$, as in the CJ fidelity analysis, Eq.~\eqref{eq:CJfid}. 

Given the fact that $K_{0}=\sqrt{p_{0}} \id$, and $F^{0} = p_{0}$, we therefore find $N= \frac{1}{p_{0}+1}$, such that Eq.~\eqref{eq:A8} reduces to
\begin{gather}
\frac{2p_0}{p_0+1}>p_0,
\end{gather}
which is always satisfied, proving the advantage of our protocol under the circumstances considered.

Consider now the case where the input, as well as the computation, involve $m>1$ qubits.  An auxiliary system of the same number of qubits ($m$) is required in this case. The same results above apply in this case. However, in order to better understand the protocol performance with multi-qubit states, we can analyze the particular case where each qubit is locally affected by the same noise as before.  Eq.~\eqref{eq:densitymatrixprevious2} becomes 
\begin{equation}
\label{eq:densitymatrixprevious3}
\begin{split}
    &\frac{1}{{2}}  \proj{+}_{\rm c} \left[\sum_{\substack{ q_{1} \dots q_{m} \\ r_{1} \dots r_{m}}} \left( L_{q_{m}} \dots L_{q_{1}} \proj{\psi_{\rm in}}_{\rm a} L^{\dagger}_{q_{1}} \dots L^{\dagger}_{q_{m}} \right) \left( L_{r_{m}} \dots L_{r_{1}} \proj{\phi_{0}}_{\rm b} L^{\dagger}_{r_{1}} \dots L^{\dagger}_{r_{m}} \right) \right] + \\ 
    &\frac{1}{{2}} \proj{+}_{\rm c}\left[\sum_{\substack{ q_{1} \dots q_{m} \\ r_{1} \dots r_{m}}} \left( L_{q_{m}} \dots L_{q_{1}} \proj{\psi_{\rm in}}_{\rm a} L^{\dagger}_{r_{1}} \dots L^{\dagger}_{r_{m}} \right) \left( L_{r_{m}} \dots L_{r_{1}} \proj{\phi_{0}}_{\rm b} L^{\dagger}_{q_{1}} \dots L^{\dagger}_{q_{m}} \right) \right] + \\
    &\frac{1}{{2}}  \proj{-}_{\rm c}\left[\sum_{\substack{ q_{1} \dots q_{m} \\ r_{1} \dots r_{m}}} \left( L_{q_{m}} \dots L_{q_{1}} \proj{\psi_{\rm in}}_{\rm a} L^{\dagger}_{q_{1}} \dots L^{\dagger}_{q_{m}} \right) \left( L_{r_{m}} \dots L_{r_{1}} \proj{\phi_{0}}_{\rm b} L^{\dagger}_{r_{1}} \dots L^{\dagger}_{r_{m}} \right) \right] + \\
    &\frac{1}{{2}} \proj{-}_{\rm c}\left[\sum_{\substack{ q_{1} \dots q_{m} \\ r_{1} \dots r_{m}}} \left( L_{q_{m}} \dots L_{q_{1}} \proj{\psi_{\rm in}}_{\rm a} L^{\dagger}_{r_{1}} \dots L^{\dagger}_{r_{m}} \right) \left( L_{r_{m}} \dots L_{r_{1}} \proj{\phi_{0}}_{\rm b} L^{\dagger}_{q_{1}} \dots L^{\dagger}_{q_{m}} \right) \right].
\end{split}
\end{equation}
By measuring the control and auxiliary registers, under the same assumptions as before, with  Kraus operators $\{K_0, K_1 \}= \{\sqrt{p_{0}} \id, \sqrt{1-p_{0}} K_1\}$, and the desired measurement outcomes, we find the fidelity of the remaining state to be 
\begin{align}
    N \left[ \sum_{\substack{ q_{1} \dots q_{m}}}  \bra{\psi_{\rm in}}_{\rm a}  \left( L_{q_{m}} \dots L_{q_{1}} \proj{\psi_{\rm in}}_{\rm a} L^{\dagger}_{q_{1}} \dots L^{\dagger}_{q_{m}} \right) \ket{\psi_{\rm in}}_{\rm a} + \bra{\psi_{\rm in}}_{\rm a} \left( L_{0} \dots L_{0} \proj{\psi_{\rm in}}_{\rm a} L^{\dagger}_{0} \dots L^{\dagger}_{0} \right) \ket{\psi_{\rm in}}_{\rm a} \right],
\end{align}
with $N=\frac{1}{1+p^{m}_{0}}$. Note that the fidelity of the incoherent process is $F^{0}=p^{m}_{0}$ now. We can then obtain the advantage infidelity ratio for the probabilistic run of the protocol, i.e. the infidelity of the incoherent process over the infidelity of the coherent one,
\begin{equation}
    \frac{1-F_{\rm CJ}^{0}}{1-F_{\rm CJ}}=\frac{1-p^{m}_{0}}{1-\frac{2p^{m}_{0}}{1+p^{m}_{0}}}=1+p^{m}_{0}.
\end{equation}
The advantage depends on $m$ for equivalent noise affecting each qubit locally.

We have shown how one can always find a probabilistic advantage, i.e., when the desired outcomes are found. As explained in Sec.~\ref{sec:standardgatebased}, this advantage can be increased by adding more branches to the superposition (and therefore making use of more auxiliary qubits) such that an asymptotically perfect computation $F \rightarrow 1$ can be achieved. This can be also seen from the simplistic example treated above.

Consider again a single-qubit input and $(d-1)$ single-qubit auxiliary systems prepared in the states $\ket{\psi_{\rm in}}_{\rm a}$ and $\bigotimes_{i} \ket{\phi_{0}}_{\rm b_{i}}$  respectively. Consider again some computation $U$, whose realistic implementation lead to some noisy implementation ${\cal E}_U$, modelled by the ideal one followed by a certain noisy channel with Kraus operators $\{K_i\}$. Moreover, a $d-$dimensional control register is prepared in the state $  \ket{+_{d}}_{\rm c} = \frac{1}{\sqrt{d}}\sum_{i=0}^{d-1} \ket{i}_{\rm c}^{d}$. The effect of the cSWAP in this case is
\begin{equation}
   \sum_{i=0}^{d-1} \frac{1}{\sqrt{d}} \ket{i}_{\rm c} \ket{\psi_{\rm in}}_{\rm b_{i}}  \bigotimes_{j \neq i}^{d-1} \ket{\phi_{0}}_{\rm b_{j}},
\end{equation}
where we relabel system $a \equiv b_{0}$ for convenience. The effect of the noisy computation can be written as
\begin{equation}
\begin{split}
    &\frac{1}{{d}} \sum_{q=0}^{d-1} \proj{q}_{\rm c} \sum_{j_{0},\dots,j_{n}} \left( L_{j_{q}}  \proj{\psi_{\rm in}}_{ b_{q}} L_{j_{q}}^{\dagger} \right) \bigotimes_{s \neq q}^{d-1}  \left(L_{j_{s}}  \proj{\phi_{0}}_{ b_{s}} L_{j_{s}}^{\dagger}\right) + \\
    &\frac{1}{{d}} \sum_{q,l=0}^{d-1} \ket{q} \bra{l}_{\rm c}  \sum_{j_{0},\dots,j_{n}} \left( L_{j_{q}} \ket{\psi_{\rm in}} \bra{\phi_{0}}_{ b_{q}} L_{j_{q}}^{\dagger} \right) \left( L_{j_{l}} \ket{\phi_{0}} \bra{\psi_{\rm in}}_{b_{l}} L_{j_{l}}^{\dagger} \right) \bigotimes_{s \neq q,l}^{d-1} \left( L_{j_{s}} \ket{\phi_{0}} \bra{\phi_{0}}_{ b_{s}} L_{j_{s}}^{\dagger} \right),
\end{split}
\end{equation}
where again we define $L_i=U K_i$ and consider identical noise in all the channels (note that this assumption can be relaxed and the protocol still works). Finally, the application of the final cSWAP leads to
\begin{equation}
\begin{split}
    &\frac{1}{{d}} \sum_{q=0}^{d-1} \proj{q}_{\rm c} \sum_{j_{0},\dots,j_{n}} \left( L_{j_{0}}  \proj{\psi_{\rm in}}_{ b_{q}} L_{j_{0}}^{\dagger} \right) \bigotimes_{s \neq 0}^{d-1}  \left(L_{j_{s}}  \proj{\phi_{0}}_{ b_{s}} L_{j_{s}}^{\dagger}\right) + \\
    &\frac{1}{{d}} \sum_{q,l=0}^{d-1} \ket{q} \bra{l}_{\rm c}  \sum_{j_{0},\dots,j_{n}} \left( L_{j_{q}} \ket{\psi_{\rm in}} \bra{\psi_{\rm in}}_{ b_{0}} L_{j_{l}}^{\dagger} \right) \left( L_{j_{0}} \ket{\phi_{0}} \bra{\phi_{0}}_{ b_{q}} L_{j_{q}}^{\dagger} \right) \left( L_{j_{l}} \ket{\phi_{0}} \bra{\phi_{0}}_{ b_{l}} L_{j_{0}}^{\dagger} \right) \bigotimes_{s \neq q,l}^{d-1} \left( L_{j_{s}} \ket{\phi_{0}} \bra{\phi_{0}}_{ b_{s}} L_{j_{s}}^{\dagger} \right).
\end{split}
\end{equation}
As before, the final step consists in measuring the control and auxiliary registers. For this analytical analysis, we focus only on one branch of the possible measurement outcomes. We remark that deterministic enhancement can also always be achieved on average, see Sec.~\ref{sec:standardgateperformance}. In this case, the fidelity of the output register reads
\begin{align}
F=  N  \frac{1}{{d}} F^{0}  + N \frac{1}{{d}} d(d-1) \sum_{i,j} \bra{\psi_{\rm out}} \left( L_{0}   \proj{\psi_{\rm in}}_{\rm a} L_{0}^{\dagger}\right) \ket{\psi_{\rm out}}. \label{eq:fidineqappendix}
\end{align}
Taking into account the normalization and that $F^{0}=p_{0}$ as before, we find
\begin{align}
F=\frac{d p_{0}}{1+p_{0}(d-1)}  \overset{d \to \infty} {\longrightarrow} 1,
\end{align}
which asymptotically converges to a regime of noiseless implementation of $U$, independently of the strength of the noise. Observe how this is also satisfied for extremal noisy ($p_{0}=1/2$) computations.

Similar results can be derived for another kind of noise associated with the computation, with a higher rank \cite{nielsen_chuang_2010}. For instance, one can see in this case how, in the right part of Eqs.~\eqref{eq:unaeq}--\eqref{eq:unaeq2} more than one term survives but, taking the normalization into account, the weight corresponding to the fidelity gets increased. Analogously, the asymptotically noiseless computation can be also achieved in this case, even in the completely depolarizing case (see main text for details).

\section{Further analytical analysis. Two-qubit case} \label{app:otheranalytics}

In this section we show a two-qubit relevant example under realistic assumptions, where we find that the protocol leads to partial, but still significant, advantage. 

In most available quantum platforms, the main source of noise comes from entangling unitaries, generally either the $CX$ or the $CZ$ gate \cite{Preskill2018}. It is therefore meaningful to analyze our protocol for those specific resources. For clarity, we consider similar settings as in the previous single qubit example. We assume that the dominating noise is dephasing, now with Kraus operators $K_0 = p_{\rm 0} \id_1 \id_2 $, $K_1 = p_{1} Z_1 Z_2$, $K_2 = \sqrt{p_0 p_1} \id_1 Z_2$ and $K_3 = \sqrt{p_0 p_1} Z_1 \id_2$. As before, $p_1$ ($p_0$) is the probability that either qubit is (not) subject to an error. Notice that, while in the previous example $p_0$ was the probability $p_{\rm ne}$ of not having an error, here $p_{\rm ne} = p_0^2$. 

For what concerns the input state and unitary $U$, we choose $\ket{\psi_{\rm in}} = \ket{++}$ and $U = CZ$. In the noiseless case this prepares an $m=2$ qubits cluster state and, as in the $m = 1$ case, it yields the minimum incoherent fidelity 
\begin{equation}\label{eq:gate_ex_2qubut_incoh}
    F^{0} = p_0^2 = p_{\rm ne}.
\end{equation}

Since $U = CZ$ is an entangling gate, the choice of the auxiliary state $\ket{\phi_{ 0}}$ requires more care. Ideally, we would like to pick $\ket{\phi_{ 0}} = \ket{++}$ (i.e., choose $X_1$ and $X_2$ as stabilizers) to have $\bra{\phi_{ 0}} U^\dagger K_i U \ket{\phi_{ 0}} = p_0 \delta_{i,0}$ for all $i = 0,\dots,3$. Under this circumstance, we could follow the same steps as in the previous example and get the same (upon substitution $p_0 \rightarrow p_0^2$) post-selection probability and fidelity as in Eqs.~\eqref{eq:gate_ex_1qubit_out}. However, the application of $U = CZ$ to $\ket{++}$ results in an entangled state that may be challenging to measure. Consequently, we analyze our protocol for $\ket{\phi_{ 0}} = \ket{+1}$, which is stabilized by $+X_1$ and $+Z_2$, and is an eigenvector of $CZ$. These stabilizers, which after application of the $CZ$ gate become $+X_1 Z_2$ and $+Z_2$, represent a state that at step~\ref{prot:gate5} can be detected with local measurements only.

While the chosen $\ket{\phi_{ 0}} = \ket{+1}$ yields, after the application of $U = CZ$, a state that can be measured with local operations, it has a downside. In fact,
\begin{equation}\label{eq:gate_ex_2_interf}
    \bra{\phi_{ 0}} U^\dagger K_i U \ket{\phi_{ 0}} = p_0\delta_{i,0}+\sqrt{p_0 p_1}\delta_{i,2}
\end{equation}
gives non-zero contributions not only for $i=0$ (corresponding to $K_0 = p_{\rm 0} \id$), but also for $i=2$ (representing $K_2 = \sqrt{p_0 p_1} Z_2$). This means that $U \ket{\phi_{ 0}}$ is not maximally sensitive to the noise (in that case we would only get the desired no-noise contribution corresponding to $i=0$). Instead, it is insensitive to the Kraus operator $K_2$, meaning that our probabilistic protocol will not be capable of completely correcting the associated decoherence. This will become clearer following the detailed analysis below.

By substituting Eq.~\eqref{eq:gate_ex_2_interf} into Eq.~\eqref{eq:gate_interf}, which in turn is plugged into Eqs.~\eqref{eq:gate_out}, we find the post-selected outcome $\rho_{\rm out}$ to be
\begin{equation}
\label{eq:gate_out_ex_2}
\begin{split}
\rho_{\rm out} & =  p_0^{d+1} U \proj{\psi_{\rm in}}_{\rm a} U^{\dagger}
        + p_0^{d-1}\left( p_1 + \frac{p_0}{d} \right)
        K_2 U \proj{\psi_{\rm in}}_{\rm a} U^{\dagger} K_2^{\dagger}
        + \frac{p_0^{d-1}}{d}\Big( K_1 U \proj{\psi_{\rm in}}_{\rm a} U^{\dagger} K_1^{\dagger}  + K_3 U \proj{\psi_{\rm in}}_{\rm a} U^{\dagger} K_3^{\dagger} \Big) \\ 
        & + p_0^{d-1} \sqrt{p_0 p_1}\frac{d-1}{d}
        \Big( K_2 U \proj{\psi_{\rm in}}_{\rm a} U^{\dagger}      + U \proj{\psi_{\rm in}}_{\rm a} U^{\dagger} K_2^{\dagger}
        \Big),
\end{split}
\end{equation}
where we used that $\mathcal{A}_d = p_0^{d-1}$ and $(p_0+p_1)^2=p_0+p_1=1$. By comparing this last equation to the one in Eq.~\eqref{eq:gate_out_noise_suppr}, we identify an important difference. The noise insensitivity of the chosen auxiliary state determines the survival of undesired terms.
These terms are exactly the ones corresponding to the Kraus operator $K_2$, of which $U\ket{\phi_{ 0}}$ is an eigenstate (and thus insensitive to the associated decoherence). All terms associated with $K_1$ and $K_3$, on the other hand, are suppressed. As in the previous example, this follows from $U\ket{\phi_{ 0}}$ being completely sensitive to them. This is better seen in the limit $d\gg 1$, in which we can approximate $\rho_{\rm out}$ in Eq.~\eqref{eq:gate_out_ex_2} as
\begin{equation}
\label{eq:gate_out_ex_2_1}
\begin{split}
\rho_{\rm out} &  = p_0^{d+1} U \proj{\psi_{\rm in}}_{\rm a} U^{\dagger}    + p_0^{d-1} p_1  K_2 U \proj{\psi_{\rm in}}_{\rm a} U^{\dagger} K_2^{\dagger} \\ 
& +  p_0^{d-1}\sqrt{p_0 p_1} \Big( K_2 U \proj{\psi_{\rm in}}_{\rm a} U^{\dagger}  + U \proj{\psi_{\rm in}}_{\rm a} U^{\dagger} K_2^{\dagger} \Big).
\end{split}
\end{equation}
Both $K_1$ and $K_3$ are absent from this last equation, implying that their associated noise contributions are asymptotically eliminated for large $d$. However, compared to the previous example where $\rho_{\rm out} \propto U \proj{\psi_{\rm in}} U^\dagger$ and $F \approx 1$ for $d \gg 1$, here undesired contributions survive, such that the resulting fidelity is upper bounded to a value lower than one (see below).

The results in Eqs.~\eqref{eq:gate_out_ex_2} and \eqref{eq:gate_out_ex_2_1} are valid no matter the input state. The associated fidelity $F$, however, do depend on the choice of $\ket{\psi_{\rm in}}$. As mentioned above, for $\ket{\psi_{\rm in}} = \ket{++}$ we obtain a lower bound on the incoherent fidelity $F^{0} = p_0^2 = 1 - p_1\left(p_0 + 1 \right)$. Since the surviving noise contribution described by $K_2$ is such that $K_2 U \ket{++}$ and $U \ket{++}$ are orthogonal, we conclude that $F$ determined with respect to $\ket{\psi_{\rm in}} = \ket{++}$ is also a lower bound that, from Eq.~\eqref{eq:gate_out_ex_2}, we calculate to be
\begin{subequations}
\label{eq:gate_ex_2qubit_out}
\begin{align}
    \mathcal{P} & = p_0^{d+1} 
    + p_0^{d}p_1\left( p_1 + \frac{p_0}{d} \right)
    + \frac{p_0^{d-1}p_1}{d} 
    ,\label{eq:gate_ex_2qubit_out_P}\\
    F & = \frac{d p_0^2}{p_1 + p_0 \left[ p_0 p_1 + d \left( p_0 + p_1^2 \right) \right]}.\label{eq:gate_ex_2qubit_out_F}
\end{align}
\end{subequations}
For completeness, in the last equation we included the post-selection probability $\mathcal{P}$.
It is interesting to take again the limit $d \gg 1$ for these quantities, which can be found directly from Eq.~\eqref{eq:gate_out_ex_2_1}, to obtain
\begin{subequations}
\label{eq:gate_ex_2qubit_out_lim}
\begin{align}
    \mathcal{P} & \underset{d \gg 1}{\longrightarrow} p_0^{d} \left( p_0 + p_1^2 \right),\label{eq:gate_ex_2qubit_out_lim_P}\\
    F & \underset{d \gg 1}{\longrightarrow} 1 - \frac{p_1^2}{p_{0} + p_1^2}.\label{eq:gate_ex_2qubit_out_lim_F}
\end{align}
\end{subequations}

In contrast to the previous example, where the auxiliary state was chosen such that $U \ket{\phi_{ 0}}$ was maximally sensitive to \textit{all} Kraus operators, here $F$ is upper bounded to a value that is less than one, no matter the number of branches $d$ one employs. However, this does not mean that our protocol is not advantageous. By comparing Eq.~\eqref{eq:gate_ex_2qubit_out_lim_F} with the incoherent fidelity $F^{0} = p_0^2$, it is possible to conclude that $F \geq F^{0}$ always, and in the experimentally relevant scenario $1 \simeq p_0 \gg p_1$ the advantage infidelity ratio becomes 
$(1-F)/(1-F^{0})\approx 2/p_1$. This improvement, which is substantial within the considered approximation, comes from the elimination of the noise associated with $K_1$ and $K_3$. 

An interesting question is whether it is possible, for generic (unknown) noise and the constraints on $\ket{\phi_{ 0}}$ described in the first part of this section, to modify our scheme to \textit{always} reach unit fidelity asymptotically for $d \gg 1$. This is what we investigate in Sec.~\ref{sec:gate_nested} with the nested protocol, where we eliminate the noise contributions from all Kraus operators in consecutive, nested application of our protocol.

\section{Proof of Eq.~(\ref{eq:finalresult1})} \label{app:Proof}
We prove here the derivation that leads to  $F>F^{0}$  from Eq.~\eqref{eq:finalresult1}, which shows that our protocol always leads to a probabilistic advantage for any kind of noise, under assumptions that guarantee  $(\omega_1, \omega_2)=(1,1)$ in Eq.~\eqref{eq:omegas}.

Given Eq.~\eqref{eq:generallambdageneral}, we first need to prove that $|\lambda_{mn} | ^{2} \leq \lambda_{mm}\lambda_{nn}$ for any coefficient of the process matrix $\lambda$ of Eq.~\eqref{eq:generalmapintro}, i.e.
\begin{equation}
    {\cal E} (\rho_{\rm in}) = \sum_{m,n} \lambda_{mn} \sigma_m \rho_{\rm in} \sigma_n^{\dagger}, \label{eq:generalmapapp}
\end{equation}
where $\sigma_i= \{\id,X,Z,Y\}$ are the Pauli matrices. Consider a Kraus decomposition of the noise, 
\begin{equation}
   {\cal E}(\rho_{\rm in}) =\sum_{i} K_{i} \rho_{\rm in} K_{i}^{\dagger},
\end{equation}
where without loss of generality we can always define the Kraus operators such that they are orthogonal to each other \cite{nielsen_chuang_2010}, ${\rm tr} [K_i^{\dagger} K_j] \propto \delta_{ij}$. We can always express each Kraus operator in terms of the Pauli matrices, i.e. 
\begin{equation}
    K_{i}= \sum_j \alpha_{i,j}  \sigma_j, \label{eq:kraus appendix}
\end{equation}
where $\sum_{i,j} |\alpha_{i,j}|^2=1$. We can directly relate these coefficients to the process matrix coefficients, $\lambda_{mn}=\sum_i \alpha_{im} \alpha^{*}_{in}$. From this decomposition, one can see how the inequality
\begin{equation}
    |\lambda_{mn} | ^{2} \leq \lambda_{mm}\lambda_{nn},
\end{equation}
reduces to the Cauchy–-Schwarz inequality, i.e. 
\begin{equation}
    |\lambda_{mn} | ^{2} = |\sum_{i} \alpha_{im} \alpha_{in}^{*} | ^{2} \leq  \sum_{i} | \alpha_{im} |^2 \sum_{i} | \alpha_{in} |^2 = \lambda_{mm}\lambda_{nn}, \label{eq:cauchy}
\end{equation}
hence proving its validity. By applying this inequality into Eq.~\eqref{eq:generallambdageneral}, we find that $N \leq 2 \lambda_{00}$, and therefore $\lambda'_{00} \geq \lambda_{00}$. 

Observe also that the equality in Eq.~\eqref{eq:cauchy} is only satisfied when $\alpha_{im}=q \alpha_{in}$ for $\forall i$ with some $q \in \mathbb{R}$. In order to have $\lambda'_{00} = \lambda_{00}$, we need the previous equality to be satisfied for $m=0$ and $n=0,1,2,3$. This is however not possible owing to the orthogonality of Kraus operators, Eq.~\eqref{eq:kraus appendix}, therefore proving that in general
\begin{equation}
    \lambda'_{00} > \lambda_{00}
\end{equation}
for any noisy channel affecting $U$, implying that $F>F^{0}$ after our protocol implementation.
Note the independence of the results on the number of qubits $m$ in the input state. 

\section{Enhanced standard circuit gate-based computation using auxiliary qubits. Analytic analysis based on the environmental formalism}\label{sec:Appendixstandard1}

\subsection{Stinespring dilation theorem and operator sum representation}
\label{sec:appendixMB-QC111}

The evolution of any quantum system is unavoidably subjected to noise and decoherence owing to interactions with the surroundings and imperfections of the apparatuses. Some information in the evolution of a quantum system gets lost from our knowledge during the process. The part of the whole system where the information is leaked out, and which we cannot observe or control, is called the environment, with an associated Hilbert space $\mathcal{H}_{e}$. Only the complete description of system and environment, i.e.  $\mathcal{H}_{s} \otimes \mathcal{H}_{e}$, gives us full information about the evolution of the state of the system. 

In particular, the dynamics of the joint system can always be described by a unitary evolution $U_{\rm se}$ acting on $\mathcal{H}_{s} \otimes \mathcal{H}_{e}$, such that $\rho=U_{\rm se} (\rho^{\rm s} \otimes \rho^{\rm e}) U_{\rm se}^{\dagger}$. The evolution of the state of the system corresponds to tracing out the environment, i.e.
\begin{equation}
\label{eq:operatorsum}
\begin{split}
    \xi (\rho_{\rm in}^{\rm s}) = \rho_{\rm out}^{\rm s} = \text{Tr}_{e} (U_{\rm se} (\rho_{\rm in}^{\rm s} \otimes \proj{e_{0}}) U_{\rm se}^{\dagger}) = \sum_{i} \bra{e_{i}} (U_{\rm se} (\rho^{\rm s}_{\rm in} \otimes \proj{e_{0}})U_{\rm se}^{\dagger})\ket{e_{i}} = 
     \sum_{i} K_{i} \rho_{\rm in}^{\rm s} K_{i}^{\dagger},
\end{split}
\end{equation}
where $\ket{e_{i}}$ are elements of an orthogonal basis of the environment, and where $K_{i}=\bra{e_{i}}U_{\rm se}  \ket{e_{0}}$ are the Kraus operators of the channel.

This allows us to write any quantum noisy channel as a unitary evolution on the larger Hilbert space $\mathcal{H}_{s} \otimes \mathcal{H}_{e}$ (Stinespring dilation theorem \cite{nielsen_chuang_2010}). 
Consider any computation  acting on certain input qubits, given by some unitary operation $U$. The unitary $U$ applied in a non-ideal way can be modeled as the perfect gate followed by certain noise. Given an arbitrary input state $\rho_{\rm in}=|\psi_{\rm in}\rangle \langle\psi_{\rm in}|$,  the noisy application of $U$ leads to the state
\begin{equation}
    \rho =  \sum_{s} K_{s} U\rho_{\rm in} U^{\dagger} K_{s}^{\dagger},
    \label{eq:rhocomp}
\end{equation}
where $\{K_{s}\}$ are the Kraus operators associated with the noisy implementation of the gate $U$. Observe that the Kraus decomposition of the noisy channel is not unique. In particular, the unitary freedom of the operator sum-representation \cite{nielsen_chuang_2010} implies that  descriptions given by the sets of Kraus operators $\{K_{i}\}$ and $\{K'_{i}\}$, where $K'_{i} = \sum_{j} u_{i,j} K_{j}$ with $u_{i,j}$ elements of some unitary matrix, lead to the same quantum map. This feature can be also interpreted as the insensitivity of the quantum channel to local operations acting on the environment during the evolution, so that two apparently different physical processes can lead to the same evolution. This can be seen considering the Stinespring dilation theorem \cite{nielsen_chuang_2010}, which allows us to describe any quantum map as unitary evolutions acting on certain pure states of a larger Hilbert space, i.e.
\begin{equation}\label{eq:Stinespring1}
\left|\psi\right\rangle \left|\epsilon_{0}\right\rangle _{\epsilon_{0}}  \rightarrow \sum_{s} K_{s} U \left|\psi\right\rangle \otimes\left|s\right\rangle _{\epsilon_{0}},
\end{equation}
where $\epsilon_{0}$ represents the state of the environment into which the information of the system is leaked out during the evolution. Note that by tracing out this environment at the end of the evolution, one recovers the description of Eq.~\eqref{eq:rhocomp}, independently of the aforementioned choice of the set of Kraus operators.

In this section, we analyze an example of our SQEM protocol using this formalism. This allows us to show how the three different implementations we analyze in this work are fundamentally different, but at the same time lead to similar behavior in terms of protocol advantage.

\subsection{Enhanced GB-QC Protocol. Simple example} \label{sec:appendixMB-QC11}

As shown in Sec.~\ref{sec:standardgatebased}, the gate-based quantum computation (GB-QC) approach we propose can be treated analytically using the density matrix formalism. It is however interesting, in order to understand the differences with respect to the other approaches, to tackle the problem from the environmental formalism point of view, which is based on the purified version of the states discussed above. For simplicity, we address here the case $d=2$.

Following Protocol~\ref{table:GBstandard}, consider an input qubit in some state $\ket{\psi_{\rm in}}_{a}$. In step~\ref{prot:gate1}, an auxiliary qubit is prepared in some suitable state $\ket{\phi_{0}}_{b}$ and a control qubit is initialized in the state $\left|+\right\rangle_{c}$. We take into account the system+environment Hilbert space in order to analyze the process dynamics, where we assign an environmental system to each qubit. The step-by-step procedure reads (see also Fig.~\ref{fig:compstandard1}):
\begin{equation}
\begin{split}
&\left|+\right\rangle_{c} \ket{\psi_{\rm in}}_{a} \ket{\phi_{0}}_{b} \ket{\varepsilon_{a}}_{\epsilon_{a}} \ket{\varepsilon_{b}}_{\epsilon_{b}}  \overset{(2)}{\rightarrow}   \frac{1}{\sqrt{2}}  \left( \ket{0}_{c} \ket{\psi_{\rm in}}_{a} \ket{\phi_{0}}_{b} +\ket{1}_{c} \ket{\psi_{\rm in}}_{a} \ket{\psi_{{\rm in}}}_{b} \right) \ket{\varepsilon_{a}}_{\epsilon_{a}} \ket{\varepsilon_{b}}_{\epsilon_{b}}  \overset{(3.a)}{\rightarrow} \\ 
&\frac{1}{\sqrt{2}}  \left( \ket{0}_{c} U\ket{\psi_{\rm in}}_{a} U\ket{\phi_{0}}_{b} +\ket{1}_{c}U \ket{\psi_{\rm in}}_{a} U\ket{\psi_{{\rm in}}}_{b} \right) \ket{\varepsilon_{a}}_{\epsilon_{a}} \ket{\varepsilon_{b}}_{\epsilon_{b}}  \overset{(3.b)} {\rightarrow} \\
&\frac{1}{\sqrt{2}} \sum_{i,j}  \left( \ket{0}_{c} K_{i} U\ket{\psi_{\rm in}}_{a} K_{j}U\ket{\phi_{0}}_{b} +\ket{1}_{c} K_{i} U \ket{\psi_{\rm in}}_{a} K_{j}U\ket{\psi_{{\rm in}}}_{b} \right) \ket{i}_{\epsilon_{a}} \ket{j}_{\epsilon_{b}}
 \overset{(4)}{\rightarrow} \\
&\frac{1}{\sqrt{2}} \sum_{i,j}  \left( \ket{0}_{c} K_{i} U\ket{\psi_{\rm in}}_{a} K_{j}U\ket{\phi_{0}}_{b} +\ket{1}_{c} K_{i} U \ket{\phi_{0}}_{b} K_{j}U\ket{\psi_{{\rm in}}}_{a} \right) \ket{i}_{\epsilon_{a}} \ket{j}_{\epsilon_{b}} 
\end{split}
\end{equation}
If one traces out the environments to analyze the final state of the  output registers, one obtains
\begin{equation}
\label{eq:finalGBenv}
\begin{split}
&\rho= \proj{0}_{c} \sum_{i,j} K_{i} (U \proj{\psi_{\rm in}}_{a} U^{\dagger} ) K_{i}^{\dagger} \otimes K_{j} (U \proj{\phi_{0}}_{b} U^{\dagger} ) K_{j}^{\dagger} + \\
&\proj{1}_{c} \sum_{i,j} K_{i} (U \proj{\psi_{\rm in}}_{a} U^{\dagger} ) K_{i}^{\dagger} \otimes K_{j} (U \proj{\phi_{0}}_{b} U^{\dagger} ) K_{j}^{\dagger}+ \\
&\ket{0}\bra{1}_{c} \sum_{i,j} K_{i} (U \proj{\psi_{\rm in}}_{a} U^{\dagger} ) K_{j}^{\dagger} \otimes K_{j} (U \proj{\phi_{0}}_{b} U^{\dagger} ) K_{i}^{\dagger} + \\
&\ket{1}\bra{0}_{c} \sum_{i,j} K_{j} (U \proj{\psi_{\rm in}}_{a} U^{\dagger} ) K_{i}^{\dagger} \otimes K_{i} (U \proj{\phi_{0}}_{b} U^{\dagger} ) K_{j}^{\dagger}.
\end{split}
\end{equation}
By analogy with the density matrix formalism case, see App.~\ref{sec:Appendix_density} from which it is direct to see how one recovers Eq.~\eqref{eq:densitymatrixprevious}, the effect of the protocol is to get the noise correlated in the off-diagonal terms of the final density matrix. Measuring the auxiliary system $b$ in a suitable basis effectively eliminates certain elements and, together with the measurement of the control register in the X basis, significantly enhances the fidelity of the computation. 

\section{Enhanced MB-QC. Analytic analysis based on the environmental formalism }
\label{sec:MB-QCappendix2}
A similar environmental-based analysis can be performed for the measurement-based (MB) implementation, introduced in Sec.~\ref{sec:enhancedMBQC}. Although this analysis is already explored in the main text, we detail and expand the derivations already presented there. 

We explain the process with two proof-of-concept examples that can be easily generalized. First, we consider a simple measurement-based teleportation process, and then we treat a more general computation in a 1D cluster state.

\subsection{Entanglement based teleportation}
Entanglement-based teleportation can be conceived as a particular basic instance of an MB-QC process, where one simply transports (or teleports) information by using Bell states as resources. We show how to enhance the fidelity of a teleported state by running the process in superposition using more than one noisy Bell copy.

\begin{figure}
    \centering
    \includegraphics[width=\textwidth]{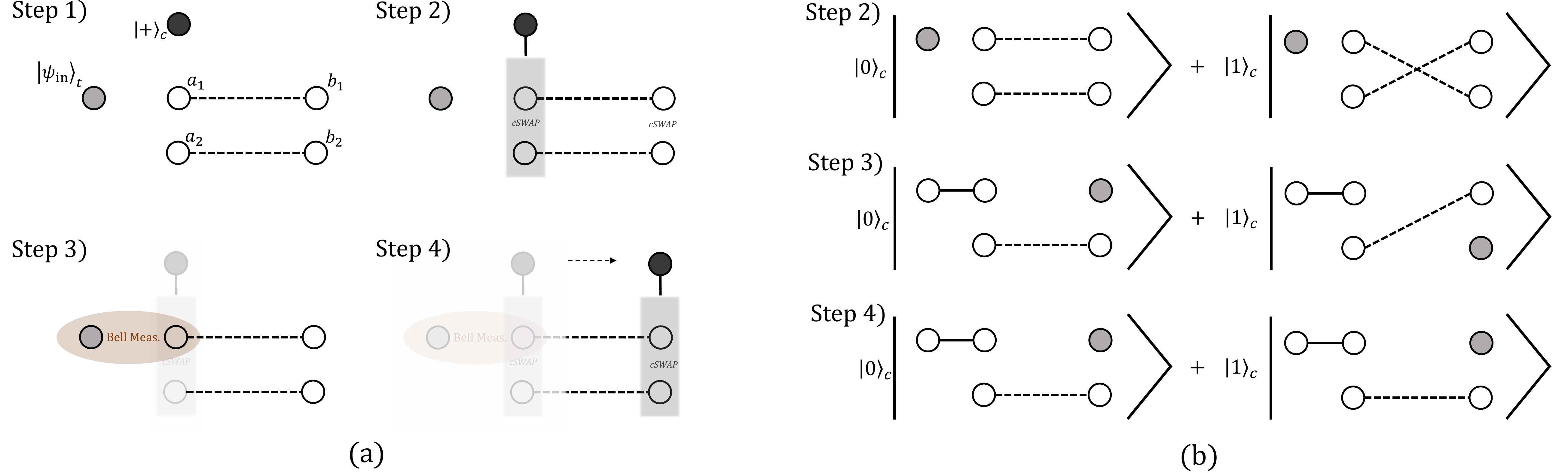}
    \caption{\label{fig:comp1} (a) Operational representation of steps for the entanglement-based teleportation protocol. Once these steps are completed, final measurements on the remaining noisy Bell state $a_{2}b_{2}$ and on the control register are performed. (b) Schematic representation of the joint state of the system after each step in (a).}
\end{figure}

Consider a 1D resource state consisting of two qubits (i.e. a Bell state), where the information of an additional qubit, in some arbitrary state $\left|\psi_{\rm in}\right\rangle=\alpha \left|0\right\rangle+ \beta \left|1\right\rangle$, is teleported to the next qubit by a Bell measurement (see Fig.~\ref{fig:comp1}). Noise in the resource state can be modeled in different ways. We assume that the noise comes from an imperfect preparation of the resource Bell state. The incoherent process reads
\begin{equation}
\left|\psi_{\rm in}\right\rangle_{t} \otimes \rho_{a_{1}b_{1}}  {\rightarrow} \sum_{q} K_{q} \left|\psi_{\rm in}\right\rangle_{b_{1}}\left\langle \psi_{\rm in}\right|  K^{\dagger}_{q} \otimes \ket{\Phi^{+}}_{t a_{1}}\bra{\Phi^{+}},
\end{equation}
up to unitary corrections, and where $ \rho_{a_{1}b_{1}}$ indicates a noisy Bell pair. A Bell measurement has been applied between qubits $t$ and $a_{1}$ and we assume the outcome $\ket{\Phi^{+}}$ is found for simplicity.

Consider now the case that two independent \textit{noisy} Bell states are available. One can achieve an equally-weighted superposition between the input state being teleported using one or the other Bell pair. This is done by using our protocol, i.e., by simply applying a controlled-SWAP operation from a control system initialized in the  $\left|+\right\rangle_{c}$ state, acting on the first qubits of each Bell pair (see Fig.~\ref{fig:comp1}), of the form
\begin{equation}
\label{eq:cswap}
G_{swap}=\left|0\right\rangle _{c}\left\langle 0\right|\otimes\mathbb{1}_{a_{1}}\otimes\mathbb{1}_{a_{2}}
 +\left|1\right\rangle _{c}\left\langle 1\right|\otimes U_{a_{1},a_{2}}^{\rm swap}.
\end{equation}
After a Bell measurement on both the $t$ and the $a_{1}$ qubits, and recombination (a second cSWAP operation) on the $b_{1}$ and $b_{2}$ qubits, the process is completed (see Fig.~\ref{fig:comp1}). Note that noise acting on individual qubits can always be described with Kraus operators acting on the joint system, i.e. $K_{i}=K^{(1)}_{i} \otimes K^{(2)}_{i} $, with $i \in \{0,...,m^n \}$, where $m$ is the number of Kraus operators and $n$ the number of cluster qubits, in this example $n=2$. As before, we make use of the Stinespring dilation theorem, which allows us to describe any quantum map as a unitary evolution acting on a certain pure state in a larger Hilbert space. The additional dimensions of the larger Hilbert space can be interpreted as the environment into which the information of the system is leaked out during the evolution. The whole protocol in detail reads (see Table \ref{table:MBstandard})
\begin{equation}
\begin{split}
&\left|+\right\rangle_{c} \left|\psi_{\rm in}\right\rangle_{t} \sum_{q} K_{q} \left|\Phi^{+}\right\rangle_{a_{1}b_{1}}  \left|q\right\rangle_{\epsilon_{1}} \sum_{j} K_{j}  \left|\Phi^{+}\right\rangle_{a_{2}b_{2}} \left|j\right\rangle_{\epsilon_{2}} \overset{(2)} {\rightarrow} \\
&\frac{1}{\sqrt{2}} \left|0\right\rangle_{c}  \left|\psi_{\rm in}\right\rangle_{t} \sum_{q} K_{q} \left|\Phi^{+}\right\rangle_{a_{1}b_{1}} \left|q\right\rangle_{\epsilon_{1}} \sum_{j} K_{j}  \left|\Phi^{+}\right\rangle_{a_{2}b_{2}} \left|j\right\rangle_{\epsilon_{2}} + 
\frac{1}{\sqrt{2}} \left|1\right\rangle_{c}  \left|\psi_{\rm in}\right\rangle_{t} \sum_{q} K_{q} \left|\Phi^{+}\right\rangle_{a_{2}b_{1}}  \left|q\right\rangle_{\epsilon_{1}} \sum_{j} K_{j}  \left|\Phi^{+}\right\rangle_{a_{1}b_{2}} \left|j\right\rangle_{\epsilon_{2}} \overset{(3)} {\rightarrow} \\
&\frac{1}{\sqrt{2}}  \left|\Phi^{+}\right\rangle_{t a_{1}} \left( \left|0\right\rangle_{c}  \sum_{q} K_{q}\left|\psi_{\rm in}\right\rangle_{b_{1}}   \left|q\right\rangle_{\epsilon_{1}} \sum_{j} K_{j}  \left|\Phi^{+}\right\rangle_{a_{2}b_{2}} \left|j\right\rangle_{\epsilon_{2}} + 
\left|1\right\rangle_{c}  \sum_{q} K_{q} \left|\Phi^{+}\right\rangle_{a_{2}b_{1}}  \left|q\right\rangle_{\epsilon_{1}} \sum_{j} K_{j}\left|\psi_{\rm in}\right\rangle_{b_{2}}  \left|j\right\rangle_{\epsilon_{2}}  \right) \overset{(4)} {\rightarrow} \\
&\frac{1}{\sqrt{2}}  \left|\Phi^{+}\right\rangle_{t a_{1}} \left( \left|0\right\rangle_{c}  \sum_{q} K_{q}\left|\psi_{\rm in}\right\rangle_{b_{1}}   \left|q\right\rangle_{\epsilon_{1}} \sum_{j} K_{j}  \left|\Phi^{+}\right\rangle_{a_{2}b_{2}} \left|j\right\rangle_{\epsilon_{2}} + 
\left|1\right\rangle_{c}  \sum_{q} K_{q} \left|\Phi^{+}\right\rangle_{a_{2}b_{2}}  \left|q\right\rangle_{\epsilon_{1}} \sum_{j} K_{j}\left|\psi_{\rm in}\right\rangle_{b_{1}}  \left|j\right\rangle_{\epsilon_{2}}  \right),
\end{split}
\end{equation}
where steps~\ref{prot:mb2}, \ref{prot:mb3} and \ref{prot:mb4} correspond to the first cSWAP, the Bell measurement, and the final cSWAP respectively. Observe that any byproduct can be corrected by a controlled unitary before the final cSWAP recombination. The illustrative effect of each step is depicted in Fig.~\ref{fig:comp1}. If we trace out the environments, we can analyze the reduced physical state of the qubits, i.e.,
\begin{multline}
\rho_{b_{1}}=  \left|0\right\rangle_{c}\left\langle 0\right| \sum_{q} K_{q}\left|\psi_{\rm in}\right\rangle_{b_{1}}\left\langle \psi_{\rm in}\right| K^{\dagger}_{q} \sum_{j} K_{j}\ket{\Phi^{+}}_{a_{2}b_{2}} \bra{\Phi^{+}} K^{\dagger}_{j} + \left|1\right\rangle_{c}\left\langle 1\right| \sum_{q} K_{q}\left|\psi_{\rm in}\right\rangle_{b_{1}}\left\langle \psi_{\rm in}\right| K^{\dagger}_{q}  \otimes \sum_{j} K_{j}\ket{\Phi^{+}}_{a_{2}b_{2}} \bra{\Phi^{+}} K^{\dagger}_{j} + \\
\left|0\right\rangle_{c}\left\langle 1\right|  \sum_{q j} K_{q}\left|\psi_{\rm in}\right\rangle_{b_{1}}\left\langle \psi_{\rm in}\right| K^{\dagger}_{j} \otimes K_{j} \ket{\Phi^{+}}_{a_{2}b_{2}} \bra{\Phi^{+}} K^{\dagger}_{q} + \left|1\right\rangle_{c}\left\langle 0\right| \sum_{q j}  K_{q}\left|\psi_{\rm in}\right\rangle_{b_{1}}\left\langle \psi_{\rm in}\right| K^{\dagger}_{j} \otimes K_{j} \ket{\Phi^{+}}_{a_{2}b_{2}} \bra{\Phi^{+}} K^{\dagger}_{q}.
\label{eq:MB-QCtraceout}
\end{multline}
Observe the equivalence of this result with the one obtained in App.~\ref{sec:Appendixstandard1} for the GB case, Eq.~\eqref{eq:finalGBenv}.

Recall that the Kraus decomposition of any single channel is not unique but invariant up to some unitary matrix. Unlike in the interferometric-based scenario (see Sec.~\ref{sec:IB_protocol} and App.~\ref{sec:Appendixvacuum1}), this invariance remains in place in the MB-QC (and in the GB-QC) setting. 

The protocol generates correlations of the environment states, initially associated with individual noisy Bell states, with each other, as well as with the remaining Bell pair $a_{2},b_{2}$ and with the state at $b_{1}$ [see Eq.~\eqref{eq:MB-QCtraceout} and Fig.~\ref{fig:compstandard2}]. Subsequent measurement of the Bell state $a_{2},b_{2}$, performed in an appropriate basis, allows one to interfere in the computational output, ending up with a state with enhanced fidelity. As an example, assume a measurement in the Bell basis is performed, where the output $\ket{\Phi^{+}}$ is obtained. The final state of qubit $b_{1}$ is then
\begin{equation}
\rho_{b_{1}}= \left|0\right\rangle_{c}\left\langle 0\right| \sum_{q} K_{q}\left|\psi_{\rm in}\right\rangle_{b_{1}}\left\langle \psi_{\rm in}\right| K^{\dagger}_{q} + \left|1\right\rangle_{c}\left\langle 1\right| \sum_{q} K_{q}\left|\psi_{\rm in}\right\rangle_{b_{1}}\left\langle \psi_{\rm in}\right| K^{\dagger}_{q} + 
\left|0\right\rangle_{c}\left\langle 1\right|  K_{0}\left|\psi_{\rm in}\right\rangle_{b_{1}}\left\langle \psi_{\rm in}\right| K^{\dagger}_{0} +\left|1\right\rangle_{c}\left\langle 0\right| 
K_{0}\left|\psi_{\rm in}\right\rangle_{b_{1}}\left\langle \psi_{\rm in}\right| K^{\dagger}_{0},
\end{equation}
obtaining the maximum protocol performance corresponding to $(\omega_1, \omega_2) = (1,1)$ in Eq.~\eqref{eq:omegas}.

Measuring the remaining resource qubits in a local way is also possible. However, since only two of the four Bell states can be discriminated with local measurements, the enhancement obtained can be reduced (particularly with rank-3 noise), although the output fidelity is still significantly increased with respect to the incoherent case. 

This proof-of-concept example shows how the performance of entanglement-based teleportation, which can also be used as a tool for encoding  quantum information into a cluster state for further processing, can be enhanced (in terms of fidelity) by our SQEM protocols. 

\subsection{Enhanced MB-QC. Arbitrary 1D computation } \label{sec:AppendixMB-QC3}
We consider now an arbitrary unitary operation. For simplicity, we take a 1D cluster state with some input state already encoded in the first qubit. With the assistance of another cluster state (or another part of the same one), a superposition of two \textit{identical} operations acting on the input can be achieved, leading to enhanced fidelity of the output state. It is enough to restrict to a $5$-qubit 1D cluster state for performing an arbitrary rotation \cite{Briegel2001}. Here we further focus on the simpler but completely analogous case of a $3$-qubit cluster state, on which the unitary $U_{\mu}={\rm exp}(\frac{-i}{2} \mu \sigma_{x})$ can be realized. We adopt the shorthands $\left|G_{0}\right\rangle_{1,2}= \frac{1}{\sqrt{2}} \left( \ket{0+} +  \ket{1-} \right)$, $\left|G_{0}\right\rangle_{1,2,3}= \frac{1}{\sqrt{2}} \left( \ket{+0+} +  \ket{-1-} \right)$ and $\left|G_{\psi_{\rm in}}\right\rangle_{1,2,3}= \frac{1}{\sqrt{2}} \left( \left(\alpha \ket{+}+\beta \ket{-} \right)  \ket{0+} +  \left(\alpha \ket{-}+\beta \ket{+} \right) \ket{1-} \right)$, the latter two denoting the initial cluster states with the $\ket{+}$ or the $\ket{\psi_{\rm in}}=\alpha \ket{+}+\beta \ket{-}$ state encoded in the first qubit. Observe that we just take as auxiliary another cluster state, i.e. $\ket{\phi_0}_{b}=\ket{+}$.

We consider again a noise model where independent but identical noise affects each qubit after the resource states are generated (i.e. after the entangling gates). The noise can be described as a function of some global Kraus operators of the form $K_{i}=K^{(1)}_{i} \otimes K^{(2)}_{i} \otimes K^{(3)}_{i}$, with $i \in \{0,...,m^n \}$ where $m$ is the number of Kraus operators and $n$ the number of cluster qubits, in this example $n=3$. 
The operation then consists of a single-qubit rotation via a $X$ measurement on the first qubit (to transport the information) and a measurement in a rotated basis in the second. The process reads
\footnotesize
\begin{align}
&\left|+\right\rangle_{c} \ket{\psi_{\rm in}}_{a}\ket{+}_{a_{2}} \sum_{q} K_{q} \left|G_{0}\right\rangle_{b_{1}c{1}}   \left|q\right\rangle_{\epsilon_{1}} \sum_{j} K_{j}  \left|G_{0}\right\rangle_{b_{2}c{2}}  \left|j\right\rangle_{\epsilon_{2}} \overset{(a)} {\rightarrow}  \nonumber \\ 
& \left( \frac{1}{\sqrt{2}}\ket{0}_{c} \ket{\psi_{\rm in}}_{a}\ket{+}_{a_{2}}+ \frac{1}{\sqrt{2}}\ket{1}_{c} \ket{+}_{a_{1}}\ket{\psi_{\rm in}}_{a_{2}} \right) \sum_{q} K_{q} \left|G_{0}\right\rangle_{b_{1}c{1}}   \left|q\right\rangle_{\epsilon_{1}} \sum_{j} K_{j}  \left|G_{0}\right\rangle_{b_{2}c{2}}  \left|j\right\rangle_{\epsilon_{2}} \overset{(b)} {\rightarrow}  \nonumber \\ 
&\frac{1}{\sqrt{2}} \left|0\right\rangle_{c}  \sum_{q} K_{q} \left|G_{\psi_{\rm in}}\right\rangle_{a_{1}b_{1}c{1}}   \left|q\right\rangle_{\epsilon_{1}} \sum_{j} K_{j}  \left|G_{0}\right\rangle_{a_{2}b_{2}c{2}}  \left|j\right\rangle_{\epsilon_{2}} + 
\frac{1}{\sqrt{2}} \left|1\right\rangle_{c}  \sum_{q} K_{q} \left|G_{0}\right\rangle_{a_{1}b_{1}c{1}}   \left|q\right\rangle_{\epsilon_{1}} \sum_{j} K_{j}  \left|G_{\psi_{\rm in}}\right\rangle_{a_{2}b_{2}c{2}}  \left|j\right\rangle_{\epsilon_{2}} \overset{(c)} {\rightarrow} \nonumber \\ 
&\frac{1}{\sqrt{2}}  a_{s}a_{r}  \ket{s_{x} r_{x}}_{a_{1}a_{2}} \left( \left|0\right\rangle_{c}  \bra{s_{x}} \sum_{q, j}^{m^{3}} K_{q} \left|G_{\psi_{\rm in}}\right\rangle_{a_{1}b_{1}c{1}}   \left|q\right\rangle_{\epsilon_{1}} \bra{r_{x}}  K_{j}  \left|G_{0}\right\rangle_{a_{2}b_{2}c{2}}  \left|j\right\rangle_{\epsilon_{2}} + 
\left|1\right\rangle_{c}  \bra{s_{x}} \sum_{q,j}^{m^{3}} K_{q} \left|G_{0}\right\rangle_{a_{1}b_{1}c{1}}   \left|q\right\rangle_{\epsilon_{1}} \bra{r_{x}}  K_{j}  \left|G_{\psi_{\rm in}}\right\rangle_{a_{2}b_{2}c{2}}  \left|j\right\rangle_{\epsilon_{2}}  \right) \equiv
\nonumber \\ 
&\frac{1}{\sqrt{2}} a_{s}a_{r}  \ket{s_{x} r_{x}}_{a_{1}a_{2}} \left( \left|0\right\rangle_{c}  \sum_{q}^{m^{2}} \tilde{K}_{q} \left|G_{\psi_{\rm in}}\right\rangle_{b_{1}c{1}}   \left|q\right\rangle_{\epsilon_{1}}  \sum_{j}^{m^{2}} \tilde{K}_{j}  \left|G_{0}\right\rangle_{b_{2}c{2}}  \left|j\right\rangle_{\epsilon_{2}} + 
\left|1\right\rangle_{c}  \sum_{q}^{m^{2}} \tilde{K}_{q} \left|G_{0}\right\rangle_{b_{1}c{1}}   \left|q\right\rangle_{\epsilon_{1}} \sum_{j}^{m^{2}} \tilde{K}_{j} \left|G_{\psi_{\rm in}}\right\rangle_{b_{2}c{2}}   \left|j\right\rangle_{\epsilon_{2}}  \right) \overset{(d)}{\rightarrow}
\nonumber \\ 
&A \ket{s_{x} r_{x} t_{\mu} u_{\mu}}_{a_{1} a_{2} b_{1} b_{2} }  \left( \left|0\right\rangle_{c} U_{\Sigma_{1}} \sum_{q}^{m} K'_{q}  U_{\mu}  \left| \psi_{\rm in} \right\rangle_{c{1}}   \left|q\right\rangle_{\epsilon_{1}} U_{\Sigma_{2}}  \sum_{j}^{m} K'_{j}  U_{\mu} \left| + \right\rangle_{c{2}}  \left|j\right\rangle_{\epsilon_{2}} +
\left|1\right\rangle_{c} U_{\Sigma_{1}} \sum_{q}^{m} K'_{q}  U_{\mu} \left| + \right\rangle_{c{1}}   \left|q\right\rangle_{\epsilon_{1}} U_{\Sigma_{2}} \sum_{j}^{m} K'_{j}  U_{\mu} \left| \psi_{\rm in} \right\rangle_{c{2}}  \left|j\right\rangle_{\epsilon_{2}}  \right) \overset{(e)}{\rightarrow}
\nonumber \\ 
&A \ket{s_{x} r_{x} t_{\mu} u_{\mu}}_{a_{1} a_{2} b_{1} b_{2} }  \left( \left|0\right\rangle_{c} U_{\Sigma_{1}} \sum_{q}^{m} K'_{q}  U_{\mu}  \left| \psi_{\rm in} \right\rangle_{c{1}}   \left|q\right\rangle_{\epsilon_{1}} U_{\Sigma_{2}}  \sum_{j}^{m} K'_{j}  U_{\mu} \left| + \right\rangle_{c{2}}  \left|j\right\rangle_{\epsilon_{2}} +
\left|1\right\rangle_{c} U_{\Sigma_{1}} \sum_{q}^{m} K'_{q}  U_{\mu} \left| + \right\rangle_{c{2}}   \left|q\right\rangle_{\epsilon_{1}} U_{\Sigma_{2}} \sum_{j}^{m} K'_{j}  U_{\mu} \left| \psi_{\rm in} \right\rangle_{c{1}}  \left|j\right\rangle_{\epsilon_{2}}  \right) .
\end{align}
\normalsize
Here $\ket{i_{n}}_{j}$ represents the outcome $i$ of the measurement of qubit $j$ in the basis $n$, and $U_{\Sigma}$ are the correction operations that depend on the measurement outcomes. Step (a) consists of a cSWAP operation between the first two main and auxiliary input qubits $a_{1}$ and $a_{2}$, followed by an entangling operation that encodes the qubits to the clusters (step (b)). Note that a cSWAP between the first two qubits of each cluster followed by a Bell measurement with the input leads to the same state. Alternatively, one can assume that the input states are already encoded in the first qubit of each cluster, and similar results are found. Step (c) shows the effect of the measurement on the $X$ basis of the first qubit of each cluster.  This measurement leads to a different effective Kraus description of the noise affecting the computational level, i.e. the remaining cluster. Step (d) represents the measurement of the second qubit of each cluster in a rotated $X$ basis that leads to the application of the unitary operation $U_{\mu}$. The constant $A$ encompasses all the normalization factors, and the new sets of Kraus operators $\{\tilde{K_{i}}\}$ and $\{K'_{i}\}$ include the noise coming from the measurements and (in the case of $\{K'_{i}\}$) commutation factors with $U_{\mu}$. Finally, step (e) shows the effect of the recombination after the final cSWAP between the remaining qubits of each cluster. Note how the SWAP gate induces again an exchange between the coupling of the qubits with the environments. Observe also how byproducts can be simply corrected by a controlled operation at this point. 

Finally, a measurement on the output auxiliary qubit is performed in a suitable basis, such that certain terms can be effectively selected on the off-diagonal elements of the reduced density operator, leading to an enhanced fidelity. For instance, with rank-2 noise, one can always find a measurement basis, such that the output state after tracing out the environment reads
\begin{equation}
\label{eq:MQBCAppendix3f}
\begin{split}
\rho= &\left|0\right\rangle_{c}\left\langle 0\right| \sum_{q} K'_{q}  U_{\mu}\left|\psi_{\rm in}\right\rangle_{b_{1}}\left\langle \psi_{\rm in}\right| U^{\dagger}_{\mu} K_{q}^{\prime \dagger} + \left|1\right\rangle_{c}\left\langle 1\right| \sum_{q} K'_{q}  U_{\mu} \left|\psi_{\rm in}\right\rangle_{b_{1}}\left\langle \psi_{\rm in}\right| U^{\dagger}_{\mu} K_{q}^{\prime \dagger} + \\
&\left|0\right\rangle_{c}\left\langle 1\right|   K'_{0(1)} U_{\mu}\left|\psi_{\rm in}\right\rangle_{b_{1}}\left\langle \psi_{\rm in}\right| U_{\mu} K_{0(1)}^{\prime \dagger} + \left|1\right\rangle_{c}\left\langle 0\right|   K'_{0(1)} U_{\mu}\left|\psi_{\rm in}\right\rangle_{b_{1}}\left\langle \psi_{\rm in}\right| U^{\dagger}_{\mu}  K_{0(1)}^{\prime \dagger},
\end{split}
\end{equation}
depending on the output (0 or 1) of the auxiliary qubit measurement. This result is similar to the ones in App.~\ref{sec:Appendixstandard1}; 
Generalization to more inputs or more general computations can be done following the same steps. In particular, in order to achieve arbitrary 1D computations, a cluster state of 5 qubits and two additional rotations are required, where analogous results and derivations apply.

\subsubsection{Behaviour for arbitrary computations} \label{sec:arbitrarycomp}
We briefly analyze here how the previous formalism can be extended to more general computations with multi-qubit input states. Comparable advantage to the single-input examples, as seen in the numerical analysis of Sec.~\ref{sec:mbqcperformance}, can be found with a constant overhead of resources. In case more resources are available, further improvement can be achieved.

Consider $m$ input qubits ${ a_{1},...,a_{m} }$ in some state $\ket{\psi_{\rm in}}_{a}$. Assume also two independent $m$-sized parts of the entangled resource state are available. We can generate a superposition between the computation being carried in the first resource, or in the second, where the input state is operated on in one or the other. A single control register in the $\left|+\right\rangle_{c}$ state is therefore required for generating such a superposition. With $m$ auxiliary qubits ${ b_{1},...,b_{m} }$ prepared in $\left|+\right\rangle$, and controlled-SWAP operations applied from the control register to each pair ${a_{i},b_{i}}$, one can obtain
\begin{align}
&\left|+\right\rangle_{c} \ket{\psi_{\rm in}}_{a} \left(\bigotimes_{j=1}^{m}\ket{+}_{b_{j}} \right) 
 \sum_{q} K_{q} \left|G_{0}\right\rangle_{a} \ket{q}_{\epsilon_{a}}   \sum_{t} K_{t} \left|G_{0}\right\rangle_{b} \ket{t}_{\epsilon_{b}}  \overset{(a)} {\rightarrow}  \nonumber \\ 
& \left( \frac{1}{\sqrt{2}}\ket{0}_{c} \ket{\psi_{\rm in}}_{a} \left(\bigotimes_{j=1}^{m}\ket{+}_{b_{j}} \right) + \frac{1}{\sqrt{2}}\ket{1}_{c}  \left(\bigotimes_{j=1}^{m}\ket{+}_{a_{j}} \right)  \ket{\psi_{\rm in}}_{b} \right) \sum_{q} K_{q} \left|G_{0}\right\rangle_{a} \ket{q}_{\epsilon_{a}}   \sum_{t} K_{t} \left|G_{0}\right\rangle_{b} \ket{t}_{\epsilon_{b}},
\end{align}
where $\left|G_{0}\right\rangle_{i}$ indicates a 2D cluster state of some depth. Next, the qubits are entangled to the resource states, i.e.,
\begin{align}
\frac{1}{\sqrt{2}}\ket{0}_{c}  \sum_{q} K_{q} \left|G_{\psi_{\rm in}}\right\rangle_{a} \ket{q}_{\epsilon_{a}}   \sum_{t} K_{t} \left|G_{0}\right\rangle_{b} \ket{t}_{\epsilon_{b}} +\frac{1}{\sqrt{2}}\ket{1}_{c}  \sum_{q} K_{q} \left|G_{0}\right\rangle_{a} \ket{q}_{\epsilon_{a}}   \sum_{t} K_{t} \left|G_{\psi_{\rm in}}\right\rangle_{b} \ket{t}_{\epsilon_{b}}.
\end{align}
An arbitrary computation is performed in each cluster state, leading to a final state of the form 
\small
\begin{align}
\frac{1}{\sqrt{2}}\ket{0}_{c}  \sum_{q} K^{\prime}_{q} U \left|{\psi_{\rm in}}\right\rangle_{a} \ket{q}_{\epsilon_{a}}   \sum_{t} K^{\prime}_{t} U  \left(\bigotimes_{j=1}^{m}\ket{+}_{b_{j}} \right)   \ket{t}_{\epsilon_{b}} +\frac{1}{\sqrt{2}}\ket{1}_{c}  \sum_{q} K^{\prime}_{q} U \left(\bigotimes_{j=1}^{m}\ket{+}_{a_{j}} \right)  \ket{q}_{\epsilon_{a}}   \sum_{t} K^{\prime}_{t} U \left|{\psi_{{\rm in}}}\right\rangle_{b} \ket{t}_{\epsilon_{b}},
\end{align}
\normalsize
up to byproducts, where $U$ is an arbitrary computation acting on the input qubits. The final cSWAP is now applied for recombining, i.e.,
\small
\begin{align}
\frac{1}{\sqrt{2}}\ket{0}_{c}  \sum_{q} K^{\prime}_{q} U \left|{\psi_{\rm in}}\right\rangle_{a} \ket{q}_{\epsilon_{a}}   \sum_{t} K^{\prime}_{t} U  \left(\bigotimes_{j=1}^{m}\ket{+}_{b_{j}} \right)   \ket{t}_{\epsilon_{b}} +\frac{1}{\sqrt{2}}\ket{1}_{c}  \sum_{q} K^{\prime}_{q} U \left(\bigotimes_{j=1}^{m}\ket{+}_{b_{j}} \right)  \ket{q}_{\epsilon_{a}}   \sum_{t} K^{\prime}_{t} U \left|{\psi_{\rm in}}\right\rangle_{a} \ket{t}_{\epsilon_{b}}.
\end{align}
\normalsize
Finally, a measurement of the remaining auxiliary qubits ($b$) is applied. Once the environmental states are traced out and the control register is measured in the $X$ basis, one recovers similar results to those in the examples previously analyzed.
Note that the measurement is not unique, and different bases and outcomes can lead to different results, all of them, in general, with enhanced fidelity, as already analyzed in the main text.

\section{MB-QC implementation of the controlled-SWAP operation}\label{sec:cswap_patterm}
\begin{figure}
    \centering
    \includegraphics[width=\textwidth]{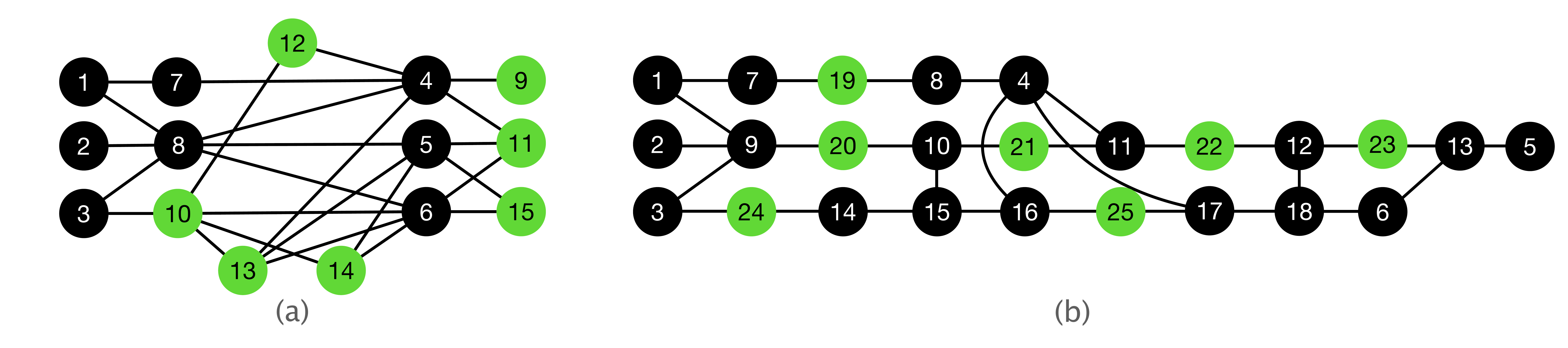}
    \caption{\label{fig:cswappattern} Graph states and measurement patterns used for the numerical simulations of the cSWAP gate in the measurement-based setting. Black qubits indicate measurements in the Pauli $X$ basis, while green qubits indicate adaptive measurements (see text).}
\end{figure}
We detail here our cSWAP implementation for the numerical simulations in the MB-QC case. We show the graph states in Fig.~\ref{fig:cswappattern}, provide the corresponding measurement patterns, and specify the bases of the adaptive measurements.

Our cSWAP pattern is based on the GB Clifford+T implementation in Ref.~\cite{Amy2013}, which requires 7 T gates. The most direct way to obtain such a graph state is concatenating the MB patterns in Ref.~\cite{Briegel2002}, and classically simulating as many nonadaptive measurements as possible following the stabilizer approach in Refs.~\cite{Aaronson2004,Nest2004}. Each of the 7 T gates contributes one adaptive measurement; in addition, the nonadaptive measurements on the input qubits and on the qubits directly entangled with the input qubits cannot be simulated classically. In the resulting graph state shown in Fig.~\ref{fig:cswappattern} (a), the input qubits (to be measured in the $X$ basis) are $1,2,3$, the output qubits are $4,5,6$, while qubits $7,8$ are measured in the $X$ basis and qubits $9$---$15$ in certain adaptive bases given below. The byproduct operator in this case reads
\begin{equation}
Z_{1}^{s_{1}+s_{9}+s_{11}+s_{12}+s_{13}}X_{1}^{s_{7}}Z_{2}^{s_{2}+s_{11}+s_{13}+s_{14}+s_{15}}X_{2}^{s_{7}+s_{8}+s_{10}+s_{12}+s_{13}+s_{14}}Z_{3}^{s_{3}+s_{11}+s_{13}+s_{14}+s_{15}}X_{3}^{s_{10}+s_{12}+s_{13}+s_{14}},
\end{equation}
and the adaptive rotation angles are
\begin{equation}
\begin{split}
\alpha_{9} & =-\left(-1\right)^{s_{7}}\frac{\pi}{4}; \quad \alpha_{10}  =-\left(-1\right)^{s_{2}+s_{3}}\frac{\pi}{4}; \quad \alpha_{11}  =\left(-1\right)^{s_{8}}\frac{\pi}{4}; \quad \alpha_{12}  =\left(-1\right)^{s_{2}+s_{3}+s_{7}}\frac{\pi}{4}; \\ 
\alpha_{13} & =-\left(-1\right)^{s_{2}+s_{3}+s_{8}}\frac{\pi}{4}; \quad \alpha_{14} =\left(-1\right)^{s_{2}+s_{3}+s_{7}+s_{8}}\frac{\pi}{4}; \quad \alpha_{15}  =-\left(-1\right)^{s_{7}+s_{8}}\frac{\pi}{4}.
\end{split}
\end{equation}

While the graph state Fig.~\ref{fig:cswappattern} (a) employs a relatively small number of qubits, the larger connectivity results in more qubits in intermediate states, causing difficulty in numerical simulations. An alternative approach we use in our simulations is shown in Fig.~\ref{fig:cswappattern} (b). In this case, at the cost of eliminating fewer nonadaptively measured qubits, we reduce the required qubit number in intermediate states. We again label the input qubits as $1,2,3$, and the output qubits as $4,5,6$. On the other hand, qubits $7$--$18$ are measured in the $X$ basis, and qubits $19$--$25$ in adaptive bases. The byproduct operator reads
\begin{align}
&  Z_{1}^{s_{1}+s_{19}+s_{20}+s_{21}+s_{25}}X_{1}^{s_{7}+s_{8}}Z_{2}^{s_{2}+s_{20}+s_{21}+s_{22}+s_{23}}
X_{2}^{s_{7}+s_{8}+s_{9}+s_{10}+s_{11}+s_{12}+s_{13}+s_{15}+s_{16}+s_{17}+s_{18}+s_{21}+s_{22}+s_{24}+s_{25}} \nonumber\\
&\times  Z_{3}^{s_{3}+s_{14}+s_{16}+s_{17}+s_{20}+s_{21}+s_{22}+s_{23}}X_{3}^{s_{15}+s_{18}+s_{21}+s_{22}+s_{24}+s_{25}},
\end{align}
while the adaptive rotation angles are:
\begin{equation}
\begin{split}
\alpha_{19} & =-\left(-1\right)^{s_{7}}\frac{\pi}{4}; \quad \alpha_{20}  =\left(-1\right)^{s_{9}}\frac{\pi}{4}; \quad
 \alpha_{21}  =-\left(-1\right)^{s_{2}+s_{3}+s_{9}+s_{10}+s_{14}}\frac{\pi}{4}; \\ 
\quad \alpha_{22}& =\left(-1\right)^{s_{2}+s_{3}+s_{7}+s_{8}+s_{9}+s_{10}+s_{11}+s_{14}}\frac{\pi}{4}; \quad
 \alpha_{23}  =-\left(-1\right)^{s_{7}+s_{8}+s_{9}+s_{10}+s_{11}+s_{12}+s_{16}+s_{17}}\frac{\pi}{4}; \\ 
\alpha_{24} & =-\left(-1\right)^{s_{2}+s_{3}}\frac{\pi}{4}; \quad \alpha_{25} =\left(-1\right)^{s_{2}+s_{3}+s_{7}+s_{8}+s_{14}+s_{16}}\frac{\pi}{4}.
\end{split}
\end{equation}
\section{Enhanced interferometric-like circuit gate-based computation. Analytic analysis based on the environmental formalism}
\label{sec:Appendixvacuum1}
We provide in this appendix an extended derivation of the interferometric-like circuit approach introduced in Sec~\ref{sec:vacuum}, making use again of the environmental formalism. The improvement arising from either the standard gate-based model or the measurement-based model assisted by ancillas does not depend on the underlying physics behind the noise. The fundamental mechanisms in the interferometric-based scenario involve very different physical processes, although similar qualitative enhancement can be achieved. The strategies we analyze here are closely related to works that analyze the superposition of trajectories in quantum communication scenarios \cite{Chiribella2019,Kristjnsson_2020,Abbott2020,Rubino2021}, where the invariance of the operator sum representation is broken. We adapt these techniques to a computational scenario, trying to understand the underlying physics behind them.  Importantly, in our approach, we do not need to deal with one of the main drawbacks of the aforementioned communication strategies, namely the assumption of  noiseless control registers. Further details will be investigated in Ref.~\cite{papercomm}.

Consider an input state that undergoes a superposition of two identical noisy gates $U$. An additional system, initialized in $|+\rangle_{c}$, acts as a control to decide which unitary (although identical) is applied, therefore generating the superposition. Note that the control can be directly encoded in some degree of freedom of the input qubit, such that this control is only needed at the beginning and the end of the process, independently of the size of the computation. In this case, an analysis based on the density matrix formalism does not provide complete information on the process because of the non-trivial role of the vacuum, and therefore we make use of the purified description of the states. A complete description can be however recovered by including global phases on the Kraus operators as stressed in Sec.~\ref{sec:vacuum}. Observe the differences with respect to the GB and MB standard models (Appendices~\ref{sec:Appendixstandard1}, \ref{sec:MB-QCappendix2}), where the role of the environments is irrelevant. The initial state then reads
\begin{align}
     \left|+\right\rangle _{c}\otimes \ket{\psi_{\rm in}} \otimes\left|\varepsilon_{0}\right\rangle _{\epsilon_{0}} \left|\varepsilon_{1}\right\rangle _{\epsilon_{1}},
\end{align}
with some initial environmental states $\ket{\varepsilon_{0}}_{\epsilon_{0}}$ and $\ket{\varepsilon_{1}}_{\epsilon_{1}}$, associated with environmental systems of each branch, where the information during the noise processes leaks out.

Superposition is generated in an interferometric way, where the system follows one or the other branch depending on the state of the control. An identical unitary computation is applied in both branches, with some noise associated described by Kraus operators $K_i$. In the noiseless case, this operation deterministically leads to the pure state $\left|+\right\rangle _{c}\otimes U\left|\psi_{\rm in}\right\rangle$. However, if the operation $U$ is not perfect, the effect of the noise can be analyzed by attending to the Stinespring dilation description of the process (see App.~\ref{sec:appendixMB-QC11}), i.e.
\begin{equation}
\left|\psi\right\rangle =\frac{1}{\sqrt{2}} \left|0\right\rangle _{c}\otimes \sum_{s}K_{s} U \left|\psi_{\rm in}\right\rangle \otimes\left|s\right\rangle _{\epsilon_{0}} \left|\varepsilon_{1}\right\rangle _{\epsilon_{1}}  + \frac{1}{\sqrt{2}} \left|1\right\rangle _{c}\otimes \sum_{s}K_{s} U \left|\psi_{\rm in}\right\rangle \otimes\left|\varepsilon_{0}\right\rangle _{\epsilon_{0}} \left|s\right\rangle _{\epsilon_{1}}.
\label{eq:Stinespring4}
\end{equation}
The remaining state of the system can be obtained by tracing out the environments, i.e.
\begin{equation}
\label{eq:appfinalinterf}
\begin{split}
\rho =&\frac{1}{2} \left|0\right\rangle _{c} \langle {0}| \sum_{i} K_{i} \rho_{f} K_{i}^{\dagger} + \frac{1}{2} \left|1\right\rangle _{c} \langle {1}| \sum_{j} K_{j} \rho_{f} K_{j}^{\dagger} + \\
&+\frac{1}{2} \left|0\right\rangle _{c} \langle{1}| \left( \sum_{i} \langle \varepsilon_{0}|i \rangle K_{i} \right) \rho_{f} \left(\sum_{j} \langle {j}| \varepsilon_{1} \rangle K_{j}^{\dagger}\right)  
+\frac{1}{2} \left|{1}\right\rangle _{c} \langle0| \left(\sum_{j} \langle \varepsilon_{1}|j \rangle K_{j}\right) \rho_{f} \left(\sum_{i} \langle  {i} | \varepsilon_{0} \rangle K_{i}^{\dagger}\right),
\end{split}
\end{equation}
with $\rho_{f}=U \rho_{\rm in} U^{\dagger}$. One can see that measuring the control register in the $X$ basis leads, in general, to some state different than the one obtained in the incoherent case Eq.~\eqref{eq:rhocomp}. The fidelity of this output state is generally enhanced, both in a probabilistic and (on average) in a deterministic way. The enhancement depends on the particular initial states of the environments that define the elements $\sum_{i} \langle \varepsilon_{0}|i \rangle K_{i}$ in the off-diagonal terms of Eq.~\eqref{eq:appfinalinterf}.

Equivalently,  Eq.~\eqref{eq:appfinalinterf} can be also derived by considering relative phases in the Kraus operators, i.e. ${{K_{j}} \rightarrow  e^{i \phi_j} {K}_{j}}$, which in the incoherent case are irrelevant. By including the vacuum in the description of the process, one can easily see how the phases of the Kraus operators analogously reproduce the $\sum_{i} \langle \varepsilon_{0}|i \rangle K_{i}$ terms in Eq.~\eqref{eq:appfinalinterf}. These can be interpreted as relative phases between the system and the vacuum, such that the Kraus operators can be described as
\begin{equation}
\tilde{K}_{j}=\begin{pmatrix} K_j & \, \\
\, &  e^{i \phi_j} \end{pmatrix},
\end{equation}
where $\tilde{K}_{j}$ are the Kraus operators associated with the system+vacuum, i.e.  $\mathcal{H}_{s} \otimes \mathcal{H}_{v}$.

Also in this case, the improvement in the fidelity can be further enhanced by increasing the number of branches in the superposition. In particular,  with rank-2 noise and one single gate, the improvement in the infidelity scales linearly 
with the number of superposition branches (see also Sec.~\ref{sec:resultsinterfero}). 

An extended analysis of the fundamentals behind these processes can be found in Ref.~\cite{papercomm}.

\textit{Multi-qubit operations.---} Generalization to multi-qubit operations is direct. Consider an arbitrary $m$-qubit quantum gate $U_{m}$, whose imperfect implementation is modeled by certain uncorrelated noise acting on each qubit after the ideal application of the gate, given by the Kraus operators $\{ K_{s_{i}}^{(i)} \}$, with $i={1,...,m}$ the corresponding qubit. We associate an environment with each qubit, initially in some state $ \left|\varepsilon_{0}\right\rangle =\otimes_{i} \left|\varepsilon_{i}\right\rangle _{\epsilon_{i}}$. The action of $k$ identical gates $U_{m}$ applied in superposition is 
\begin{equation}
\left|\psi\right\rangle =\frac{1}{\sqrt{k}} \sum^{k-1}_{i=0} \left|i\right\rangle _{c} \sum_{j} K^{(j)} U_{m} \left|\psi_{\rm in}\right\rangle\left|j\right\rangle^{i}   \otimes_{r\neq i}\left|\varepsilon_{0}\right\rangle ^{r},
\label{eq:Stinespringcomp2}
\end{equation}
where $\left|\psi_{\rm in}\right\rangle$ is the initial state of $m$ qubits and $K^{(j)} = \otimes_i K_{s_{i}}^{(i)}$ is a global Kraus operator comprising the composition of the individual ones. As before, by tracing out the environments and measuring the control register in the generalized $X$ basis, an outcome with enhanced fidelity is found.

\textit{Concatenation of gates.---}  Consider standard circuit computations consisting of sequential applications of quantum operations, each one with certain noise associated. We can write the dynamics of this process as
\begin{equation}
\sum_{i_{1} \cdots i_{m}} K^{(m)}_{i_{m}} U_{m}  \cdots K^{(0)}_{i_{1}} U_{1} \left|\psi_{\rm in}\right\rangle \left|i_{1}\right\rangle _{\epsilon_{1}} \otimes \cdots \otimes \left|i_{m}\right\rangle _{\epsilon_{m}}.
\label{eq:Stinespring2}
\end{equation}
If we trace out the environments we recover the expected action on the reduced state corresponding to the output qubit, i.e. 
\begin{equation}
\rho=\bigcirc_{s=1}^{m}\left[{\hat{\xi}_{s}}\circ\hat{U}_{s}\right]\left(\rho_{\rm in}\right),
\label{eq:Stinespring3}
\end{equation}
where $\hat{A}(\sigma)=A\sigma A^{\dagger}$ and $\hat{\xi}_{s}$ defines the noisy channel associated with the gate $s$ with Kraus operators $K_{i_s}$.

Consider now the case that two identical sequences of a concatenation of several single-qubit gates are applied in superposition. Eq.~\eqref{eq:Stinespring4} generalizes to
\begin{equation}
\begin{split}
\frac{1}{\sqrt{2}} \left|0\right\rangle _{c} \sum_{i_{1} \cdots i_{m}} K^{(m)}_{i_{m}} U_{m}  \cdots K^{(0)}_{i_{1}} U_{1} \left|\psi_{\rm in}\right\rangle \left|i_{1}\right\rangle _{\epsilon^{(0)}_{1}} \otimes \cdots \otimes \left|i_{m}\right\rangle _{\epsilon^{(0)}_{m}} \otimes \left|\varepsilon^{(1)}_{1}  \right\rangle _{\epsilon^{(1)}_{1}} \otimes \cdots \otimes \left|\varepsilon^{(1)}_{m}\right\rangle _{\epsilon^{(1)}_{m}} \nonumber\\
+\frac{1}{\sqrt{2}} \left|1\right\rangle _{c} \sum_{j_{1} \cdots j_{m}} K^{(m)}_{j_{m}} U_{m}  \cdots K^{(0)}_{j_{1}} U_{1} \left|\psi_{\rm in}\right\rangle  \left|j_{1}\right\rangle _{\epsilon^{(1)}_{2}} \otimes \cdots \otimes \left|j_{m}\right\rangle _{\epsilon^{(1)}_{m}} \otimes \left|\varepsilon_{1}^{(0)}\right\rangle _{\epsilon^{(0)}_{1}} \otimes \cdots \otimes \left|\varepsilon_{m}^{(0)}\right\rangle_{\epsilon^{(0)}_{m}}.
\label{eq:Stinespring6}
\end{split}
\end{equation}
The output state then reads
\begin{equation}
\begin{split}
\rho &=\frac{1}{2} \left|0\right\rangle _{c} \langle {0}|  \otimes \bigcirc_{s=1}^{m}\left[{\hat{\xi}^{(0)}_{s}}\circ\hat{U}_{s}\right]\left(\rho_{\rm in}\right) + \frac{1}{2} \left|1\right\rangle _{c} \langle {1}|  \otimes \bigcirc_{s=1}^{m}\left[{\hat{\xi}^{(0)}_{s}}\circ\hat{U}_{s}\right]\left(\rho_{\rm in}\right) \nonumber\\
&+\frac{1}{2} \left|0\right\rangle _{c} \langle{1}| \left( \sum_{i} \langle \varepsilon^{(0)}_{m}|i \rangle K^{(m)}_{i} U_{m} \right) \cdots \left( \sum_{i} \langle \varepsilon^{(0)}_{1}|i \rangle K^{(0)}_{i} U_{1} \right) \rho_{\rm in} \left( U_{1}^{\dagger} \sum_{i} \langle {i}| \varepsilon^{(1)}_{1} \rangle B_{i}^{\dagger}\right) \cdots \left(U_{m}^{\dagger} \sum_{i} \langle \varepsilon^{(1)}_{m}|i \rangle K^{(m)}_{i}  \right)\nonumber\\
&+\frac{1}{2} \left|1\right\rangle _{c} \langle{0}| \left( \sum_{i} \langle \varepsilon^{(1)}_{m}|i \rangle K^{(m)}_{i} U_{m} \right) \cdots \left( \sum_{i} \langle \varepsilon^{(1)}_{1}|i \rangle K^{(0)}_{i} U_{1} \right) \rho_{\rm in} \left( U_{1}^{\dagger} \sum_{i} \langle {i}| \varepsilon^{(0)}_{1} \rangle A_{i}^{\dagger}\right) \cdots \left(U_{m}^{\dagger} \sum_{i} \langle \varepsilon^{(0)}_{m}|i \rangle K^{(m)}_{i}  \right).
\label{eq:Stinespring7}
\end{split}
\end{equation}
By measuring the control state in the $X$ basis one obtains again an output state with generally enhanced fidelity with respect to the average over the possible outcomes of the measurement. The improvement depends again on the initial environmental states. In a continuum set of choices, a varying degree of advantage is found, where the maximum and the minimum advantages correspond to some discrete choices for the initial environmental states. In the worst situation, one recovers the incoherent result of Eq.~\eqref{eq:Stinespring3}.

We refer to Ref.~\cite{papercomm} for further details.

\end{document}